\documentclass[iop,twocolumn,tighten,numberedappendix,twocolappendix,revtex4]{emulateapj}

\graphicspath{{./}{Updated_Plots/}}
\usepackage{subfigure}
\usepackage{graphicx}	% Including figure files
\usepackage{amsmath}	% Advanced maths commands
\usepackage{amssymb}	% Extra maths symbols

\usepackage{threeparttable}
\usepackage{comment}
\usepackage{booktabs}

\shorttitle{COSMOS-XS: The cosmic star formation history}
\shortauthors{D. van der Vlugt et al.}

\begin{document}
%TC:ignore
\title{An ultra-deep multi-band VLA survey of the faint radio sky (COSMOS-XS): New constraints on the cosmic star formation history}

\author{D. van der Vlugt\altaffilmark{1}}
\altaffiltext{1}{Leiden Observatory, Leiden University, P.O. Box 9513, 2300 RA Leiden, the Netherlands}
\email{dvdvlugt@strw.leidenuniv.nl}

\author{J. A. Hodge\altaffilmark{1}} 

\author{H. S. B. Algera\altaffilmark{2,3}} 
\altaffiltext{2}{Hiroshima Astrophysical Science Center, Hiroshima University, 1-3-1 Kagamiyama, Higashi-Hiroshima, Hiroshima 739-8526, Japan}
\altaffiltext{3}{National Astronomical Observatory of Japan, 2-21-1 Osawa, Mitaka, Tokyo, Japan}

\author{I. Smail\altaffilmark{4}}
\altaffiltext{4}{Centre for Extragalactic Astronomy, Durham University, Department of Physics, South Road, Durham DH1 3LE, UK}

\author{S. K. Leslie\altaffilmark{1}} 

\author{J. F. Radcliffe\altaffilmark{5,6}}
\altaffiltext{5}{Jodrell Bank Centre for Astrophysics, The University of Manchester, SK11 9DL. United Kingdom}
\altaffiltext{6}{Department of Physics, University of Pretoria, Lynnwood Road, Hatfield, Pretoria 0083, South Africa}

\author{\\D. A. Riechers\altaffilmark{7}}
\altaffiltext{7}{I. Physikalisches Institut, Universit\"at zu K\"oln, Z\"ulpicher Strasse 77, D-50937 K\"oln, Germany}

\author{H. R\"{o}ttgering\altaffilmark{1}}

\def\baselinestretch{1.0}
%TC:endignore
\begin{abstract}
We make use of ultra-deep 3\,GHz Karl G. Jansky Very Large Array observations of the COSMOS field from the multi-band COSMOS-XS survey to infer radio luminosity functions (LFs) of star-forming galaxies (SFGs). Using $\sim$1300 SFGs with redshifts out to $z\sim4.6$, and fixing the faint and bright end shape of the radio LF to the local values, we find a strong redshift trend that can be fitted by pure luminosity evolution with the luminosity parameter given by $\alpha_L \propto (3.40 \pm 0.11) - (0.48 \pm 0.06)z$. We then combine the ultra-deep COSMOS-XS data-set with the shallower VLA-COSMOS $\mathrm{3\,GHz}$ large project data-set over the wider COSMOS field in order to fit for joint density+luminosity evolution, finding evidence for significant density evolution. By comparing the radio LFs to the observed far-infrared (FIR) and ultraviolet (UV) LFs, we find evidence of a significant underestimation of the UV LF by $21.6\%\, \pm \, 14.3 \, \%$ at high redshift ($3.3\,<\,z\,<\,4.6$, integrated down to $0.03\,L^{\star}_{z=3}$). We derive the cosmic star formation rate density (SFRD) by integrating the fitted radio LFs and find that the SFRD rises up to $z\,\sim\,1.8$ and then declines more rapidly than previous radio-based estimates. A direct comparison between the radio SFRD and a recent UV-based SFRD, where we integrate both LFs down to a consistent limit ($0.038\,L^{\star}_{z=3}$), reveals that the discrepancy between the radio and UV LFs translates to a significant ($\sim$1\,dex) discrepancy in the derived SFRD at $z>3$, even assuming the latest dust corrections and without accounting for optically dark sources.
\end{abstract}
%TC:ignore
\keywords{galaxies: evolution - galaxies: star formation - cosmology: observations - radio continuum: galaxies}

%%%%%%%%%%%%%%%%%%%%%%%%%%%%%%%%%%%%%%%%%%%%%%%%%%

%%%%%%%%%%%%%%%%% BODY OF PAPER %%%%%%%%%%%%%%%%%%

\section{Introduction}
\label{sec:Introduction}
Over the past two decades, impressive progress has been made in constraining the star formation rate density (SFRD) over cosmic time using a multitude of star formation rate (SFR) tracers \citep[e.g., review by][]{Madau_2014}, providing vital information for understanding galaxy evolution. There is a reasonable consensus regarding the shape of the SFRD in recent history ($z \, < \, 2$). However, above $z \, \sim \, 3$, the differences in the SFRD still encompass very different predictions from galaxy evolution models \citep[e.g.,][]{Gruppioni_2015, Henriques_2015, Lacey_2016, RowanRobinson_2016, Casey_2018, Moster_2018, Behroozi_2019}. An accurate mesurement of the evolution of the SFRD is thus vital for the understanding of galaxy evolution.

Several tracers can be used to trace the SFRD. In principle, ultraviolet (UV) light is the most direct tracer of SFR in dust free environments which originates mainly from massive stars. UV light thus directly traces young stellar populations and can be used to constrain the unobscured star formation out to very high redshifts ($z \, \simeq \, 9$; e.g., \citealt{Mclure_2013, Bouwens_2015, Bowler_2015, Finkelstein_2015, McLeod_2015, Bouwens_2016, Parsa_2016, Metha_2017, Ono_2018, Oesch_2018, Bouwens_2021}). However, UV observations need significant and uncertain corrections for dust obscuration and are unable to detect the most extreme star-forming galaxies (SFGs) in which star formation is known to be enshrouded in dust \citep[e.g.,][]{Smail_1997, lutz_2011, Riechers_2013, Casey_2014, Dudzeviciute_2020}. Therefore, knowledge on how the dust attenuation evolves with redshift is mandatory to study the redshift evolution of the SFRD, particularly as the cosmic epoch $z \, \lesssim \, 4$ may be dominated by dust obscured star formation \citep{Casey_2018, Bouwens_2020}.

Dust, heated by young massive stars, re-emits the absorbed UV light at longer wavelengths and can thus be studied in the far-infrared (FIR) or sub-millimeter (sub-mm) to trace the SFR. Current FIR observations are able to constrain the dust content and SFRD up to a redshift $z \, < \, 6$ \citep[e.g.,][]{Rodighiero_2010, Gruppioni_2013, RowanRobinson_2016, Koprowski_2017, Dudzeviciute_2020, Lim_2020}. However, the constraints beyond $z \simeq 3$ are uncertain as the measurement of the FIR LF becomes more challenging. Source confusion and blending limit the ability to detect faint objects in low resolution \emph{Herschel}/SPIRE observations at $z \, \simeq \, 3 - 4$. Such observations are thus biased towards an unrepresentative population of bright sources. In addition, these observations can be significantly contaminated by active galactic nuclei (AGN) as these sources are more numerous at high redshift \citep{Gruppioni_2013, Symeonidis_2021}.

Ground-based sub-mm/mm continuum observations of dusty galaxies can help to overcome some of the problems in FIR observations \citep[e.g.,][]{Chapman_2005, Hodge_2013, Swinbank_2014, Dunlop_2017, Dudzeviciute_2020, Zavala_2021}. In particular, ground-based interferometic arrays (e.g., ALMA) offer high-resolution observations and hence do not suffer from source blending. Sub-mm surveys are also less susceptible to AGN contamination as they are predominantly sensitive to the cool-dust in the star-forming population at high redshift \citep{Hodge_2020}. In addition, the dust-unbiased tracer [CII] was recently used in several studies conducted with ALMA to study the SFR \citep{Gruppioni_2020, Khusanova_2021, Loiacono_2021}. But even with these advantages, sub-mm observations are still impractical to carry out large surveys that would overcome cosmic variance, which can have a strong impact on any counting statistic \citep[e.g.,][]{Moster_2011, Simpson_2019, Gruppioni_2020, Loiacono_2021}, because of the small field of view. Cosmic variance in sub-mm can be overcome by combining a wide-field single dish observation with expensive interferometric follow-up observations \citep{Simpson_2020}. 

Radio continuum emission is also an end-product of the formation of the most massive stars. Synchrotron radiation originates from the shocks produced by the supernova explosions \citep[e.g.,][]{Sadler_1989, Condon_1992, Clemens_2008, Tabatabaei_2017}. Radio emission triggered by star formation is empirically found to correlate well with the far-IR (FIR) emission of SFGs: the FIR-radio correlation. Radio-SFR calibrations most often rely on this empirical FIR-radio correlation, which appears to hold across more than five magnitudes in luminosity and persists out to high redshifts \citep[e.g][]{Helou_1985, Yun_2001, Bell_2003}, albeit with ill-constrained redshift evolution \citep[e.g.,][]{Sargent_2010, Magnelli_2015, Calistro_2017, Delhaize_2017}. However, there is some discussion whether the redshift evolution can be ascribed to selection biases \citep{Sargent_2010, Algera_2020b, Smith_2021, Molnar_2021, Delvecchio_2021}. In addition, AGN activity will cause strong deviation from the local FIR-radio correlation \citep{Molnar_2018} as accreting super-massive black holes (SMBH) in AGN also accelerate the electrons that produce synchrotron emission.

Radio emission is a tracer of star formation which is, unlike UV, not attenuated by dust. In contrast to FIR observations, radio observations have a high spatial resolution and can cover larger areas of the sky than interferometric sub-mm observations with high angular resolution. Radio observations in the synchrotron regime ($\sim \, \mathrm{GHz}$ frequencies) therefore offer a unique opportunity to study the star formation history of the Universe \citep[e.g.,][]{Seymour_2008, Smolcic_2009_a, Jarvis_2015, Rivera_2017, Novak_2017, Leslie_2020, Matthews_2021}.

Besides being used to calibrate radio luminosity as a tracer of SFR, the FIR-radio correlation is also often used for the classification of galaxies. A sample used for constraining the SFRD should only consist of sources with radio emission originating from star formation. Therefore one would ideally quantify the emission coming from SF and AGN in all sources. It is, however, easier to simply remove sources that show an excess in radio emission compared to what is expected from the FIR-radio correlation \citep[radio-excess AGN, e.g.,][]{DelMoro_2013, Delvecchio_2017, Algera_2020}. Radio-loud AGN are easily removed by this method, as these sources show a large offset from the FIR-radio correlation. A major uncertainty is the ability to distinguish composite sources, which emit low-level AGN emission, from SFGs \citep[e.g.,][]{Padovani_2009, Bonzini_2013}. 

Radio studies to-date have observed radio LFs but struggled to reach the knee of the LF ($L_{\star}$) at $z \, > \, 1$. Because these studies are most sensitive to the SFG population above the knee, the density and luminosity evolution parameters may become degenerate preventing a precise estimate of the knee location. The radio studies from \citet{Smolcic_2009_a} and \citet{Novak_2017} thus assumed pure luminosity evolution rather than luminosity and density evolution \citep{Condon_1984} in order to fit the radio LF out to $z \, \sim \, 5$. Recently, \citet{Malefahlo_2022} used a Bayesian approach to reach below the 5$\sigma$ detection limit of \citet{Novak_2017} but only constrained pure luminosity evolution. \citet{Enia_2022} used $1.4 \, \text{GHz}$-selected sample to constrain the evolution of the radio LF up to $z\sim3.5$ by fitting a modified Schechter function (equivalent to fitting both luminosity and density evolution).

We have taken advantage of the upgraded capabilities of the Karl G. Jansky Very Large Array (VLA) to conduct an ultra-deep, matched-resolution survey in both X- and S-band ($\mathrm{10 \, GHz}$ and $\mathrm{3 \, GHz}$, \citealp{vanderVlugt_2021}; hereafter Paper I). In \citet{Algera_2020} (hereafter Paper II), the radio catalogs obtained from Paper I were matched with the rich multi-wavelength data available in the COSMOS field \citep{Scoville_2007} to distinguish between AGN and SFG. In this work, we use the 3 GHz star-forming sample to constrain the faint end of the LF with the faintest SFGs that can currently be probed at high redshift with radio surveys. We also leverage the combined power of the COSMOS-XS survey and the $\mathrm{3 \, GHz}$ VLA-COSMOS Large Project \citep{Smolcic_2017}, which covers a larger 2 deg$^2$ area to a shallower depth of $\sigma \, \sim 2.3 \, \mathrm{\mu Jy \, beam^{-1}}$, in order to increase our dynamic range and constrain the form and evolution of the LF -- and thus ultimately the dust-unbiased SFRD -- out to a redshift of $z \, \sim \, 4.6$.

This paper is organized as follows: in Section~\ref{sec:data and sample selection} we summarize the data and selection methods. In Section~\ref{sec:derivation of the LF}, we present the method of constraining the LFs with redshift. We compare our derived radio LFs to the literature in Section~\ref{sec:LF}. In Section~\ref{sec:Discussion} we discuss possible biases that need to be taken into account in the derivation of the LF. In Section~\ref{sec:CSFH} we use the most appropriate LF to calculate the evolution of the cosmic star formation rate density. Finally, Section~\ref{sec:Conclusions} summarizes and concludes this work. Throughout this paper, the spectral index, $\alpha$, is defined as $S_{\nu} \, \propto \, \nu^{\alpha}$, where $S_{\nu}$ is the source flux density, and $\nu$ is the observing frequency. We use a $\Lambda$CDM cosmology with parameters $H_{0}=70 \text{km s}^{-1} \text{Mpc}^{-1}$, $\Omega_m =0.3$, $\Omega_{\Lambda}=0.7$ \citep{Bennett_2013}. We assume a radio spectral index of $-0.7$ unless otherwise stated. We assume the \citet{Chabrier_2003} initial mass function (IMF) to calculate SFRs.

\section{Data and sample selection}
\label{sec:data and sample selection}
\subsection{Radio data}
\label{sec:radio}
The COSMOS-XS survey consists of two overlapping ultra-deep single VLA pointings in the COSMOS field at 3 and $\mathrm{10 \, GHz}$ of, respectively, $\simeq \, 90$ and $\simeq \, 100$ h of observation time. Further details on these observations can be found in Paper I but a short summary of the survey follows. The 3 and $\mathrm{10 \, GHz}$ observations reach a depth of $0.53 \, \mathrm{\mu Jy\,beam^{-1}}$ and $0.41 \, \mathrm{\mu Jy\,beam^{-1}}$ at their respective pointing centres. Both frequencies have a near-equal resolution of $\sim \, 2.0''$ ($2.14'' \, \times \, 1.81''$ at $\mathrm{3 \, GHz}$ and $2.33''\times2.01''$ at  $\mathrm{10 \, GHz}$) which is large enough to avoid resolving out faint SF sources.

Details on how the source extraction was performed in both images can be found in Paper I and Paper II. Sources were identified by {\sc{PyBDSF}} \citep{Mohan_2015} in the $\mathrm{3 \, GHz}$ image and resulted in identification of 1540 radio sources.

\subsection{Counterparts}
\label{sec:counterparts}
The counterpart matching method to cross-match the radio sources is fully described in Paper II and briefly summarized below. Counterparts of radio sources were found using a symmetric nearest neighbor algorithm. Counterparts were assigned within a given matching radius. This matching radius was determined through cross-matching with mock versions of the appropriate catalog containing the same sources with randomized sky coordinates. 

\subsubsection{Radio counterparts}
\label{sec:radio counterparts}
The 10 and $\mathrm{3 \, GHz}$ data were cross-matched using a matching radius of $0\farcs9$, which yields 91 matches with a false match rate (FMR) of $\lesssim$ 0.7\%. The radio sample was also matched to the VLA COSMOS $\mathrm{1.4 \, GHz}$ catalog \citep{Schinnerer_2007} using a matching radius of $1\farcs2$ (FMR $\lesssim$ 0.1\%). This generated 185 matches, with 12 sources being detected at all three frequencies (1.4, 3 and $\mathrm{10 \, GHz}$).

\subsubsection{Optical and near-infrared counterparts}
\label{sec:optical_NIR}
As described in Paper II, the radio observations were complemented with near-UV to FIR-data from various multi-wavelength catalogs: i) the Super-deblended mid- to far-infrared catalog \citep{jin_2018} containing photometry ranging from IRAC $\mathrm{3.6 \, \mu m}$ to $\mathrm{20 \, cm}$ ($\mathrm{1.4 \, GHz}$) radio observations. Blended galaxies in low-resolution FIR images are partly disentangled using priors on sources positions from high resolution images and point spread function fitting; ii) the $z^{++}YJHK_s$-selected catalog compiled by \citet{laigle_2016} (hereafter COSMOS2015) and iii) the $i$-band selected catalog by \citet{capak_2007}. 

For each source, we searched for a counterpart in the Super-deblended catalog with a matching radius of $0\farcs9$. To complement the Super-deblended matches with optical and near-IR photometry, we also matched with the COSMOS2015 catalog followed by the $i$-band catalog with matching radii of $0\farcs7$ and $0\farcs9$, respectively. Sources not in the Super-deblended catalog were matched with the COSMOS2015 catalog using a matching radius of $0\farcs7$. Sources which still lacked a counterpart were matched with the the $i$-band selected catalog with a matching radius of $0\farcs9$. A flowchart of the matching process can be found in Fig. 3 of Paper II. Overall, 70 sources ($4.5\%$) did not have any optical and NIR counterparts. These sources are not included in the subsequent analysis. An analysis on the properties of these sources can be found in Section 5.3 of Paper II. 1470 sources could be matched to a counterpart in at least one multi-wavelength catalog. Based on the matching radii used, we expect a false match rate of $\lesssim$ 3\%, corresponding to $\sim$ 40 sources.

Spectroscopic redshifts were obtained from the COSMOS master catalog (M. Salvato et al.; available internally in
the COSMOS collaboration). A spectroscopic redshift with a quality factor $Q_f > 3$ was available for 584 radio sources. If a source could be matched within $1\farcs4$ to an X-ray source, the photometric redshift from the \emph{Chandra} X-ray catalog was used \citep{Civano_2016}. Otherwise photometric redshifts from the Super-deblended catalog were used. If a Super-deblended redshift is unavailable, we instead used the photometric redshift from COSMOS2015 or the $i$-band selected catalog, in that order. 1437 sources have a counterpart and a reliable redshift. 33 sources have no redshift information and are removed from the sample. Out to $z \, \sim \, 1$, nearly two-thirds of our redshifts are spectroscopic. This fraction drops dramatically toward higher redshift (Fig. 4 in Paper II shows the distribution of the photometric and spectroscopic redshift). 

The accuracy of photometric redshifts is estimated by comparing the photometric and spectroscopic redshift of the 584 sources with a spectroscopic redshift. The median of this comparison is $\sigma(z) = | z_\text{spec} - z_\text{phot} | / (1 + z_\text{spec}) = 0.008$ at all redshifts. The catastrophic failure rate ($\sigma(z) > 0.15$) is found to be 4.8$\%$. 

\subsection{Sample selection}
\label{sec:sample selection}
To estimate the LF of SFGs, we need to select sources with their radio emission originating solely from star formation. As radio emission can also originate from accreting black holes, we thus need to remove sources that have their radio emission dominated by an AGN. We use the FIR-radio correlation to select the SFGs, where sources with their radio emission dominated by an AGN will be offset from the FIR-radio correlation. The method to remove AGN from the sample is fully described in Paper II and briefly summarized below.

The FIR-radio correlation is defined as the logarithmic ratio of a galaxy's total FIR-luminosity $L_\text{FIR}$, measured between (rest-frame) $8$--$1000 \, \mathrm{\mu m}$, and its monochromatic radio luminosity at rest-frame $\mathrm{1.4 \, GHz}$ ($L_{1.4 \, \text{GHz}}$, following e.g., \citealt{Bell_2003,Magnelli_2015,Delhaize_2017,Calistro_2017}):

\begin{align}
	q_\text{TIR} = \log_{10}\left(\frac{L_\text{FIR}}{3.75\times 10^{12} \text{ W}}\right) - \log_{10} \left(\frac{L_{1.4 \, \text{GHz}}}{\text{W Hz}^{-1}}\right) \, .
	\label{eq:qtir}
\end{align}

The factor $3.75\times 10^{12}$ is the central frequency of the total-FIR continuum ($8-1000 \, \mathrm{\mu m}$) in Hz and serves as the normalization. Each galaxy in the sample is fitted using the SED fitting code {\sc{magphys}} \citep{Cunha_2008, Cunha_2015}, and the total FIR-luminosites are obtained from the best-fitted SEDs.  

Rest-frame $\mathrm{1.4 \, GHz}$ luminosities are determined in Paper II using the measured spectral index for the required K-corrections if available. When only a single radio flux is available, a spectral index of $\alpha=-0.7$ is assumed instead. The luminosities are then calculated through 

\begin{align}
    L_{1.4 \,\text{GHz}} = \frac{4\pi D_L^2 }{(1 + z)^{1+\alpha}} \left( \frac{1.4 \, \text{GHz}}{3 \,\text{ GHz}}\right)^\alpha S_{3 \, \text{GHz}}  \, .
    \label{eq:Luminosity}
\end{align}

\begin{figure}
\centering
\includegraphics[width=\columnwidth]{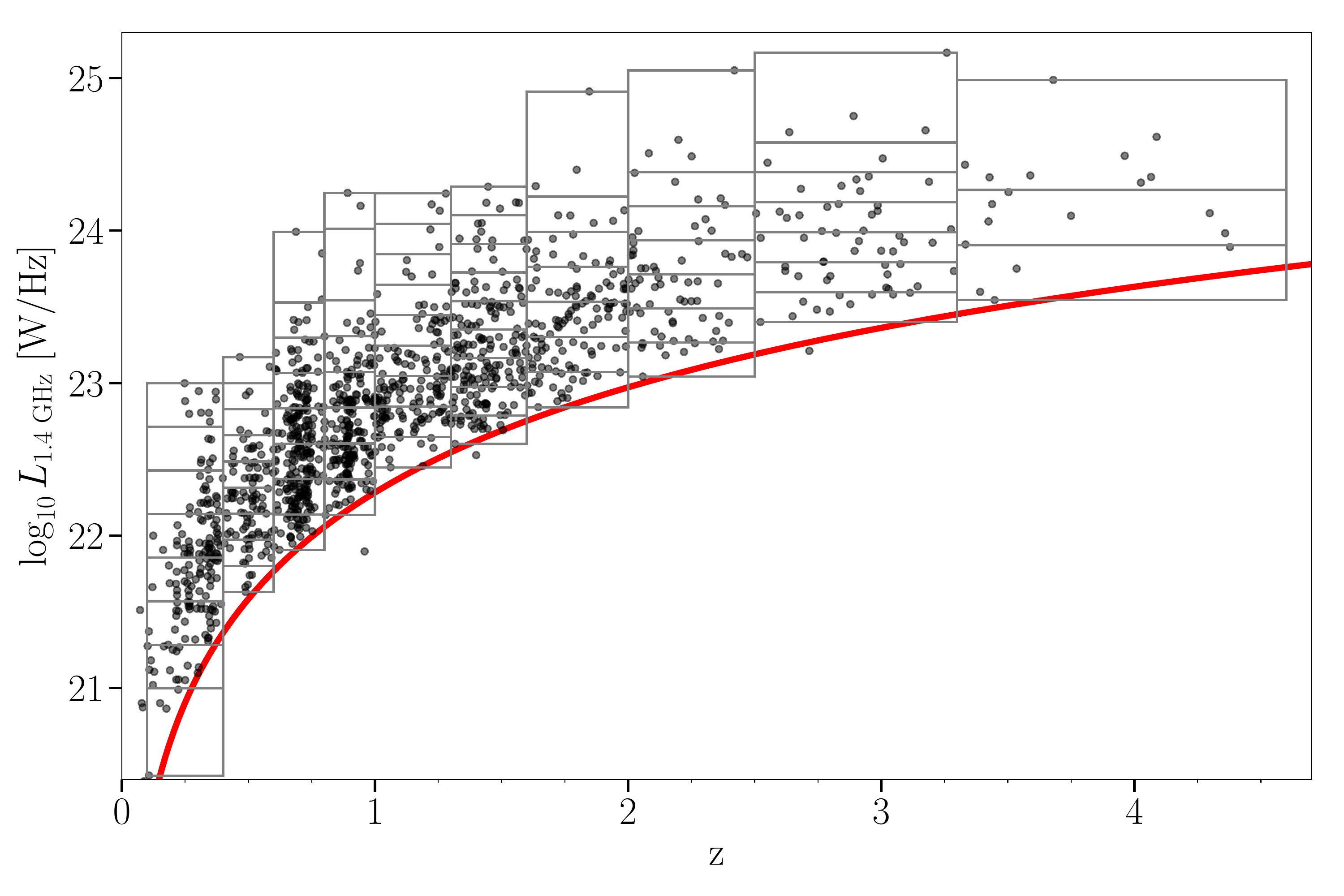}
\caption{The coverage in the COSMOS-XS survey of the luminosity-redshift plane. The gray solid lines depict the redshift and luminosity bins used in the LF analysis. The red line indicates the detection limit of 5$\sigma$, where $\sigma \, = \, 0.53 \, \mathrm{\mu Jy\,beam^{-1}}$ at $\mathrm{3 \, GHz}$ and a fixed spectral index of $\alpha \, = \, -0.7$ is assumed. Sources that fall below the detection limit exist within a region with a low local r.m.s. or have a shallower spectral index.}
\label{fig:covarage}
\end{figure}

Here $D_L$ is the luminosity distance at redshift $z$ and $S_\text{3 GHz}$ is the observed flux density at $\mathrm{3\,GHz}$. The luminosities calculated as a function of redshift are shown in Fig.~\ref{fig:covarage}.

In order to quantify the FIR-radio correlation and find outliers, we adopt the redshift and mass-dependent $q_\text{TIR}(\text{M}_{\star}, z)$ determined by \citet{Delvecchio_2021}. In order to use this $q_\text{TIR}(\text{M}_{\star}, z)$, we need to have a mass for the sample. We used the mass given by the COSMOS2015 catalog for the sources that could be matched with this catalog. For sources without a mass, we used the derived mean mass per redshift bin, ranging from $10^{10.18} \, \text{M}_{\odot}$ to $10^{10.70} \, \text{M}_{\odot}$. When the $q_\text{TIR}(\text{M}_{\star}, z)$ of a source deviates more than 3$\sigma$ from the relation from \citet{Delvecchio_2021}, it is defined as a radio-excess source, i.e.,
\begin{align}
	q_\text{TIR}(\text{M}_{\star}, z) \, < \, & 2.646 \times (1+z)^{-0.023} \notag \\ 
	&- 0.148 \times (\log_{10} \frac{\text{M}_{\star}}{\text{M}_{\odot}}-10) - (3 \times \sigma)\,.
	\label{eq:threshold_q}
\end{align}
where $\sigma=0.22$. Such a cut identifies 130 radio-excess sources in total. Recent studies suggest a different evolution, including even a non-evolving $q_\text{TIR}(z)$, may be more appropriate \citep{Molnar_2018, Smith_2021, Molnar_2021} and we test the effect of such an assumption in Section~\ref{sec:contamination}. 

An additional criterion to identify radio-excess sources is established in Paper II, as only 50\% of our sample is detected in the far-infrared at $\geq 3\sigma$. For \emph{Herschel}-undetected sources, a conservative FIR-luminosity at the $2\sigma$ level is calculated, assuming the FIR-radio correlation as determined by \citet{Delhaize_2017}. The calculated FIR-luminosity is compared with the empirically determined detection threshold of \emph{Herschel}. Sources with a calculated FIR-luminosity above the threshold are then identified as ``inverse radio-excess'' AGN, as they should have been observed with \emph{Herschel} if their radio emission originated solely from star formation. The additional criterion enables us to identify 62 ``inverse radio-excess'' sources, of which only 17 were not already identified by the threshold in Eq.~\ref{eq:threshold_q}. We thus find 147 radio-excess sources in total, leaving a total star-forming galaxy sample consisting of 1290 radio sources. The redshift distribution of the sample is shown in Fig.~\ref{fig:covarage}\footnote{Sources with $z \, > \, 4.6$ like AzTEC-3 are not included because there are too few to give meaningful constraints on the LF.}.

\subsection{VLA-COSMOS $\mathrm{3 \, GHz}$ Large Project}
\label{sec:VLA-COSMOS 3 GHz Large Project}
\citet{Novak_2017} studied the SFRD using the VLA-COSMOS $\mathrm{3 \, GHz}$ Large Project data. This project provided data over the entire $2 \, \rm{deg}^2$ COSMOS field allowing for the detection of typical SFGs ($\text{SFR} \, \lesssim \, 100 \text{M}_{\odot} \rm{yr}^{-1}$) out to $z \, \sim \, 1.5$. The COSMOS-XS survey is $\sim \, 5$ times deeper than the VLA-COSMOS $\mathrm{3 \, GHz}$ Large Project and when we combine the VLA-COSMOS $\mathrm{3 \, GHz}$ Large Project data set over the whole field with our deep COSMOS-XS pointing, we obtain a survey ``wedding-cake'' with sufficient dynamic range to enable a meaningful measurement of the form and evolution of the LF. 

The radio-excess diagnostics by \citet{Novak_2017} and Paper II are similar, the overall number of radio-excess sources identified is very similar and the overlap between these two samples is substantial. However, there are a few differences that are addressed in the appendix of Paper II and will be summarized below. Firstly, Paper II used the improved FIR photometry from the Super-deblended catalog \citep{jin_2018} with a more detailed deblending technique and photometry up to 1.2 mm. Secondly, \citet{Novak_2017} used the method from \citet{Delvecchio_2017}, which uses $\log_{10}(L_{1.4 \, \text{GHz}}/\text{SFR}_{\text{IR}})$ to separate radio-excess sources from SF sources. The $\text{SFR}_{\text{IR}}$ is correlated with the $L_{1.4 \, \text{GHz}}$ because the $\text{SFR}_{\text{IR}}$ is calculated with SED fitting from the $L_{\text{IR}}$ and owing to the FIR-radio correlation. Therefore $\log_{10}(L_{1.4 \, \text{GHz}}/\text{SFR}_{\text{IR}})$ is equal to $q_\text{TIR}$ up to a constant.

\citet{Delvecchio_2017} then define radio-excess sources when the $\log_{10}(L_{1.4 \, \text{GHz}}/\text{SFR}_{\text{IR}})$ of a source deviates by more than 3$\sigma$ from the peak of the distribution as a function of redshift. Although this results in a small difference in the total number of radio-excess sources identified in both surveys, we decided to use a consistent criterion for radio-excess sources. We used our threshold which is the $q_\text{TIR}(\text{M}_{\star}, z)$ determined by \citet{Delvecchio_2021} minus 3$\sigma$, as described in Section~\ref{sec:sample selection}, to select SFGs using the $L_{1.4 \, \text{GHz}}$, $L_{\text{IR,SF}}$ and $\text{M}_{\star}$ from the $\mathrm{3 \, GHz}$ radio catalog \citep{Delvecchio_2017}. This results in a data-set with 5822 star-forming sources.

\section{Analyses}
\label{sec:derivation of the LF}
The LF describes the volume density of galaxies as a function of their intrinsic luminosity. We first discuss the method of determining the rest-frame $1.4 \, \text{GHz}$ LF from the COSMOS-XS survey. We then show how the data can be fitted with a modified-Schechter function assuming different ``fixed parameters''. Finally, we will consider the addition of the VLA-COSMOS $\mathrm{3 \, GHz}$ Large Project data to constrain the LF over a larger dynamic range.

\subsection{Estimating the LF}
\label{sec:estimating LF}
The radio LFs are derived using the $1/V_{\text{max}}$ method \citep{Schmidt_1968}. In each redshift bin, we have computed the co-moving volume available to each source in that bin, defined as $V_{\text{max}} = V_{z_{\text{max}}} - V_{z_{\text{zmin}}}$, where $z_{\text{zmin}}$ is the lower boundary of the redshift bin and $z_{\text{max}}$ is the maximum redshift at which the source could be seen given the flux density limit of the sample. The maximum value of $z_{\text{max}}$ corresponds to the upper limit of the redshift bin. For each luminosity bin, the LF is then given by:
\begin{align}
	\Phi(L,z) =  \frac{1}{\Delta \log_{10} L} \sum_{i} \frac{1}{\frac{\Omega}{4 \pi} \times V_{\text{max}, i} \times w_{i}(z)} \, , 
	\label{eq:LF_vmax}
\end{align}
where $V_{\text{max}}$ is the co-moving volume over which the ith galaxy could be observed, $\Omega$ is the observed area of $350 \, \mathrm{arcmin^2}$, $\Delta \log_{10} L$ is the size of the luminosity bin, and $w_i$ is the completeness correction factor of the $i$th galaxy. The parameter $w_i$ takes into account the observed area and sensitivity limit and mitigates completeness issues
\begin{align}
	 w_{i}(z) =& f_{\text{flux}}(S_{\nu_i}(z)) \times f_{\text{res}}(S_{\nu_i}(z)) \notag \\
	 &\times f_{\text{ctrpt}}(S_{\nu_i}(z)) \times o_{i}(z) \, ,
	\label{eq:correction}
\end{align}
where $f_\text{flux}$ is the flux density completeness of our radio catalog, $f_\text{res}$ is a correction for resolution bias and $f_\text{ctrpt}$ is the fraction of sources, which we have obtained reliable non-radio counterparts, for $i$th galaxy with flux density $S_{\nu_i}$. $o_{i}(z)$ is the overdensity factor derived as discussed in Appendix~\ref{sec:Cosmic variance}.

The completeness ($f_{\text{flux}}$) of the COSMOS-XS radio catalog is shown and tabulated in Paper I. The completeness is based on Monte Carlo simulations where mock sources were inserted and extracted from the image. These simulations take into account the effect of the primary beam and the non-uniform r.m.s.. To correct for the resolution bias, we take the values tabulated in Paper I. These resolution bias corrections ($f_\text{res}$) were calculated using the analytic method as used in \cite{Prandoni_2001} assuming a radio size for faint sources. As discussed in Paper II, 6.7\% of our radio sources were not assigned a counterpart. To correct for this incompleteness, we use the counterpart completeness ($f_{\text{ctrpt}}$) of the COSMOS-XS radio catalog which is shown as a function of flux density in Fig 5 from Paper II. The completeness in all bins is upwards of 90\%, and no trend with radio flux density can be seen, indicating that the association of counterparts to our radio sources is not limited by the depth of the multi-wavelength photometry.

The error of the LF in each redshift and luminosity bin is calculated as in \citet{Marshall_1985}:

\begin{align}
	\sigma_{\Phi}(L,z) \, = &\, \frac{1}{\Delta \log_{10} L} \notag \\ &\sqrt{\sum_{i} \left(\frac{1}{\frac{\Omega}{4 \pi} \times V_{\text{max}, i} \times w_{i}(z)}\right)^{2}} \, . 
	\label{eq:LF_error}
\end{align}

If there are $\leq \, 10$ sources in a luminosity bin, the error is calculated using the tabulated values from \citet{Gehrels_1986}; we take the tabulated upper and lower 84\% confidence interval as $\sigma_{\text{N}}$ and calculate the upper and lower error on the LF as $\sigma_{\Phi}(L,z) \, = \, \Phi(L,z) \, \times \, \sigma_{\text{N}}$. We take the average value of the upper and lower error as the final error on the sparsely populated bins. 

\def\baselinestretch{1.1}
\begin{deluxetable*}{ccccc}
	\tablecolumns{4}
	\tabletypesize{\footnotesize}
	\tablewidth{\textwidth}
	\tablecaption{Parameter values describing the pure luminosity evolution fit and the density+luminosity evolution fits.}
	\tablehead{
		 &
		\colhead{COSMOS-XS\tablenotemark{a}} &
		\multicolumn{3}{c}{COSMOS-XS + VLA-COSMOS 3 GHz\tablenotemark{b}} \\
		\cmidrule{3-5}
		\colhead{Redshift range} &
		\colhead{$\alpha_{\text{L}}$} &
		\colhead{$\alpha_{\text{L}}$} &
		\colhead{[$\alpha_{\text{L}}$} \, &
		\colhead{$\alpha_{\text{D}}$]} 
	}
	
	\startdata
	 
	\vspace{-1.0ex}\\
    0.1 $ < \, z \, < $ 0.4  &  3.26 $^{+  0.51 }_{-  0.52 }$ &  1.53 $^{+  0.17 }_{-  0.18 }$ & [ 4.36 $^{+  0.36 }_{-  0.35 }$ &  $-2.41$ $^{+  0.28 }_{-  0.28 }$ ]  \\
    0.4 $ < \, z \, < $ 0.6  &  2.73 $^{+  0.31 }_{-  0.32 }$ &  2.39 $^{+  0.09 }_{-  0.09 }$ & [ 3.27 $^{+  0.27 }_{-  0.28 }$ &  $-1.0$ $^{+  0.29 }_{-  0.29 }$ ]  \\
    0.6 $ < \, z \, < $ 0.8  &  3.17 $^{+  0.14 }_{-  0.15 }$ &  2.78 $^{+  0.05 }_{-  0.05 }$ & [ 2.46 $^{+  0.21 }_{-  0.21 }$ &  $0.45$ $^{+  0.29 }_{-  0.29 }$ ]  \\
    0.8 $ < \, z \, < $ 1.0  &  3.2 $^{+  0.13 }_{-  0.13 }$ &  3.13 $^{+  0.04 }_{-  0.04 }$ & [ 3.17 $^{+  0.2 }_{-  0.2 }$ &  $-0.05$ $^{+  0.26 }_{-  0.25 }$ ]  \\
    1.0 $ < \, z \, < $ 1.3  &  2.86 $^{+  0.1 }_{-  0.11 }$ &  2.75 $^{+  0.03 }_{-  0.03 }$ & [ 3.19 $^{+  0.16 }_{-  0.17 }$ &  $-0.64$ $^{+  0.24 }_{-  0.22 }$ ]  \\
    1.3 $ < \, z \, < $ 1.6  &  2.91 $^{+  0.08 }_{-  0.08 }$ &  2.68 $^{+  0.03 }_{-  0.03 }$ & [ 2.59 $^{+  0.15 }_{-  0.14 }$ &  $0.16$ $^{+  0.24 }_{-  0.24 }$ ]  \\
    1.6 $ < \, z \, < $ 2.0  &  2.52 $^{+  0.07 }_{-  0.07 }$ &  2.63 $^{+  0.02 }_{-  0.02 }$ & [ 2.87 $^{+  0.11 }_{-  0.11 }$ &  $-0.4$ $^{+  0.18 }_{-  0.18 }$ ]  \\
    2.0 $ < \, z \, < $ 2.5  &  2.27 $^{+  0.07 }_{-  0.08 }$ &  2.48 $^{+  0.02 }_{-  0.02 }$ & [ 2.99 $^{+  0.13 }_{-  0.12 }$ &  $-0.87$ $^{+  0.2 }_{-  0.2 }$ ]  \\
    2.5 $ < \, z \, < $ 3.3  &  1.99 $^{+  0.06 }_{-  0.07 }$ &  2.25 $^{+  0.02 }_{-  0.02 }$ & [ 2.96 $^{+  0.12 }_{-  0.12 }$ &  $-1.24$ $^{+  0.2 }_{-  0.2 }$ ]  \\
    3.3 $ < \, z \, < $ 4.6  &  1.63 $^{+  0.1 }_{-  0.15 }$ &  1.83 $^{+  0.03 }_{-  0.04 }$ & [ 2.76 $^{+  0.21 }_{-  0.2 }$ &  $-1.77$ $^{+  0.35 }_{-  0.34 }$ ]  \\
	\enddata
    
    \tablenotetext{a}{Parameter value describing the pure luminosity evolution fit to the	COSMOS-XS data.}
    \tablenotetext{b}{Parameter values describing the pure luminosity evolution fit (third column) and the density+luminosity evolution fits (right two columns) to the combined COSMOS-XS + VLA-COSMOS 3 GHz data-sets. The parameters $\alpha_{\text{L}}$ and $\alpha_{\text{D}}$ shown in the table within brackets are fitted simultaneously.}
	\label{tab:parameters_both}
\end{deluxetable*}

\subsection{Constraining the LF}
\label{sec:evolving the LF}
\begin{figure}
\centering
\includegraphics[width=\columnwidth]{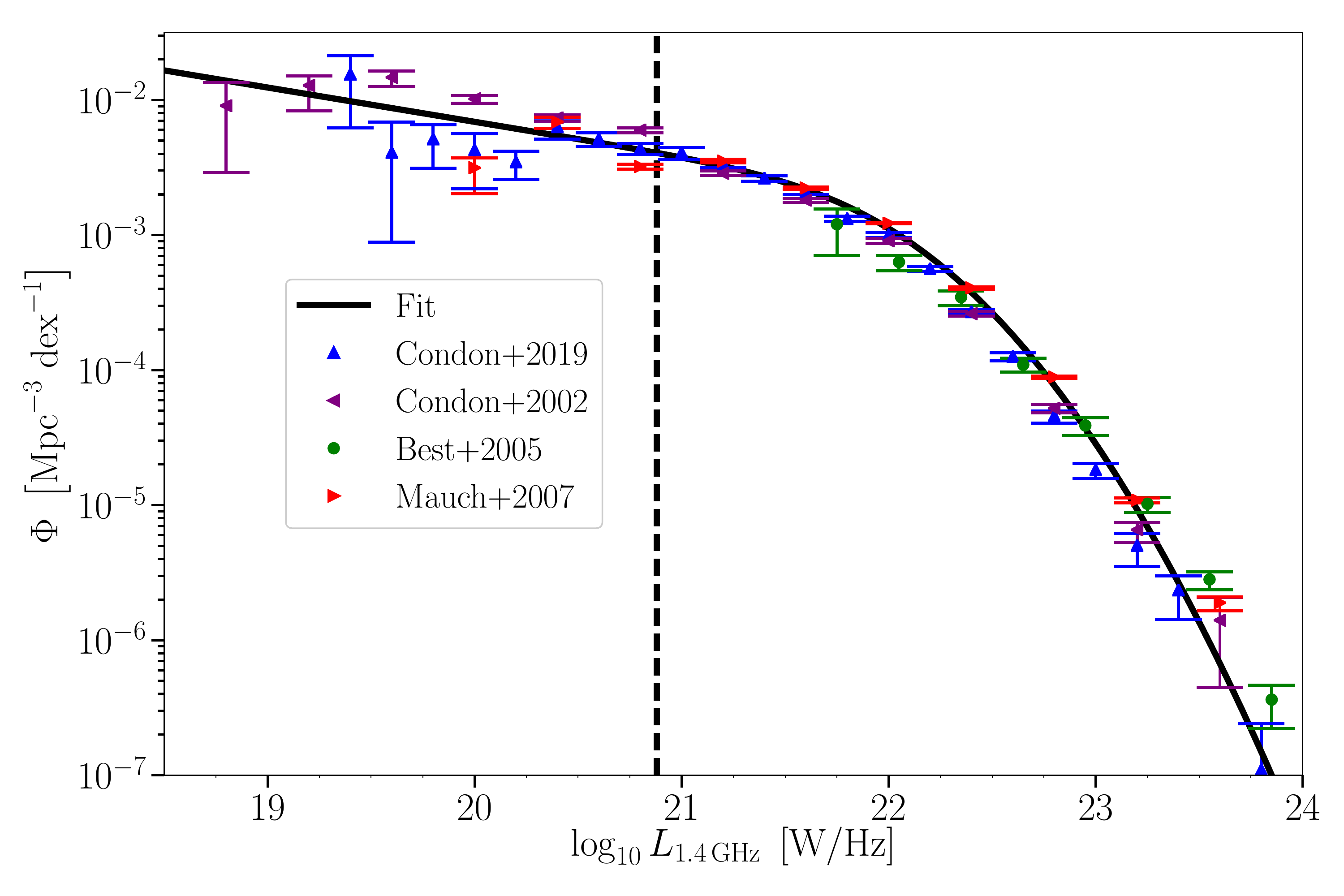}
\caption{Local radio LF of SF galaxies from several surveys with different observed areas and sensitivities. Our modified-Schechter function fit to the combined data is shown with the solid line. The dashed line indicates the depth of the COSMOS-XS survey at $0.1 \, < \, z \, < \, 0.4$.}
\label{fig:Local_LF}
\end{figure}

In order to study the evolution of the radio LF, we derive a parametric estimate of the LF at different redshifts. We assume a modified-Schechter function \citep[e.g][]{Saunders_1990, Smolcic_2009_a, Gruppioni_2013} for the shape of the LF:

\begin{align}
	\Phi_{0}(L)\, \mathrm{d}(\log_{10} L) &\, = \,   \Phi_{\star}\left(\frac{L}{L_{\star}}\right)^{1-\alpha}\notag\\ &\exp\left[-\frac{1}{2\sigma^{2}}\log^{2}(1+\frac{L}{L_{\star}})\right] \, \mathrm{d}(\log_{10} L) \, .
	\label{eq:local_LF}
\end{align}

This function behaves as a power-law for $L \, < \, L_{\star}$ and as a Gaussian in $\log_{10} L$ for $L \, > \, L_{\star}$. Four parameters are used to describe the shape of the LF: $L_{\star}$ describes the position of the turnover of the distribution, $\Phi_{\star}$ is used for the normalization and $\alpha$ and $\sigma$ are used to fit, respectively, the faint and bright end of the distribution. Following previous work \citep[e.g.,][]{Novak_2017}, the values of $\alpha$ and $\sigma$ will be frozen at the values found for the local LF. In reality, $\alpha$ and $\sigma$ may both change with redshift. 

To find the parameters of the local LF, we used the Markov chain Monte Carlo (MCMC) algorithm, available in the Python package {\sc{emcee}} \citep{Foreman_2013} to fit a modified-Schechter function to data of the local SFGs from \citet{Condon_2002, Best_2005, Mauch_2007, Condon_2019}. The fit is shown in Fig.~\ref{fig:Local_LF}. The obtained best fit parameters are: $L_{\star}  \, =  \, 2.93 \, _{-0.20}^{+0.21} \, \times \, 10^{21} \, \text{W}\text{Hz}^{-1}$, $\Phi_{\star} \, =  \, 2.93 \, _{-0.11}^{+0.10} \, \times \, 10^{-3} \, \text{Mpc}^{-3}\text{dex}^{-1}$, $\alpha  \, =  \, 1.25 \, _{-0.02}^{+0.01}$ and $\sigma  \, =  \, 0.57 \, _{-0.01}^{+0.01}$. These values lie close to the values assumed in the studies from \citet{Gruppioni_2013} and \citet{Novak_2017}.

\subsection{COSMOS-XS: Pure luminosity evolution}
\label{sec:pure luminosity evolution}
When we fit the LF to the COSMOS-XS data, we only assume the position of the turnover ($L_{\star}$, characteristic luminosity) to change with redshift. As we are not able to constrain both $L_{\star}$ and $\Phi_{\star}$ for the higher redshift bins ($z \, > \, 0.4$), we choose to keep $\Phi_{\star}$ at the local LF value. In reality, $\Phi_{\star}$ may also change with redshift. We assume the shape of the LF to remain unchanged. This pure luminosity evolution can be expressed as
\begin{align}
	\Phi(L, z, \alpha_L) \, = \, \Phi_{0}\left(\frac{L}{(1+z)^{\alpha_L}}\right) \, ,
	\label{eq:LF_L}
\end{align}
where $\alpha_L$ corresponds to the pure evolution parameter and $\Phi_{0}$ is given in Eq.~\ref{eq:local_LF}.

\begin{figure*}
\centering
\includegraphics[width=0.90\textwidth]{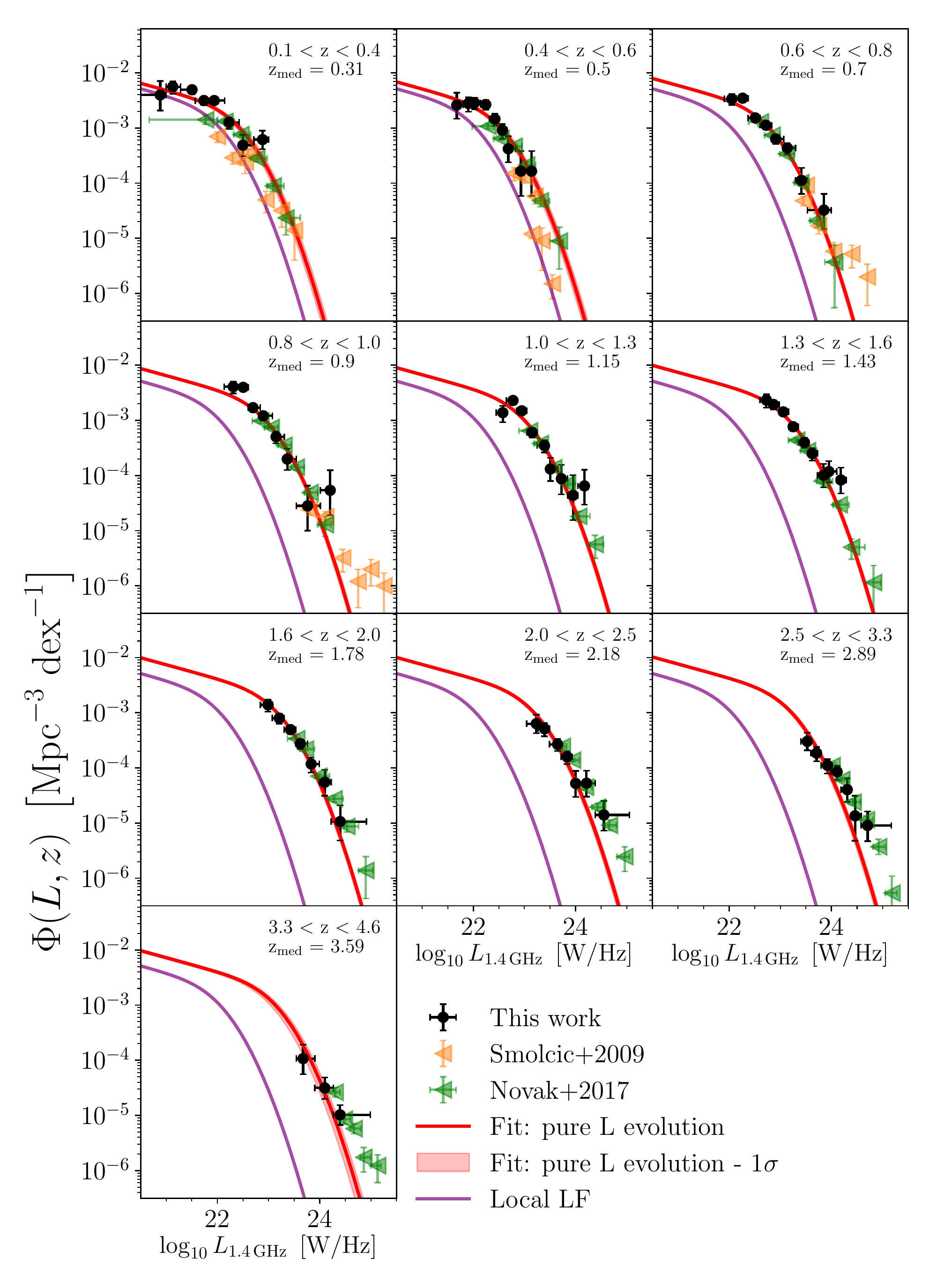}
\caption{Radio LFs of SFGs in COSMOS-XS in different redshift bins. The best-fit pure luminosity function in each redshift bin is shown by the red lines, and the shaded area shows the 1$\sigma$ confidence interval. The local radio LF is shown for reference as a solid purple line. We compare our data with the radio LFs from \citet{Smolcic_2009_a} and \citet{Novak_2017}. The redshift range and median redshift are given in each panel.}
\label{fig:radio_LF}
\end{figure*}

The range of luminosities and redshifts for which the LFs were calculated were determined from the coverage of the luminosity-redshift plane shown in Fig.~\ref{fig:covarage}. All sources are distributed into equally spaced luminosity bins spanning the observed luminosity range. Bins which contain fewer than two sources are merged with the lower L consecutive bin. The gray solid lines in Fig.~\ref{fig:covarage} show the redshift and luminosity bins used. The LFs calculated with the $V_{\text{max}}$ method are shown in Fig.~\ref{fig:radio_LF} and tabulated in Table~\ref{tab:LF_values} in Appendix~\ref{sec:LF of SFGs}. As noted in Section~\ref{sec:estimating LF}, the LFs are calculated using the $\mathrm{1.4 \, GHz}$ rest-frame luminosity for easier comparison with previous studies. The black circles show the median luminosity of all sources in the corresponding luminosity bin. The horizontal error bars show the width of the bin. The vertical errors correspond to the errors calculated using Eq.~\ref{eq:LF_error}. The data points were fitted with the analytical form from Eq.~\ref{eq:LF_L} using the MCMC algorithm assuming flat priors \footnote{$\alpha_L \in [1.0, 7.0]$ and $\alpha_D \in [-7.0, 7.0]$}. The redshift used in this expression is the median redshift of all the sources in the redshift bin. This value is given in the panels of Fig.~\ref{fig:radio_LF}. The best-fit values for $\alpha_L$ are tabulated in Table~\ref{tab:parameters_both} and the best-fit pure luminosity evolved function is shown with the red line in Fig.~\ref{fig:radio_LF}. Fig.~\ref{fig:par_ev} shows $\alpha_L$ as a function of redshift. We find that $\alpha_L$ remains roughly constant at $z \, < \, 1.8$, thereafter $\alpha_L$ decreases with $z$.

\subsection{COSMOS-XS + VLA-COSMOS 3 GHz samples: Luminosity and density evolution}
\label{sec:luminosity and density evolution}
Up until now, we have been considering pure luminosity evolution as we lacked sensitivity to constrain both the luminosity and density evolution. To constrain a LF with both luminosity and density evolution, we need both the contribution of the brightest and faintest sources to the LF, otherwise the two evolution parameters become degenerate. As shown in Fig.~\ref{fig:radio_LF}, the LF data from \citet{Novak_2017} is more sensitive to the most luminous SFGs, whereas our data extend to the low-luminosity sources. \citet{Novak_2017} found that significant density evolution could not be properly constrained by their observations alone as the faint end was not well-sampled. However, combining the two data-sets offers the possibility of jointly constraining the luminosity and density evolution.  

To combine the VLA-COSMOS $\mathrm{3 \, GHz}$ Large Project data with the COSMOS-XS survey, we select the SFGs from the VLA-COSMOS $\mathrm{3 \, GHz}$ Large Project data, as discussed in Section~\ref{sec:VLA-COSMOS 3 GHz Large Project} using the criterion described in Section~\ref{sec:sample selection}. We then treat the two data-sets as two separate regions. This means we mask out the observed area of COSMOS-XS in the VLA-COSMOS $\mathrm{3 \, GHz}$ Large Project. We then combine the two data-sets by means of the \citet{Avni_1980} method for coherent analysis of independent data-sets. The depth of the whole sample is not constant throughout the region, as COSMOS-XS is $\sim \, 5$ times deeper than the VLA-COSMOS $\mathrm{3 \, GHz}$ Large Project. A source in the VLA-COSMOS $\mathrm{3 \, GHz}$ Large Project area will therefore be detectable over the  whole joint area, while fainter sources detected in COSMOS-XS are only detectable in the COSMOS-XS area. This means that the maximum volume of space ($V_{\text{max}, i}$) available for an object in the joint sample is defined by
\begin{align}
	V_{z_{\text{max}}, i} =\begin{cases}       \frac{\Omega_{\rm V}}{4\pi}V_{z_{\text{max}}}^{\rm V} + &\frac{\Omega_{XS}}{4\pi}V_{z_{\text{max}}}^{\rm XS} \text{(if $z_{\text{max},i} \, \leq \, z_{\text{max}}^{\text{V}}$)} \\
	\\
	\frac{\Omega_{\rm XS}}{4\pi}V_{z_{\text{max}}}^{\rm XS} &\text{(if $ z_{\text{max}}^{\text{V}} \, < \, z_{\text{max},i} \, \leq \, z_{\text{max}}^{\text{XS}}$)} \, ,
	\end{cases}
	\label{eq:LF_novak}
\end{align}
where $V_{z_{\text{max}}}^{\text{fld}}$ (with fld = $\rm{V}$, $\rm{XS}$ corresponding to VLA-COSMOS $\mathrm{3 \, GHz}$ Large Project and COSMOS-XS, respectively) is the co-moving volume available to each source in that field, in a given redshift bin, while $\Omega_{\text{fld}}$ is the area observed ($1.673 \, \text{deg}^{2}$ and $0.097 \, \text{deg}^{2}$ for VLA-COSMOS $\mathrm{3 \, GHz}$ Large Project and COSMOS-XS, respectively). 

For each luminosity and redshift bin, the LF is given by Eq.~\ref{eq:LF_vmax} with $o_{i} = 1$ for the VLA-COSMOS $\mathrm{3 \, GHz}$ Large Project sources. The completeness correction $w_{i}$ for these sources consists of a completeness correction for the radio catalog and a counterpart completeness correction. These corrections are derived and described in, respectively, \citet{Smolcic_2017} and \citet{Novak_2017}.

\begin{figure}
\centering
\includegraphics[width=\columnwidth]{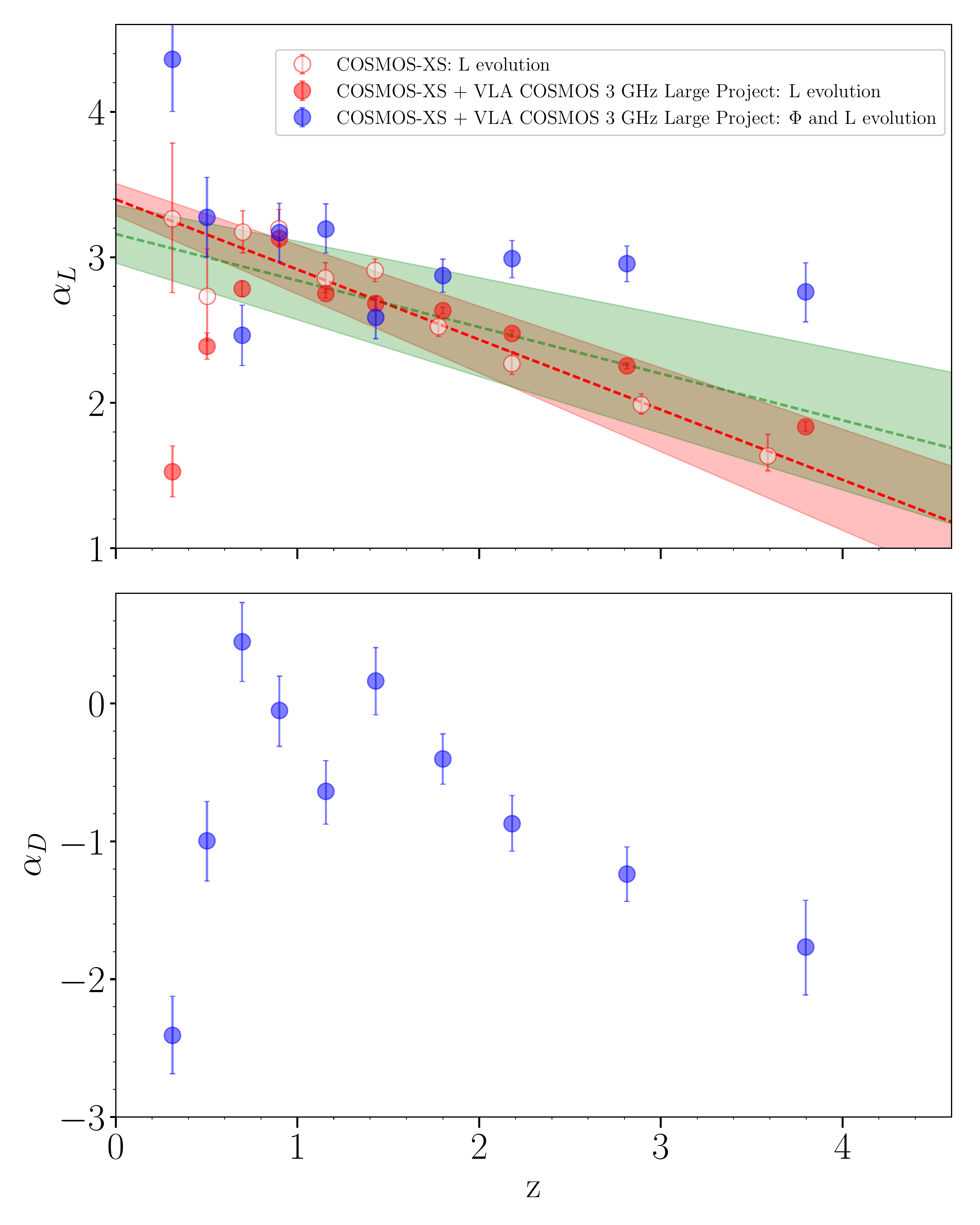}
\caption{The best-fit parameters for the luminosity functions as a function of redshift. The upper panel shows the evolution of the luminosity parameter $\alpha_L$. The lower panel shows the evolution of the density parameter $\alpha_D$. Open circles correspond to pure luminosity evolution for the COSMOS-XS survey. The red dashed line shows the fitted evolution to these points of the form $\alpha_L+z\beta$. The green dashed line shows the simple pure luminosity evolution model described by \citet{Novak_2017}. The luminosity parameter shows a similar evolution as the evolution that \citet{Novak_2017} described.
\newline The filled symbols correspond to the joint COSMOS-XS + VLA-COSMOS 3 GHz data. The red and blue filled symbols correspond to the best-fit parameters found for, respectively, the pure luminosity evolution and the luminosity and density evolution fitted to the joint sample. The density parameter shows a strong evolution while we observe little evolution in the luminosity parameter.}
\label{fig:par_ev}
\end{figure}

\begin{figure*}
\centering
\includegraphics[width=0.88\textwidth]{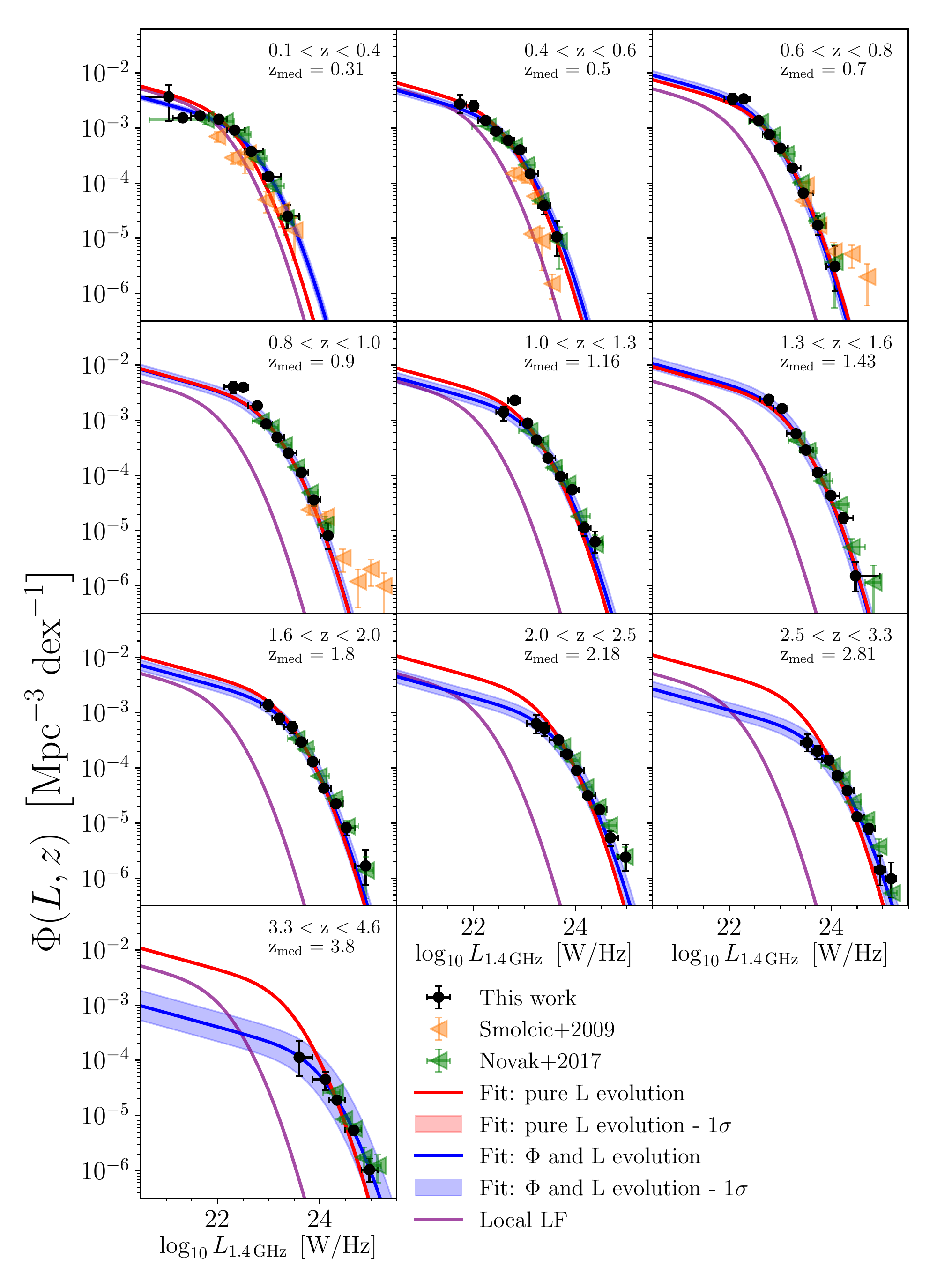}
\caption{Radio LFs of SFGs in different redshift bins from the combined sample of the COSMOS-XS + VLA-COSMOS $\mathrm{3 \, GHz}$ data-sets compared with the radio LFs from \citet{Smolcic_2009_a} and \citet{Novak_2017}. Our best-fit pure luminosity function and best-fit density + luminosity function in each redshift bin are shown with solid red and blue lines, respectively, where the shaded areas show the 1$\sigma$ confidence interval for the best-fit functions. The local radio luminosity function is shown as the purple line for reference. The redshift range and median redshift are given in each panel.}
\label{fig:radio_LF_inc_novak}
\end{figure*}

For comparison, we first fit the LF described by the analytical expression from Eq.~\ref{eq:LF_L} (i.e., pure luminosity evolution) to the joint COSMOS-XS + VLA-COSMOS 3 GHz data points using the method described in Section~\ref{sec:evolving the LF}. The best-fit values for $\alpha_L$ are tabulated in Table~\ref{tab:parameters_both} and the best-fit pure luminosity evolved function is shown with the red line in Fig.~\ref{fig:radio_LF_inc_novak}. Fig.~\ref{fig:par_ev} shows $\alpha_L$ as a function of redshift. At $z \, > \, 1.8$, we find that $\alpha_L$ decreases similarly to what was found when pure luminosity evolution was fitted to the COSMOS-XS data-set alone. 

With the larger dynamic range probed by the combination of the COSMOS-XS + VLA-COSMOS 3 GHz data-sets, we can now fit not only the position of the turnover with redshift, but also the normalization. This luminosity and density evolution can be described as

\begin{align}
	\Phi(L, z, \alpha_L, \alpha_D) =   (1+z)^{\alpha_D}\Phi_{0}\left(\frac{L}{(1+z)^{\alpha_L}}\right) \, .
	\label{eq:LF_D_L}
\end{align}

Because the joint COSMOS-XS + VLA-COSMOS 3 GHz data-sets constrain both the high and low luminosity ends, the evolution parameters ($\alpha_D$ and $\alpha_L$) are less degenerate. The fit with luminosity and density evolution is shown in Fig.~\ref{fig:radio_LF_inc_novak}. Fig.~\ref{fig:all_corner_plots} shows the two dimensional posterior probability distributions of $\alpha_L$ and $\alpha_D$ at each redshift. Fig.~\ref{fig:par_ev} shows the fitted parameters $\alpha_L$ and $\alpha_D$ as a function of redshift. We find that, when allowing for both luminosity and density evolution, $\alpha_L$ decreases and $\alpha_D$ increases to $z \, \sim \, 1$, above which $\alpha_L$ is constant with $z$ while $\alpha_D$ decreases.

\pagebreak
\section{A comparison with luminosity functions from the literature}
\label{sec:LF}

In the following section we compare our results to literature LFs derived from radio, FIR and UV observations. 

\subsection{Radio}
\label{sec:Radio}
Fig.~\ref{fig:radio_LF} and Fig.~\ref{fig:radio_LF_inc_novak} show the determination of the radio LF of \citet{Smolcic_2009_a} and \citet{Novak_2017}. \citet{Smolcic_2009_a} derived the radio LF up to $z \, < \, 1.3$ using 340 galaxies from the VLA-COSMOS $\mathrm{1.4 \, GHz}$ survey conducted over the $2 \, \rm{deg}^2$ COSMOS field \citep{Schinnerer_2007}. Our data generally lies slightly above the data from \citet{Smolcic_2009_a}, which could be due to the different selection criteria. Specifically, \citet{Smolcic_2009_a} only used rest-frame optical colors to select SFGs. However, at $z \, > \, 0.6$, the high luminosity bins ($\log_{10} L_{1.4 \, \text{GHz}} \, \gtrsim \, 24 \text{W/Hz}$) from \citet{Smolcic_2009_a} lie above our data. This could be due to contamination of their sample from AGN \citep{Smolcic_2009_a}, as they used a different AGN selection method.

The VLA-COSMOS $\mathrm{3 \, GHz}$ Large Project (\citealt{Smolcic_2017}) was also conducted over the COSMOS field and yielded about four times more radio sources compared to the $\mathrm{1.4 \, GHz}$ data of \citet{Schinnerer_2007}. This resulted in LFs up to $z \, \lesssim \, 5.7$ using 5915 SFGs selected as described in Section~\ref{sec:VLA-COSMOS 3 GHz Large Project}. Overall, our radio LFs generally agree very well with those derived by \citet{Novak_2017} based on this data-set. Because of the large field of view of the VLA-COSMOS $\mathrm{3 \, GHz}$ Large Project, the LF data from \citet{Novak_2017} is more sensitive to the most luminous SFGs, especially for $z \, < \, 1.6$. On the other hand, as Fig.~\ref{fig:radio_LF} shows, the COSMOS-XS data extends towards lower luminosity and adds in almost every redshift bin two low-luminosity data points. Given the good agreement between the COSMOS-XS and VLA-COSMOS $\mathrm{3 \, GHz}$ data-sets and the larger constraining power of the combination (Section~\ref{sec:luminosity and density evolution}), we  use the radio LFs derived from the combined COSMOS-XS $+$ VLA-COSMOS $\mathrm{3 \, GHz}$ data-sets for the comparison with the LFs derived from the IR and UV in the following sections. 

\subsection{Far-infrared}
\label{sec:IR}
\begin{figure*}
\centering
\includegraphics[width=0.86\textwidth]{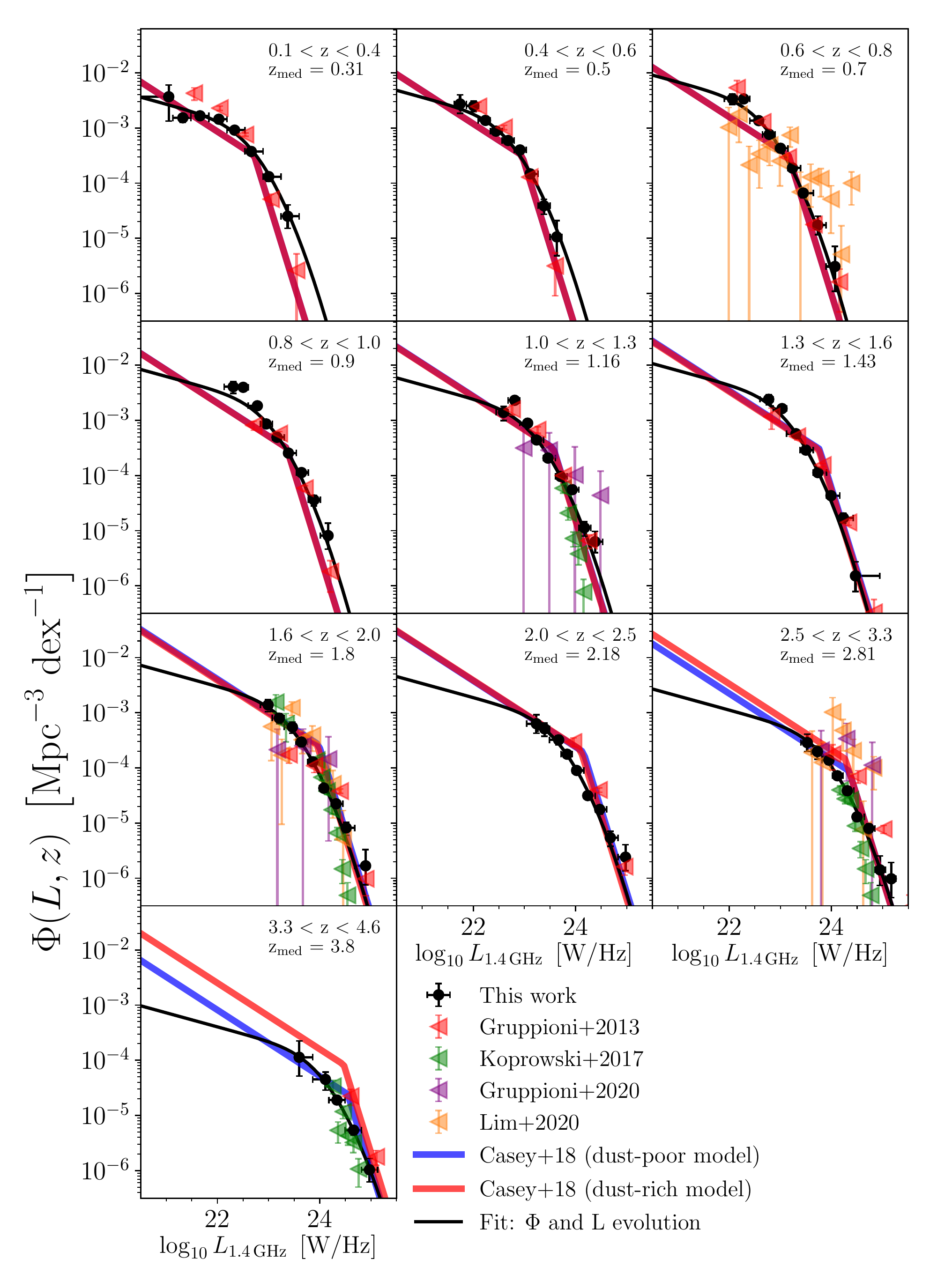}
\caption{Radio LFs of SFGs in different redshift bins for the combined COSMOS-XS + VLA-COSMOS $\mathrm{3 \, GHz}$ data-sets compared to various FIR LFs from the literature. Our best-fit luminosity + density function in each redshift bin is shown with the black lines. We show FIR LFs from \citet{Gruppioni_2013}, \citet{Koprowski_2017}, \citet{Gruppioni_2020} and \citet{Lim_2020}. The LFs of the theoretical study by \citet{Casey_2018} for the dust-poor model and the dust-rich model are also shown. The redshift range and median redshift are given in each panel. Our best-fit luminosity + density function is roughly consistent, within the error bars, with the FIR LFs. At $z \, > \, 2$ our LFs are systematically lower than the FIR studies expect for the result from \citet{Koprowski_2017}. Our best-fit function is most consistent with the dust-poor model from \citet{Casey_2018}}
\label{fig:radio_LF_IR}
\end{figure*}

If the FIR-radio correlation is linear (see Section~\ref{sec:sample selection}), both FIR and radio LFs should follow each-other well. In Fig.~\ref{fig:radio_LF_IR}, we compare our results with the FIR LFs from \citet{Gruppioni_2013}, \citet{Koprowski_2017}, \citet{Gruppioni_2020} and \citet{Lim_2020}. To adapt their results with our redshift bins, we simply plot the value of $\Phi$ for which the mean $z$ is within our redshift bin.

\cite{Gruppioni_2013} used the data-sets from the \emph{Herschel} PEP Survey, in combination with the HerMES imaging data to derive the evolution of the FIR LFs up to $z \, \sim \, 4$. \cite{Koprowski_2017} found their total FIR LF measurements based on SCUBA-2 $850 \, \mathrm{\mu m}$ observations. \citet{Gruppioni_2020} determined the total FIR LF using the non-target ALPINE sources observed with ALMA. These 56 sources were blindly detected at $860 \, \mathrm{\mu m}$ within the fields of targeted galaxies of the ALPINE survey, and the total FIR was derived using SED fitting using semi-empirical templates. Finally, \citet{Lim_2020} used a SCUBA-2 $450 \, \mathrm{\mu m}$ map in the COSMOS field covering a area of $300 \, \mathrm{arcmin^2}$ to construct the FIR LF. 

To convert the total FIR LF given by \citet{Gruppioni_2013}, \citet{Gruppioni_2020} and \citet{Lim_2020} to a radio LF, we use the FIR-radio correlation as described in Eq.~\ref{eq:qtir}, with $q_{\text{TIR}}$ as the FIR-radio correlation from \citet{Delhaize_2017} and rewritten as: 

\begin{align}
	\log_{10} L_{1.4 \, \text{GHz}} &= \log_{10}\left(\frac{L_\text{FIR}}{3.75\times 10^{12} \text{ W}}\right) - q_\text{TIR}(z) \nonumber \, , \\
	\log_{10} L_{1.4 \, \text{GHz}} &= \log_{10}\left(\frac{L_\text{FIR}}{3.75\times 10^{12} \text{ W}}\right) \nonumber  \\ 
	&- (2.88 \times (1+z)^{-0.19}) \, .
	\label{eq:qtir_inv}
\end{align}

To find the total FIR LF for \citet{Koprowski_2017}, we used the $L_{250\mu{\rm m}}/L_{\rm{FIR}}$ ratio given by the \citet{Michalowski_2010} template to convert the rest-frame $250 \, \mathrm{\mu m}$ LF from the SCUBA-2 data to total FIR LF, which is then converted to a radio LF using Eq.~\ref{eq:qtir_inv}.

Similar to what \cite{Novak_2017} found, our data agree well with these FIR surveys. However, at $z \, > \, 2$, our LFs are systematically lower than \citet{Gruppioni_2013}. We find that the more recent studies from \citet{Gruppioni_2020} and \citet{Lim_2020} are also higher than our data, although these data-sets are more uncertain due to the low number of sources per bin. The offset between our data and the studies from \citet{Gruppioni_2013}, \citet{Gruppioni_2020} and \citet{Lim_2020} at $z \, > \, 2$ may be partly attributed to the presence of AGN in the FIR selected sample. While we start from a radio sample that excludes AGNs, as described in Section~\ref{sec:data and sample selection}, \citet{Gruppioni_2013} and \citet{Gruppioni_2020} derive the total FIR LF and thus include sources powered by AGN. In addition, the fraction of AGN is found to increase with redshift: \citet{Gruppioni_2013} find that AGN largely dominate the FIR luminosity density at $z \, \gtrsim \, 2.5$. 
However, \citet{Gruppioni_2020} find that the large majority of the SEDs of their sources are best fitted by star-forming or composite templates. In contrast, the \citet{Lim_2020} study excludes sources identified as AGN based on their X-ray, mid-IR or radio-emission
and finds a low AGN fraction compared to literature studies due to their deep observations. These probe a faint sub-mm galaxies (SMGs) population which are less likely to host an AGN. The difference can thus not solely be explained by the presence of AGN.
Some of the difference could therefore be due to the evolving $q_\text{TIR}(z)$ used in the conversion from FIR to radio. This will be discussed in more depth in Section~\ref{sec:IR-radio conversion}.
In addition, there are a lot of uncertainties in measuring the FIR luminosity from a few data points which is reinforced by discussion of \citet{Gruppioni_2019} on the study of \citet{Koprowski_2017}. We find that the \citet{Koprowski_2017} LFs are systematically lower than the other FIR studies over the whole luminosity range and match our data at $z \, > \, 2$. \citet{Gruppioni_2019} explained the discrepancy to other FIR studies by attributing the difference to a choice of sub-mm SED and sample incompleteness. 

In Fig.~\ref{fig:radio_LF_IR} we also compare our results with the observationally motivated sub-mm LF models from \citet{Casey_2018}. They developed an evolutionary model based on existing measurements of sub-mm number counts, redshift distributions, and multi-band flux information to study the shape and behavior of the FIR LF out to high redshift ($z \, > \, 4$). They considered two extreme cases: a dust-poor model, where the abundance of very dust-rich dusty star-forming galaxies (DSFGs) relative to UV-bright galaxies is low ($< \, 10 \, \%$ at $z \, = \, 4$), and a dust-rich model, where DSFGs dominate and contribute $> \, 90 \, \%$ to the star formation at $z \, = \, 4$. Both models include a ``turning point'' redshift at which the knee of the LF ($L_{\star}$) and the characteristic number density of the LF ($\Phi_{\star}$) are transitioning in their evolution. For example, $\Phi_{\star}$ might evolve like $(1+z)^{-2.8}$ up to $z \, \sim \, 1.5$, and then gradually transition to $(1+z)$ by a redshift of $z \, \sim \, 3.5$. The turning point for the dust-poor and dust-rich model lies at, respectively, $z \, = \, 2.1$ and $z \, = \, 1.8$. Before this redshift the models use the same evolution parameters. Thereafter, they will evolve at different rates.  

The dust-poor model is similar to the often adopted evolutionary scenario in the rest-frame UV literature. It represents the model that the dust-formation timescale is longer than the timescale for the formation of UV-bright galaxies. This means that DSFGS are rare at $z \, > \, 4$ in this model and only dominate the star formation at $z \, \sim \, 2$. The dust-rich model is quite extreme and suggests that most star formation at high redshift was isolated to rare starbursts with very high SFR and that DSFGS would dominate the star formation at $z \, > \, 1.5$. \citet{Casey_2018} showed that both models were consistent with the sub-mm data that existed at that time.

The predictions of the FIR LFs by \citet{Casey_2018} shown in Fig.~\ref{fig:radio_LF_IR} are converted as discussed above. The converted LFs are consistent with our measurements at $z \, < \, 2.5$ and from $z \, > \, 2.5$ the models start to deviate from each other. At $z \, > \, 2.5$ the dust-rich model over-predicts our data, while the dust-poor model matches quite well, as also seen with the VLA-COSMOS $\mathrm{3 \, GHz}$ Large Project data alone \citep{Novak_2017}. 

In summary, we find that our radio LFs are roughly consistent, within the error bars, with the FIR LFs. At $z \, > \, 2$ our LFs are systematically lower than \citet{Gruppioni_2013}, which we attribute at least partly due to AGN contamination. In addition, we find that the radio data is most consistent with the dust-poor model from \citet{Casey_2018}.

\subsection{UV}
\label{sec:UV}

\begin{figure*}
\centering
\includegraphics[width=1.0\textwidth]{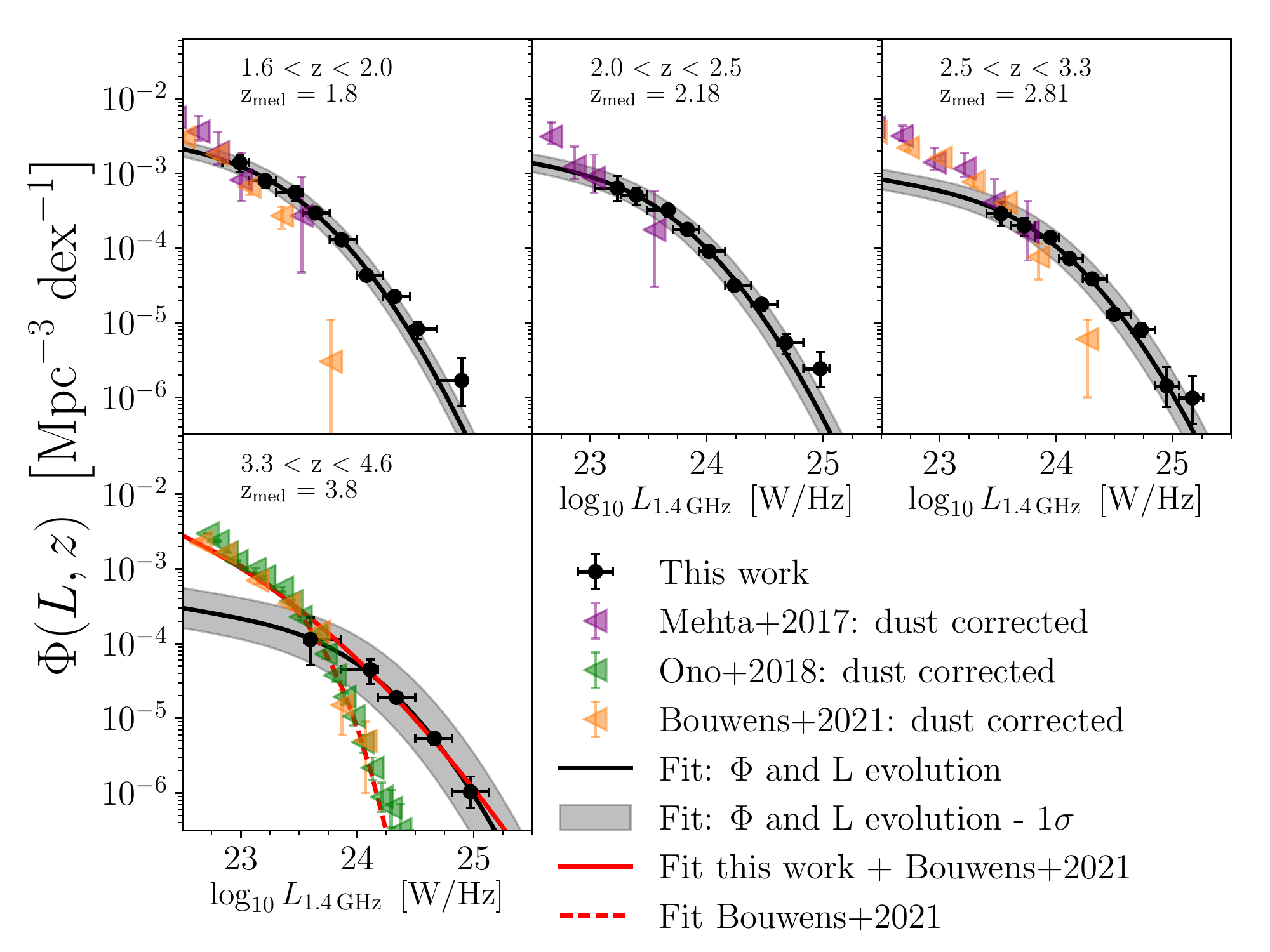}
\caption{Radio LFs of SFGs in different redshift bins compared to UV LFs from the literature. Our best-fit luminosity + density function in each redshift bin is shown with the solid black line. The shaded areas shows the $1\sigma$ confidence interval for the best-fit functions. We compare our data with the UV LFs from \citet{Metha_2017}, \citet{Ono_2018} and \citet{Bouwens_2021}. The redshift range and median redshift are given in each panel. The best-fit pure local LF for the data from \citet{Bouwens_2021} at $z \, \sim \, 4.6$ is shown with the dashed red line, while the solid red line shows the joint fit to the UV+radio data discussed in Section~\ref{sec:UV}. The comparison of these two fits in the last panel suggests that at the highest redshifts probed here ($3.3 \, < z \, < \, 4.6$), the UV data underestimate the integrated LF by at least $21.6 \, \pm \, 14.2 \, \%$ where the LF is integrated down to $L^{\star}_{z=3}$.}
\label{fig:radio_LF_UV}
\end{figure*}

It is also interesting to compare our radio LFs with previous UV LFs studies. The UV probes fainter sources at higher redshift and therefore offers a comparison sample complementary to that of FIR-based studies. In addition, the SFR calibrations from \citet{Kennicutt_1998} are self-consistent which means that all SFR tracers should result in roughly the same SFR estimate. The UV and radio both trace SF where radio is mostly sensitive to SFGs with a high SFR and UV is probing emission from SF not obscured by dust. The UV and radio LFs should thus follow each-other well if the UV can be fully corrected for dust extinction. In Fig.~\ref{fig:radio_LF_UV} we compare our results with the UV LFs from \citet{Metha_2017}, \citet{Ono_2018} and \citet{Bouwens_2021}.

\citet{Metha_2017} used deep NUV imaging data as part of the \textit{Hubble} Ultra-Violet Ultra Deep Field program to find the rest-frame 1500$\text{\AA}$ UV LF at $z \, \sim \, 1.7$, 2.2 and 3.0. 
\citet{Ono_2018} conducted the GOLDRUSH project with the  optical images taken by the HSC-SSP which cover a large area of $\sim$ 100 deg$^2$. The sample is constructed using the so-called drop-out technique. In this case the sample consisted of a total of $\sim \, 580,000$ Lyman break galaxies at $z \, \sim \, 4 - 7$. The UV LF is then derived by combining the LFs from the HSC Subaru program with the LFs from the ultra-deep \textit{Hubble} Space Telescope legacy surveys. 
\citet{Bouwens_2021} derived UV LFs at $z \, \sim \, 2 - 10$ based on the \textit{Hubble} data from various legacy fields covering an area of $\sim \, 0.3 \text{deg}^2$ which contains $> \, 24,000$ sources.

The conversion needed to compare LFs at radio and UV wavelengths is derived by \citet{Novak_2017} following \citet{Kennicutt_1998}:
\begin{equation}
    L_{1.4 \, \text{GHz}} = 16.556 - 0.4(M_\text{1600,AB} - A_\text{UV}) - q_\text{TIR} \, ,
\end{equation}
where $M_{1600,AB}$ is rest-frame UV, $A_{UV}$ is the extinction given by $4.43 - 1.99\beta$ with $\beta$ the UV spectral slope and $q_\text{TIR}$ is the FIR-radio correlation defined by \citet{Delhaize_2017}. To correct the UV data for dust extinction, we used the UV spectral slope $\beta$ as tabulated as a function of magnitude by \citet{Bouwens_2009} ($z \, \sim \, 2.5-4$) and \citet{Bouwens_2014b} ($z \, \sim \, 4-8.5$). Following \citet{Viironen_2018}, we added a small correction of $\Delta M_{\rm UV} \, = \, +0.035$ to the luminosity values of \citet{Metha_2017}, in order to scale them from $1500 \, \text{\AA}$ to $1600 \, \text{\AA}$. This was done by roughly defining the average $\beta$-slopes for the sources ($\beta \sim -1.7$) and deriving the correction from there. 

To adapt the UV LF results to our redshift bins, we simply plot the value of $\Phi$ for which the mean $z$ is within our redshift bin. We find that our LFs predict an excess of bright sources compared to \citet{Bouwens_2021}, \citet{Metha_2017} and \citet{Ono_2018} at $z \, \sim \, 2.9$ and $z \, \sim \, 3.6$. The excess is especially striking at high luminosity at $z \, \sim \, 3.6$, where the UV dust correction is most severe. Although the UV LFs have been corrected for dust extinction, they still seem to miss a part of the galaxies with dust obscured SF, as previously noted by \citet{Novak_2017} and \citet{Viironen_2018}. Based on their radio LFs, \citet{Novak_2017} estimated that \citet{Bouwens_2014b} underestimated the obscured SFR observed in UV by 15--20 \%.

\def\baselinestretch{1.1}
\begin{deluxetable}{lcc}
	\tabletypesize{\footnotesize}
	\tablewidth{\columnwidth}
	\tablecaption{Best fit parameters of the local luminosity function, as described in Eq.~\ref{eq:local_LF}, fitted to the luminosity function from \citet{Bouwens_2020} and to our radio luminosity function + the luminosity function from \citet{Bouwens_2020}. Both fits are shown in Fig~\ref{fig:radio_LF_UV}.}
	
	\tablehead{
		\colhead{ } &
		\colhead{Bouwens+2021} &
		\colhead{This work + Bouwens+2021}
	}
	
	\startdata
	 
	\vspace{-1.0ex}\\
    $L_{\star}$ ($\times \, 10^{23}$ & 3.79 $_{-1.50}^{+2.44}$ & 0.22 $_{-0.09}^{+0.13}$ \\
    $\, [\text{W}\text{Hz}^{-1}]$)  &   &   \\
    $\Phi_{\star}$ ($\times \, 10^{-4}$ & 4.76 $_{-1.61}^{+2.52}$ & 39.0 $_{-12.27}^{+17.63}$  \\
    $[\text{Mpc}^{-3}\text{dex}^{-1}]$) &  &   \\
    $\alpha$ & 1.71 $_{-0.03}^{+0.03}$ & 1.67 $_{-0.05}^{+0.04}$\\
    $\sigma$ & 0.21 $_{-0.06}^{+0.07}$ & 0.94 $_{-0.06}^{+0.06}$\\
	\enddata
    \label{tab:parameters_luminosity_bouwens}
\end{deluxetable}

To determine the UV underestimation of the obscured SFR suggested by our data, we fitted the local LF, as described in Eq.~\ref{eq:local_LF}, to the dust corrected data from \citet{Bouwens_2021} with all parameters unconstrained. The obtained best fit parameters are tabulated in Table~\ref{tab:parameters_luminosity_bouwens}. The fit is shown in Fig.~\ref{fig:radio_LF_UV}. 
We then fitted the local LF in the same way to a combination of the radio data and the dust corrected data from \citet{Bouwens_2021}. We disregarded the three most luminous LF points from \citet{Bouwens_2021} and took the radio data points instead. The obtained best fit parameters are tabulated in Table~\ref{tab:parameters_luminosity_bouwens} and the fit is shown in Fig.~\ref{fig:radio_LF_UV}. 
We then integrated the two fits from $L^{\star}_{z=3}$, as defined by \citet{Bouwens_2021}, which corresponds to $\log_{10} L_{1.4\,\text{GHz}} \, = \, 21.14 \,  \text{W} \text{Hz}^{-1}$ to $\infty$, to find the difference between the two. We find the UV data presented in \citet{Bouwens_2021} underestimate the integrated LF by $21.6 \, \pm \, 14.2 \, \%$ at these redshifts ($3.3 \, < z \, < \, 4.6$). We can interpret this estimate as a lower limit, as the mean redshift of the UV sample presented in the last panel of Fig.~\ref{fig:radio_LF_UV} is 3.8, slightly higher than the median redshift of the radio sample and we expect the UV LF to increase between $z \, = \, 3.7$ and $z \, = \, 3.8$.

We additionally note that the radio LFs displayed in Fig.~\ref{fig:radio_LF_UV} do not include any of the ``optically dark'' sources as described in Paper II. These 70 sources were not matched to a counterpart in any of the catalogs used in the counterpart matching as described in Section~\ref{sec:counterparts}. Some of these sources could be spurious detections but most of these ``optically dark'' sources are expected to be real; we expect only $\sim$ 20 spurious sources. As discussed in Section~\ref{sec:estimating LF}, we do correct for the counterpart completeness with $f_\text{ctrpt}$. This small correction as a function of flux density is done over the whole redshift range. However, the method used in Paper II, which finds 29 robust ``optically dark'' sources, shows that these sources are likely to have a redshift of $z \, \gtrsim \, 4$, similar to what was found in ALMA follow up of sources without an optical counterpart \citep[e.g][]{Dudzeviciute_2020, Smail_2021}. The LF at $z \, \sim \, 4$ including these ``optically dark'' sources will be higher than shown in Fig.~\ref{fig:radio_LF_UV}. 

Different works have already identified ``optically dark'' sources, extreme SFGs heavily obscured by dust which lack an optical or near-IR counterpart, out to high redshift ($z \, \simeq \, 5$) \citep[e.g][]{Dannerbauer_2008, Walter_2012, Riechers_2020}. \citet{Wang_2019} reported the results from the ALMA follow-up of a population of optically dark galaxies, and found a fraction of them to be massive dusty galaxies at high-redshift. They concluded that this population constitutes a significant fraction of the SFRD at $z \, > \, 3$. In addition, \citet{Talia_2021} estimated that dust-obscured star-forming galaxies, found based on their emission at radio wavelengths and the lack of optical counterparts, have a contribution to the SFRD which can be as high as 40\% of the previously known UV-SFRD. More recently, \citet{Enia_2022} estimated the contribution of ``optically dark'' sources ($H$--dark galaxies) to the SFRD using 8 ``optically dark'' galaxies found at $z \, \sim \, 3$, and finding they contribute $7 - 58 \%$ to the UV-based SFRD.
The discrepancy between our radio LF and the UV LFs will thus also be greater with the inclusion of the ``optically dark'' sources. The derivation of the radio LF including these sources and implications that follow will be further discussed in a future paper.

In summary, we find our radio observations show an excess above the UV LFs for $z \, > \, 2.9$ even without including the ``optically dark'' sources. Although the UV LFs have been corrected for dust extinction, we estimate that they miss at least $21.6 \, \pm \, 14.2 \, \%$ of the star formation traced by the integrated radio LF. 

\subsection{Radio vs. FIR vs. UV}
As discussed above, the LF can be constrained by using different tracers: radio, FIR and UV. Each tracer may be affected by different biases. Radio observations can be contaminated by AGN. FIR and sub-mm observations lack, respectively, high resolution and large field of view observations. In addition, these bands have a limited sensitivity to galaxies at $z \, > \, 3$ and FIR observations can be significantly affected by AGN. UV observations need significant corrections for dust-obscuration and are unable to uncover the most extreme SFGs. By comparing all three tracers, we are able to find which bias is most impactful. 

As discussed in Section~\ref{sec:IR}, the radio data presented here are roughly in agreement with the FIR observations. In addition, Fig.~\ref{fig:radio_LF_UV} shows that our radio observations show an excess above the UV observations at $z \,  > \, 2.9$. The current radio data thus confirm a discrepancy that exists between the FIR and UV data. This was already suggested by the work of \citet{Novak_2017}, and the new analysis of the combined data strengthens the evidence for the discrepancy and suggest an underestimation of the UV LF. Although radio and FIR observations share the risk of AGN contamination, these AGN are observed at a different wavelengths and thus have different methods of removal. Seeing that the radio and FIR observations are moderately consistent suggests that the most significant issue is with UV observations and their dust corrections 

The IRX-$\beta$ relation \citep{Meurer_1999} is used in UV studies to attempt to correct for dust extinction. Their relation consists of the ratio of total FIR to UV luminosity ($L_{\text{FIR}}/L_{\text{UV}} = \text{IRX}$), a proxy for extinction, and the UV spectral slope ($\beta$), which depends on the column density along the line of sight that is attenuating the UV light. The relation is therefore sensitive to a range of ISM properties including dust geometries, dust-to-gas ratios, dust grain properties, and the spatial distribution of dust.

\citet{Mancuso_2016} have built an intrinsic SFR function and find that, even when corrected for dust absorption with the IRX-$\beta$ relation, UV observations underestimate the intrinsic SFR for galaxies with a SFR $> 30 \text{M}_{\odot} \rm{yr}^{-1}$. Their result suggests a galaxy population at $z \, \gtrsim \, 4$ with large dust-obscured SFR of $\gtrsim 100 \text{M}_{\odot} \rm{yr}^{-1}$, the higher redshift counterparts to the dusty SF population observed by FIR observations at $z \, \lesssim \, 3$. In addition, several studies have already shown that low redshift luminous infrared galaxies -- so-called luminous and ultra-luminous galaxies ($10^{11} \, \leq \, L_{\text{IR}} \, < 10^{13} L_{\odot}$, LIRGs and ULIRGs) and high redshift dusty SFGs ($L_{\text{IR}} \, \geq \, 10^{12-13} L_{\odot}$, DSFGs) -- are offset from the nominal UV spectral slope \citep{Goldader_2002, Howell_2010, Casey_2014b, Bourne_2017}. Furthermore, \citet{Khusanova_2020} recently concluded that the brightest Ly$\alpha$ emitters at $z \, > \, 5$ are very diverse and found that these galaxies have large scatter in observed $\beta$ values. These studies show that UV observations miss a part of the galaxies with dust obscured SF and question the existing IRX-$\beta$ relation as a method of dust correction. 

In particular, we know the reliability of the IRX-$\beta$ relation for high-redshift galaxies has several issues.
Firstly, the shape of the FIR SED at high-redshift is poorly constrained due to a lack of sampling of the SED peak. This means that the FIR luminosity is derived from FIR SED models that are fitted at lower redshift. We also know that the dust temperature ($T_{\text{dust}}$) is crucial for the derivation of $L_{\text{IR}}$, with an incorrectly assumed $T_{\text{dust}}$ changing the $L_{\text{IR}}$ by as much as an order of magnitude \citep[e.g.,][]{Hodge_2020}. Unfortunately, $T_{\text{dust}}$ is typically highly uncertain for lower luminosity high-redshift galaxies and might depend on various galaxy properties \citep[e.g.,][]{Chapman_2003, Magnelli_2014}. In addition, the distribution of dust could be more patchy in high-redshift galaxies due to their turbulent nature. The UV slope is then dominated by the least obscured part of the galaxies, leading to an under-prediction of the necessary correction \citep{Faisst_2017}. These issues indicate that different dust corrections for bright and highly star-forming galaxies at high redshift are necessary, and we may thus need a different approach to correctly estimate dust corrections for these galaxies.

\subsection{Evolution parameters}
\label{sec:evolution parameters}
In this section, we compare the implied evolution of our LF parameters (Fig.~\ref{fig:par_ev}) with previous multi-wavelength works from the literature.
The FIR studies from \citet{Gruppioni_2013}, \citet{Koprowski_2017} and \citet{Lim_2020}, and the UV study from \citet{Bouwens_2021}, describe the position of the turnover in the FIR and UV LF with $L^\star$ and $M^\star$, respectively. The normalization of the LF is described by $\Phi^\star$. In these studies, $L^\star$/$M^\star$ and $\Phi^\star$ are simultaneously fitted. The FIR studies find the position of the turnover to evolve to higher luminosities. \citet{Bouwens_2021} also find the characteristic luminosity $M^\star$ to increase to $z \, \sim \, 3$, but thereafter they find it to remain relatively fixed over the redshift range $z \, \sim \, 3 - 8$. This kind of evolution can also be seen in the study by \citet{Gruppioni_2013}, who describe the luminosity evolution of $L^\star$ up to $z \, \sim \, 1.85$ as $L^\star \propto (1+z)^{3.55 \pm 0.10}$. Thereafter they find a somewhat slower evolution of $L^\star \propto (1+z)^{1.62 \pm 0.51}$ up to $z \, \sim \, 4$. The normalization of the LF was found to decrease with redshift by \citet{Gruppioni_2013}, \citet{Koprowski_2017} and \citet{Bouwens_2021}. \citet{Lim_2020} also found this once the faint-end slope $\alpha$ was fixed. \citet{Gruppioni_2013} describe the normalization evolution again with a break. They find $\Phi^\star$ to slowly decrease as $\Phi^\star \propto (1+z)^{-0.57 \pm 0.22}$ up to $z \, \sim \, 1.1$, followed by a quick decrease $\Phi^\star \propto (1+z)^{-3.92 \pm 0.34}$ up to $z \, \sim \, 4$.

As shown in Fig.~\ref{fig:par_ev}, we find a strong evolution of the luminosity parameter, with a clear break at $z \, \sim \, 1$, when we fit the COSMOS-XS survey and the combined data-sets for pure luminosity evolution. The evolution at $z \, > \, 1$ can roughly be fitted with $(3.40 \pm 0.11) - (0.48 \pm 0.06) \times z$, shown with the red dashed line in Fig.~\ref{fig:par_ev}. This agrees with the evolution that was found by \citet{Novak_2017}. The green dashed line in Fig.~\ref{fig:par_ev} shows the simple pure luminosity evolution model described by \citet{Novak_2017}, where they fit an evolution of $(3.16 \pm 0.2) - (0.32 \pm 0.07) \times z$. In addition, we clearly see an increase of the position of the turnover, as seen before in UV, FIR, and radio studies. 

When we instead fit simultaneously for luminosity and density evolution, we find a strong evolution of the density evolution parameter, whereas the evolution in the luminosity parameter remains relatively fixed. While the evolution of these parameters could be influenced by the need to fix the bright and faint end shapes of the distribution to the local values (Section~\ref{sec:evolving the LF}), we note that the same caveat applies to all studies, regardless of the LF form fitted, that fix these parameters \citep[e.g.,][]{Novak_2017, Enia_2022}. 
We will see that this density+luminosity evolution has an effect on the cosmic star formation history in Section~\ref{sec:CSFH}. 

\section{Potential biases and additional caveats}
\label{sec:Discussion}
Before we discuss the implications of our derived radio LFs for the cosmic star formation rate history, we first discuss the possible biases and additional caveats that need to be taken into account when deriving and interpreting the radio LF.

\subsection{AGN contamination}
\label{sec:contamination}
A recent paper by \citet{Symeonidis_2021} investigated the difference between the flatter high luminosity slope seen in the FIR LF compared to the UV LF. They constrained the AGN LF using X-ray observations and then converted the X-ray AGN LF to the FIR AGN LF. This AGN LF was then compared to the total FIR LF, which corresponds to emission from dust heated by stars and AGN. \citet{Symeonidis_2021} claim that at $z \, < \, 2.5$, the high luminosity tail of the AGN FIR LF and total FIR LF converge, suggesting that the most FIR-luminous galaxies are AGN-powered. They conclude from this that the flatter high-luminosity slope seen in the FIR LF compared to that in the UV and optical can be attributed to the increasing fraction of AGN-dominated galaxies with increasing total FIR luminosity. The AGN FIR LF and total FIR LF can be used to find the maximum value of SFR that would be believable if computed from the FIR luminosity. The range of maximum SFRs is between 1,000 and 4,000 $\text{M}_{\odot} \text{yr}^{-1}$ at the peak of cosmic star formation history ($1 \, < \, z \, < \, 3$). When converted to radio luminosities, this gives a range of $\log_{10} L_{1.4 \, \text{GHz}} \sim 23.6 - 24.1$. This suggests that the brightest bins in the radio LF in this redshift range could be contaminated with sources powered by AGN. 

\begin{figure}
	\centering
	\includegraphics[width=1.0\columnwidth]{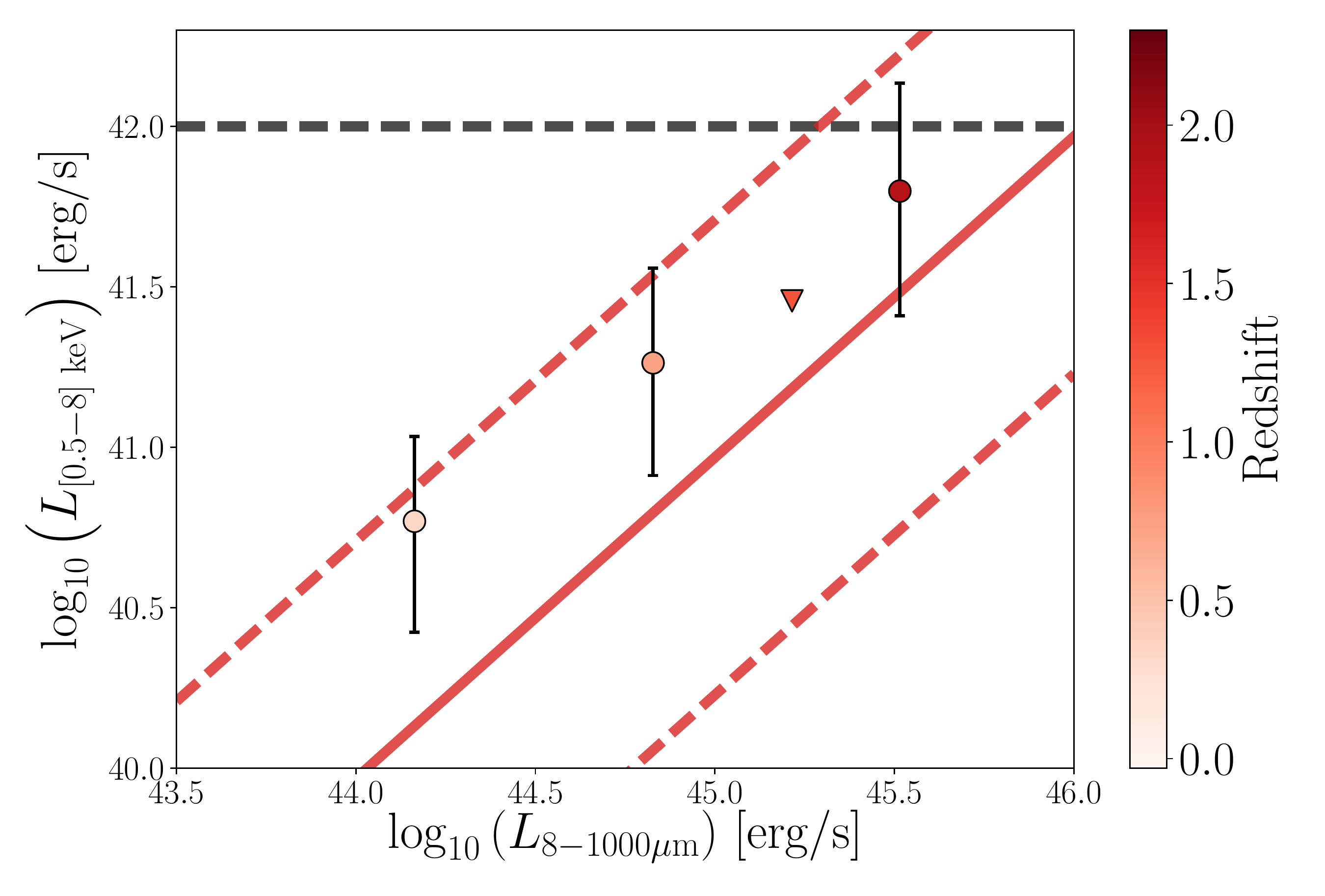}
	\caption{X-ray luminosity derived via X-ray stacking versus FIR-luminosity, a proxy for star formation rate, for the star-forming sample, binned in four redshift bins. Circles and triangles represent detections and upper limits, respectively. The solid line represents the $L_{\text{X}} - L_\text{IR}$ relation from \citet{Symeonidis_2014}. The dashed line shows the $2\sigma$ scatter around this relation. The binned data fall within the scatter range from the trend from \citet{Symeonidis_2014}, indicating no appreciable contribution from AGN.}
	\label{fig:xraysfg}
\end{figure}

To assess to what extent our SFG sample is contaminated by AGN, we divide our data into four equally populated redshift bins and stack the X-ray images. The stacking is done with the online available tool {\sc{CSTACK}}, which utilizes a mean-stacking method\footnote{{\sc{CSTACK}} was developed by Takamitsu Miyaji and can be found at \url{http://cstack.ucsd.edu/}.}. X-ray luminosities are calculated from the stacks assuming a power law spectrum with a slope of $\Gamma \, = \, 1.4$. Fig.~\ref{fig:xraysfg} shows the X-ray luminosities as a function of FIR-luminosities, where the error-bars represent the bootstrapped spread on the median. The solid line shows the median trend found by \citet{Symeonidis_2014}, and the dashed line constitutes the $2\sigma$ scatter. We find little excess in the X-ray compared to the typical X-ray -- star-formation relations; the stacked data matches the trend from \citet{Symeonidis_2014} within the scatter. Thus, we conclude that our star-forming sample is not substantially contaminated by AGN.

In addition to our examination of the contamination of unidentified AGN in our radio LFs, we want to assess the influence of our SFG selection criteria. As discussed in Section~\ref{sec:sample selection}, we used the following selection criterion to select SF sources:

\begin{align}
	q_\text{TIR}(\text{M}_{\star}, z) \, > \, & 2.646 \times (1+z)^{-0.023} \notag \\ 
	&- 0.148 \times (\log_{10} \frac{\text{M}_{\star}}{\text{M}_{\odot}}-10) - (3 \times 0.22)\,.
	\label{eq:threshold_q_inv}
\end{align}

\begin{figure*}
\centering
\includegraphics[width=0.83\textwidth]{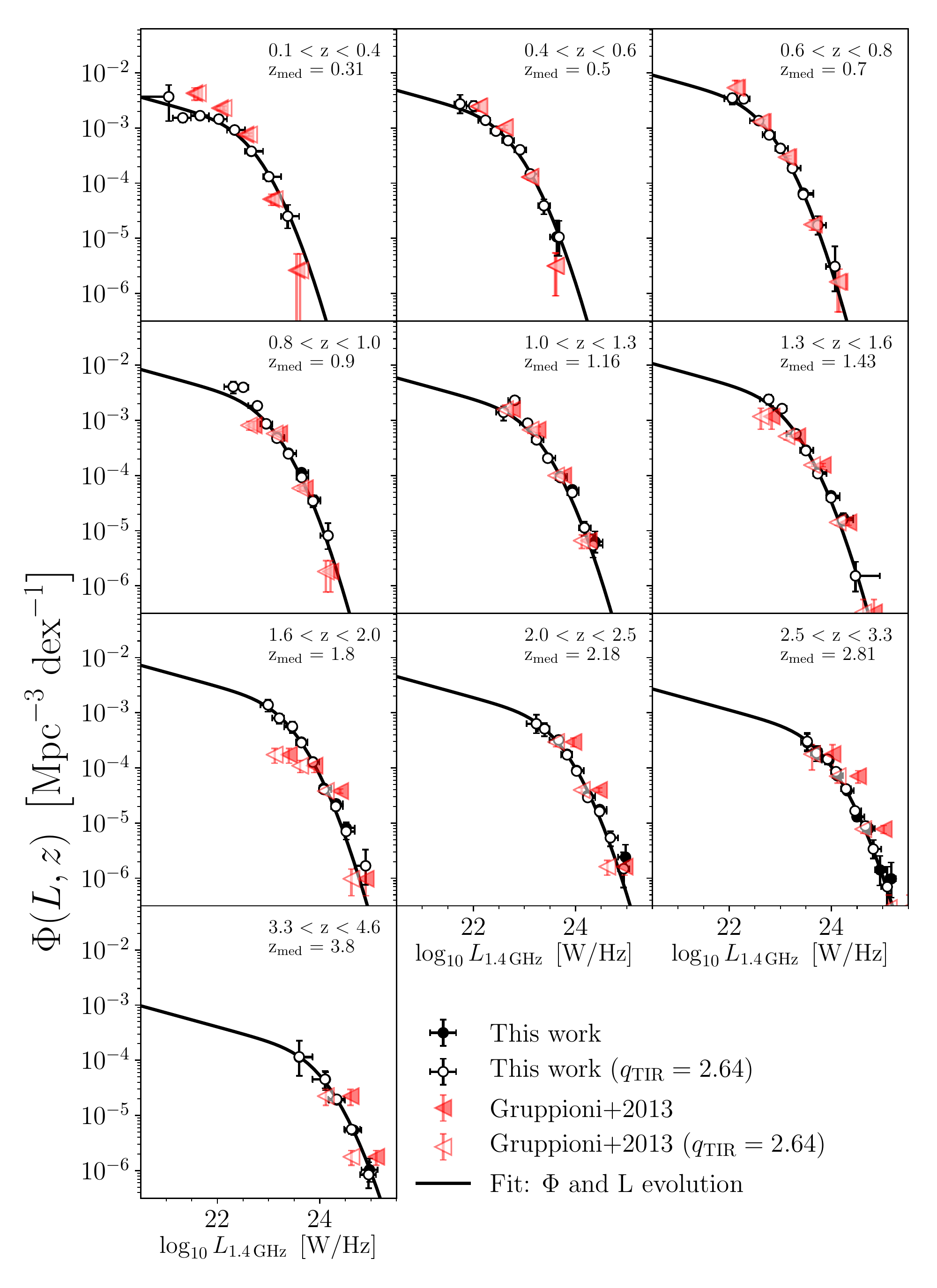}
\caption{Radio LF of SFGs in different redshift bins for the combined COSMOS-XS + VLA-COSMOS $\mathrm{3 \, GHz}$ data-sets derived using SFGS selected with an evolving FIR-radio correlation. Our best-fit density+luminosity function in each redshift bin is shown with solid lines. The redshift range and median redshift are given in each panel. The open circles show the radio LF of SFGs selected with a constant FIR-radio correlation. The small difference between the open and closed circles shows that the influence of our selection criterion used to select SFGs on the derived LF is small.
\newline We also show the LFs from \citet{Gruppioni_2013}, converted as described in Section~\ref{sec:IR} to radio LFs. The open symbols are converted assuming a constant FIR-radio correlation of 2.64 \citep{Bell_2003} and the filled symbols are converted assuming an evolving FIR-radio correlation. The difference between the open and filled symbols shows the influence of the FIR-radio correlation on the comparison between the radio LF and FIR LF. The FIR-radio correlation remains the largest uncertainty in this comparison.}
\label{fig:radio_q}
\end{figure*}

These sources do not show an excess in radio emission with respect to their FIR emission and are likely powered by SF. To assess the impact of this criterion, we also investigated using a non-evolving local value as defined by \citet{Bell_2003}:
\begin{align}
    q_\text{TIR}(z) >  2.64 - (3 \times \sigma) \, ,
    \label{eq:threshold_q_bell_inv}
\end{align}
where $\sigma=0.26$ is the 1$\sigma$ scatter in FIR-radio relation as found by \citet{Bell_2003}. This resulted in a sample containing 187 fewer SFGs than the original sample. The number of sources excluded by this new criterion is thus not much larger than excluded by Eq.~\ref{eq:threshold_q_inv}. This can also be seen in Fig.~\ref{fig:radio_q}, where the difference between the original sample and the sample derived with the new criterion is small. The biggest impact can be seen in the last two redshift bins, where the high luminosity points differ slightly in the new sample. We thus conclude that the influence of our selection criterion used to select SFGs is small.

\subsection{Radio spectral indices}
\label{sec:Radio spectral indices}
Where possible, we calculate the spectral index of our sources using the other radio data available over the field. In particular, we find that 8\% and 6\% of our sources have a spectral index calculated with the $\mathrm{1.4 \, GHz}$ data and the $\mathrm{10 \, GHz}$ data, respectively. However, we were unable to measure the spectral index for 86\% of our sample, as these sources were only detected at $\mathrm{3 \, GHz}$. Because our survey is $\sim \, 19$ times deeper than the $\mathrm{1.4 \, GHz}$ survey ($\sigma \, \sim \, 10 \, \mathrm{\mu Jy\,beam^{-1}}$, \citealt{Schinnerer_2010}), this induces a bias towards steeper spectra. Sources at the limit of our survey would need to have a spectral index of $\alpha \, = \, -3.9$ to be observed in the $\mathrm{1.4 \, GHz}$ survey. The median spectral index of sources matched at $\mathrm{1.4 \, GHz}$ is $\alpha \, = \, -0.91$. Because the $\mathrm{3 \, GHz}$ survey is matched in depth with the $\mathrm{10 \, GHz}$ survey, this bias does not exist for sources matched with the $\mathrm{10 \, GHz}$ data. The median spectral index of these sources is $\alpha \, = \, -0.63$. For the bulk of our sample, we therefore assume a standard spectral index of $\alpha \, = \, -0.7$, which is consistent with that typically found for SFGs \citep{Condon_1992, Kimball_2008, Murphy_2009, Smolcic_2017}.

An uncertainty in the spectral index of $\Delta \alpha = 0.1$ would change $L_{1.4\,\text{GHz}}$ by 0.08 dex and 0.11 dex at $z \, = \, 2$ and $z \, = \, 5$, respectively \citep{Novak_2018}. Assuming the canonical spectral index of $\alpha \, = \, -0.7$ thus adds a large uncertainty to the measured LF. However, the observed spread in spectral indices is symmetric ($\sigma \approx 0.35$; e.g., \citealt{Kimball_2008, Smolcic_2017}) and therefore expected to cancel out statistically. 

When we derive spectral indices, we assume the radio SED to be well described by a single power-law. However, there are processes which can alter the shape of the radio spectrum. For example, if thermal free-free emission substantially contributes to the radio emission \citep[e.g.,][]{Tabatabaei_2017, Tisanic_2019} the spectrum will flatten and the single power-law will not hold. 
Recent work by \citet{Algera_2021} using COSMOS-XS and COLDz on the radio spectra of high-redshift star-forming galaxies finds thermal fractions and synchrotron spectral indices typical of local star-forming galaxies, suggesting this is not a major source of uncertainty. Future deep, multi-frequency radio observations of larger samples will be necessary to study the radio SEDs of SFGs and understand the physical processes shaping them across cosmic redshift. 

\section{Cosmic star formation rate history}
\label{sec:CSFH}

In this section, we first discuss how to calculate the SFRD from the radio LFs (Section~\ref{sec:calculating_SFR}). We then discuss how the form of the LF fitted and FIR-radio conversion can affect the results (Sections~\ref{sec: different models} and \ref{sec:IR-radio conversion}, respectively). Finally, we compare our results to literature results derived from radio, FIR and UV observations (Section~\ref{sec:literature}).

\subsection{Calculating the SFRD}
\label{sec:calculating_SFR}
Having constructed the rest-frame $\mathrm{1.4 \, GHz}$ LF, it is now possible to establish the redshift evolution of the star formation rate density. To convert luminosity density into a star formation rate density, we use the functional form given in \citet{Delvecchio_2021}: 
\begin{align}
	\text{SFR}(L_{1.4\,\text{GHz}}) = f_{\text{IMF}} \, 10^{-24} \, 10^{q_\text{TIR}(z)} \, L_{1.4\,\text{GHz}} \, ,
	\label{eq:SFR}
\end{align}
where $\text{SFR}$ is the star formation rate in units of $\text{M}_{\odot}/\text{yr}$, $f_{\text{IMF}}$ is a factor accounting for the IMF ($f_{\text{IMF}} \, = \, 1$ for a Chabrier IMF and $f_{\text{IMF}} \, = \, 1.7$ for a Salpeter IMF) and $L_{1.4\, \text{GHz}}$ is the rest-frame $\mathrm{1.4 \, GHz}$ luminosity in units of $\text{W Hz}^{-1}$. \citet{Novak_2017} stresses that since low-mass stars do not contribute significantly to the total light of the galaxy, only the mass-to-light ratio is changed when the Chabrier IMF is used. Following \citet{Novak_2017}, we therefore used the Chabrier IMF.

The SFRD can then be estimated by taking the luminosity-weighted integral of the analytical form of the fitted LF and converting the luminosity in the integral to SFR. The integral of the SFRD can thus be written as:

\begin{align}
	\text{SFRD} = \int_{L_{\text{min}}}^{L_{\text{min}}} \Phi(L,z,\alpha_{L}) \times \text{SFR}(L_{1.4\,\text{GHz}}) \, \mathrm{d} \log_{10} L \,.
	\label{eq:SFRD}
\end{align}

This integral gives the SFRD of a given epoch. Unless stated otherwise, all results show the SFRD obtained by integrating the fitted LF from 0.0 to $\to \infty$. Our errors are estimated from the fitting parameters uncertainties through boostrapping whereby the uncertainties in $q_{\text{TIR}}(z)$ are taken into account. The quoted errors do not account for any systematic errors due to cosmic variance.

\subsection{Luminosity evolution vs. density and luminosity evolution}
\label{sec: different models}
\begin{figure*}
    \centering
        \subfigure
    {
        \includegraphics[width=1.0\columnwidth]{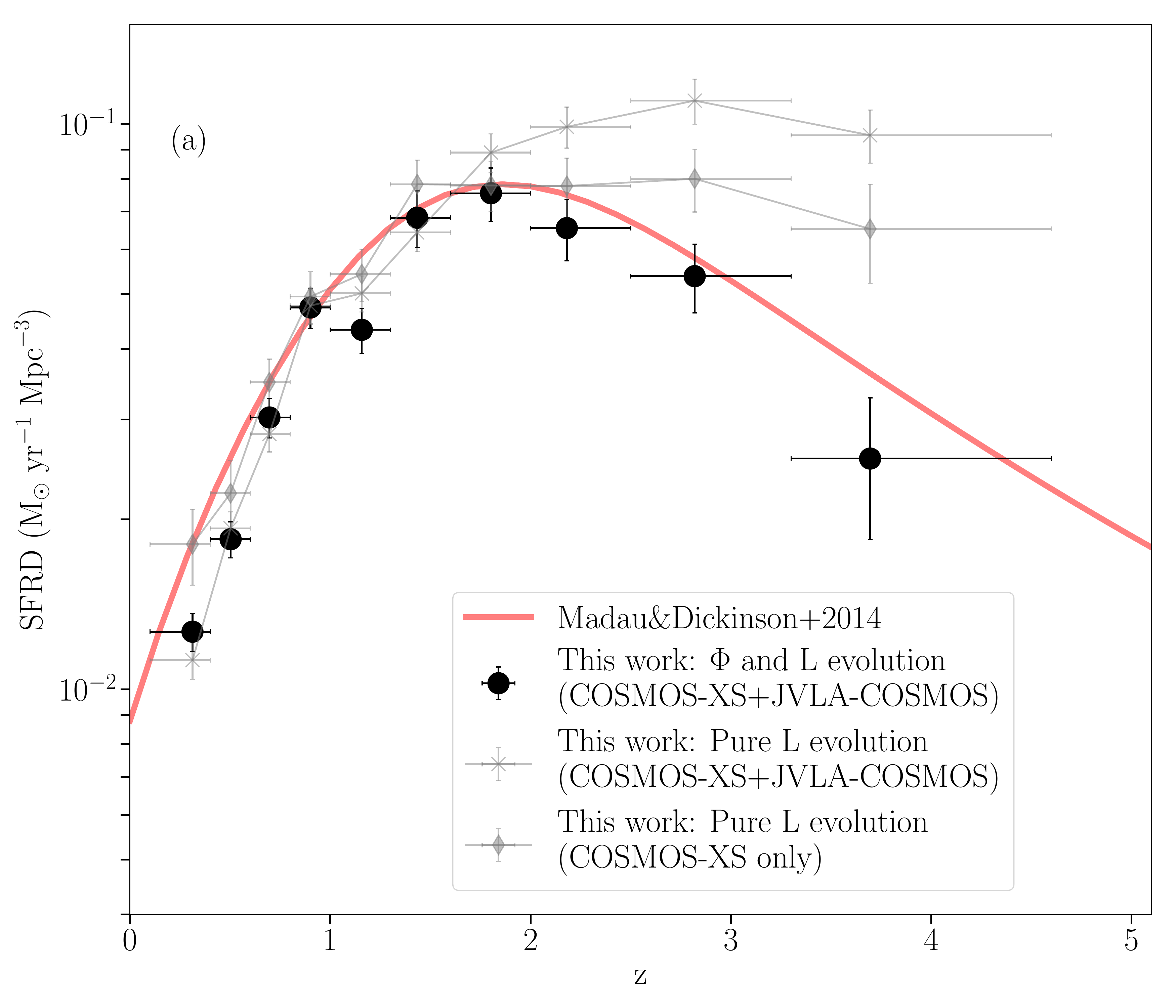}
        \label{fig:SFRD_D_L}
    }
        \subfigure
    {
        \includegraphics[width=1.0\columnwidth]{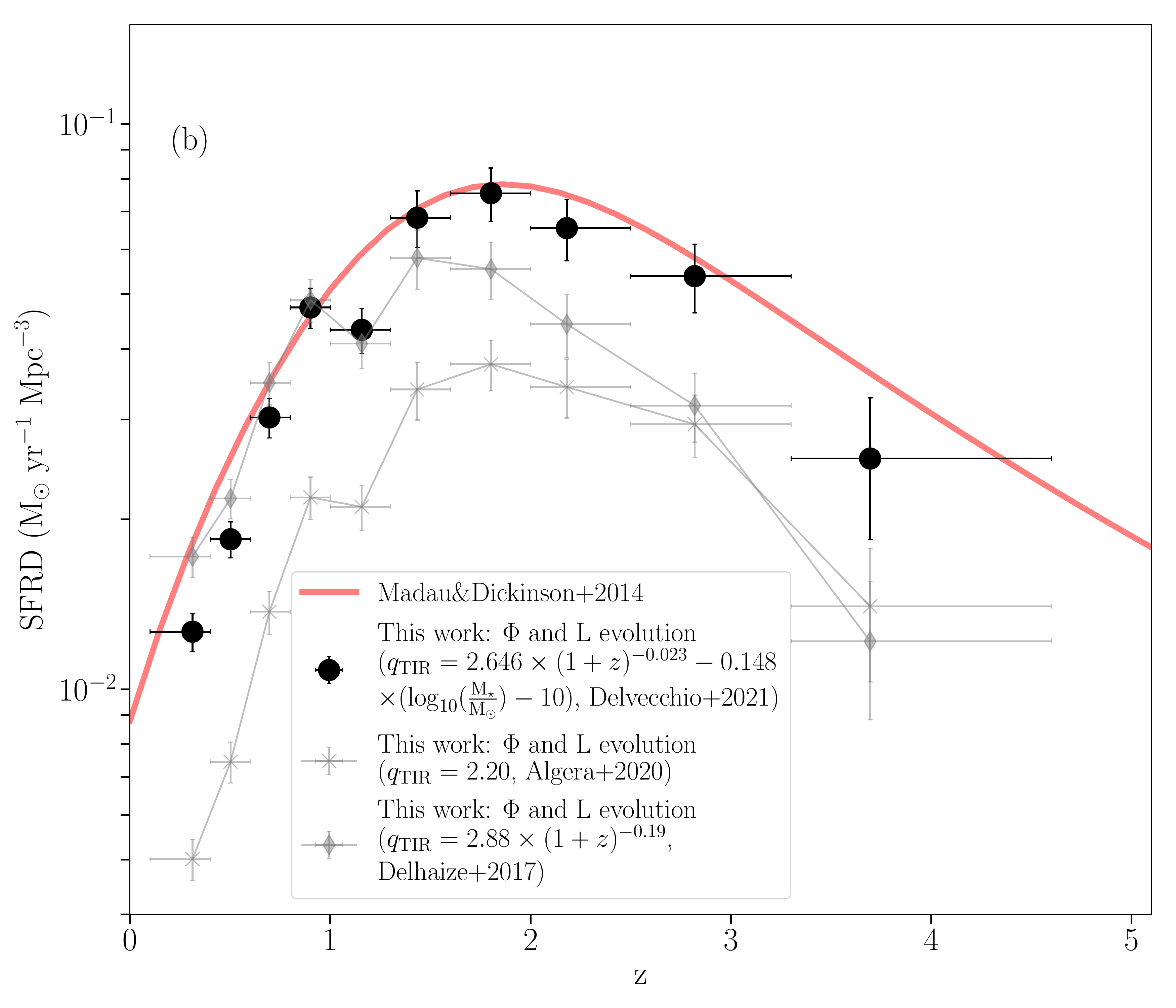}
        \label{fig:SFRD_q}
    }
    \caption{The impact of the fitted LF form and assumed FIR-radio correlation on the derived cosmic star formation rate density (SFRD). The left panel shows the effect of fitting pure luminosity evolution compared to the preferred model of density+luminosity evolution (assuming the same FIR-radio correlation). 
    \newline The right panel shows the SFRD obtained from the combined COSMOS-XS + VLA-COSMOS 3 GHZ data-sets (assuming density+luminosity evolution) but for different assumed FIR-radio correlations. This gives an indication of the impact an assumed FIR-radio correlation has. The study of \citet{Madau_2014} is shown as a red line in both panels. In the remainder of the paper, we convert our radio LFs to SFRD using the \citet{Delvecchio_2021} FIR-radio correlation.}
    \label{fig:all_SFRD}
\end{figure*}

Fig.~\ref{fig:SFRD_D_L} shows the SFRD computed using the different fits to the radio LF discussed in Section~\ref{sec:pure luminosity evolution} and Section~\ref{sec:luminosity and density evolution}, and using only the COSMOS-XS data compared to the combination of the COSMOS-XS + VLA-COSMOS $\mathrm{3 \, GHz}$ data-sets. The combination enables us to fit not only pure luminosity evolution, but also to constrain the joint density+luminosity evolution. 

When we compare all three results, we find that they all roughly agree up to $z \, \sim \, 1.8$. At that point, both pure luminosity evolution model fits show an elevated SFRD at high redshift compared to the \citet{Madau_2014} curve \citep[as also seen for the VLA-COSMOS 3 GHz data alone][]{Novak_2017}. However, when we fit density+luminosity evolution to the combined data-sets, as favored by the data (Fig.~\ref{fig:radio_LF_inc_novak}), we instead find that the SFRD falls below the \citet{Madau_2014} curve at $z \, \gtrsim \, 1.8$. In the following sections, we will use the SFRD derived from the combined COSMOS-XS $+$ VLA-COSMOS $\mathrm{3 \, GHz}$ data-sets using density+luminosity evolution for the comparison with the SFRD derived from radio, FIR and UV observations. 

\subsection{FIR-radio conversion}
\label{sec:IR-radio conversion}
As Eq.~\ref{eq:SFR} shows, the calibration of the SFR depends on $q_\text{TIR}(z)$ (see Section~\ref{sec:sample selection}). Therefore not only is the FIR-radio correlation one of the uncertainties in the conversion from FIR luminosities to radio luminosities, as discussed in Section~\ref{sec:IR}, but it is also one of the main uncertainties in the SFRD calculation. Current observations do not favor a constant $q_\text{TIR}(z)$ \citep{Magnelli_2015, Delhaize_2017, Calistro_2017}, although there is some discussion as to whether this evolution can be ascribed to AGN activity \citep{Molnar_2018} or selection biases such as the sampling of high mass galaxies at high redshift \citep{Smith_2021} and/or a redshift-dependent sampling of different parts of a non-linear FIR/SFR relation \citep{Molnar_2021}. To illustrate the impact of the assumed FIR-radio correlation on the comparison between the FIR LF and radio LF, we show in Fig.~\ref{fig:radio_q} the data from \citet{Gruppioni_2013} converted using an evolving $q_\text{TIR}(z)$ (Eq.~\ref{eq:qtir_inv}) and using the local constant value for the FIR-radio correlation: $q_\text{TIR} \, = \, 2.64$ \citep{Bell_2003}. The difference between the two samples increases with redshift as expected due to the growing difference between the evolving and non-evolving $q_\text{TIR}$. Fig.~\ref{fig:radio_q} also shows that if we assume $q_\text{TIR}(z)$ to be constant at the local value of 2.64, our radio LFs would match the FIR LFs better. 

The impact of the $q_\text{TIR}(z)$ on the SFRD derived from the radio LFs is shown in Fig.~\ref{fig:SFRD_q}.All three curves show density+luminosity evolution fitted to the combined COSMOS-XS + VLA-COSMOS 3 GHz sample, but with different values of $q_\text{TIR}(z)$. These values are derived by \citet{Delhaize_2017}, \citet{Algera_2020b} and \citet{Delvecchio_2021}. Fig.~\ref{fig:SFRD_q} also shows the fit from \citet{Madau_2014} based on a collection of previously published UV and FIR data. The first FIR-correlation we consider is from \citet{Delhaize_2017}. They constrained the evolution $q_\text{TIR}(z)$ using a doubly censored survival analysis on $\sim \, 10,000$ SFGs. To prevent from biases towards low and high average $q_\text{TIR}(z)$ measurements, these star-forming sources are jointly-selected in radio observations at $\mathrm{3 \, GHz}$ and FIR observations. Assuming an average spectral index of $-0.7$, \citet{Delhaize_2017} find that $q_\text{TIR}(z)$ decreases with redshift as:

\begin{align}
	q_\text{TIR}(z) \, = \, (2.88 \pm 0.03) \times (1+z)^{-0.19 \pm 0.01}\,.
	\label{eq:qtir_delhaize}
\end{align}

Fig.~\ref{fig:SFRD_q} shows that this adopted $q_\text{TIR}(z)$ has a large impact on the evolution of the SFRD due to the steep evolution of $q_\text{TIR}(z)$ with redshift. The SFRD matches the fit from \citet{Madau_2014} at $z \, < \, 1$ well, after which there is an increasing and systematic discrepancy with redshift toward low implied SFRD values. 

We next consider the FIR-radio relation from the recent study by \citet{Delvecchio_2021}. They calibrated $q_\text{TIR}(z)$ with a stacking analysis in the radio/FIR of a mass-selected sample of more than 400,000 SFGs in the COSMOS field. \citet{Delvecchio_2021} find that $q_\text{TIR}(z)$ evolves primarily with $\text{M}_{\star}$. A secondary, weaker dependence on redshift is also observed. The $q_\text{TIR}(\text{M}_{\star}, z)$ is quantified as:

\begin{align}
	q_\text{TIR}(\text{M}_{\star}, z) \, = \, &(2.646 \pm 0.024) \times (1+z)^{-0.023 \pm 0.008} \notag \\ 
	&- (0.148 \pm 0.013) \times (\log_{10} \frac{\text{M}_{\star}}{\text{M}_{\odot}}-10) \,.
	\label{eq:qtir_delvecchio}
\end{align}

In order to use Eq.~\ref{eq:qtir_delvecchio} to derive the SFRD, we need to have a mass for the sample used to derive the radio LF as shown in Fig.~\ref{fig:radio_LF_inc_novak}. We used the mass given by the COSMOS2015 catalog for the sources that could be matched with this catalog. We then derived the mean mass per redshift bin, ranging from $10^{10.18} \, \text{M}_{\odot}$ to $10^{10.70} \, \text{M}_{\odot}$, 
to find $q_\text{TIR}(\text{M}_{\star}, z)$. Fig.~\ref{fig:all_SFRD} shows that the SFRD derived with $q_\text{TIR}(\text{M}_{\star}, z)$ described in Eq.~\ref{eq:qtir_delvecchio} has a weaker dependence on redshift compared to \citet{Delhaize_2017} and results in the best match with the compilation from \citet{Madau_2014} over the whole redshift range. 

Lastly, we consider the FIR-radio correlation from \citet{Algera_2020b}, which focuses on a luminosity-limited sample SMGs. They find $q_\text{TIR}(z) \, = \, 2.20 \, \pm \, 0.03$, where they have addressed the incompleteness in the radio observations through a stacking analysis, and they find no evidence of evolution between $1.5 \, \leq \, z \, \leq \, 4.0$. We note that the SMG sample is not well matched to the radio sample observed by the COSMOS-XS survey and the VLA-COSMOS $\mathrm{3 \, GHz}$ Large Project. However, at $z \, \gtrsim \, 2$ the sample would be a better match as our sample traces SFGs with a high SFR. In addition, the derived $q_\text{TIR}(z)$ is free from some of the biases that come into play in $q_\text{TIR}(z)$ estimates from studies based on radio-selected samples. As expected, Fig.~\ref{fig:all_SFRD} shows that the SFRD calculated with $q_\text{TIR}(z) \, = \, 2.20$ does not match the fit from \citet{Madau_2014} at $z \, < \, 1$, and we see that the SFRD values are in fact systematically low at all redshifts. 

In summary, we show that the assumed FIR-radio relation has a significant impact on the derived SFRD. We find that the recent study by \citet{Delvecchio_2021}, which constitutes the first calibration of the FIR-radio correlation as a function of both stellar mass and redshift, shows the best agreement with the multi-wavelength compilation from \citet{Madau_2014}, while the other two FIR-radio relations explored result in under-predicted SFRDs at high redshift. To be consistent with our sample selection described in Section~\ref{sec:sample selection} and given that the \citet{Delvecchio_2021} result was also derived with a large unbiased sample using some of the deepest radio and FIR images available over the same field as our observations, we will use the FIR-radio correlation from \citet{Delvecchio_2021} to convert our radio LFs to SFRD in the following.

\subsection{Comparison with the literature}
\label{sec:literature}
\begin{figure*}
    \centering
    \subfigure
    {
        \includegraphics[width=0.94\columnwidth]{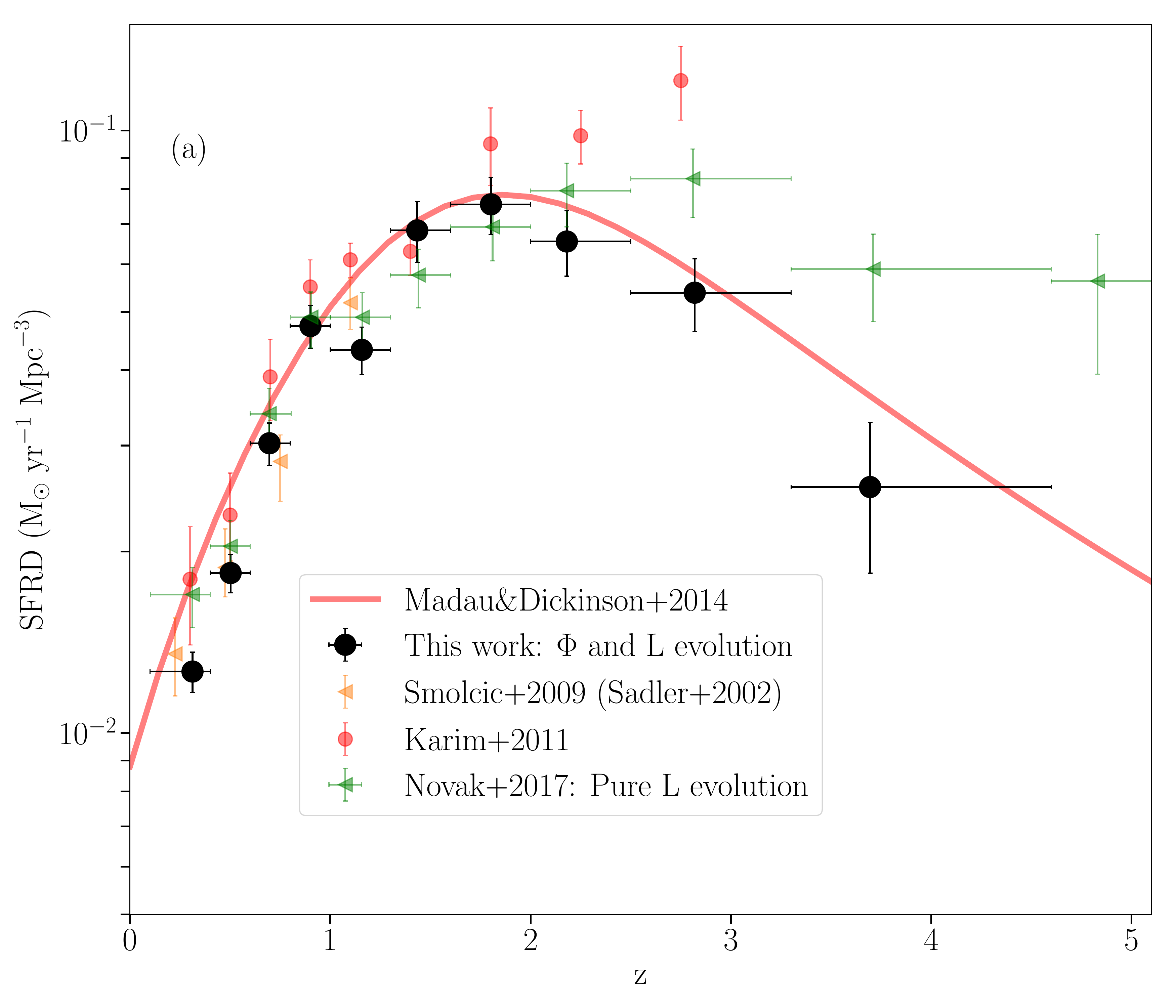}
        \label{fig:SFRD_radio}
    }
    \subfigure
    {
        \includegraphics[width=0.94\columnwidth]{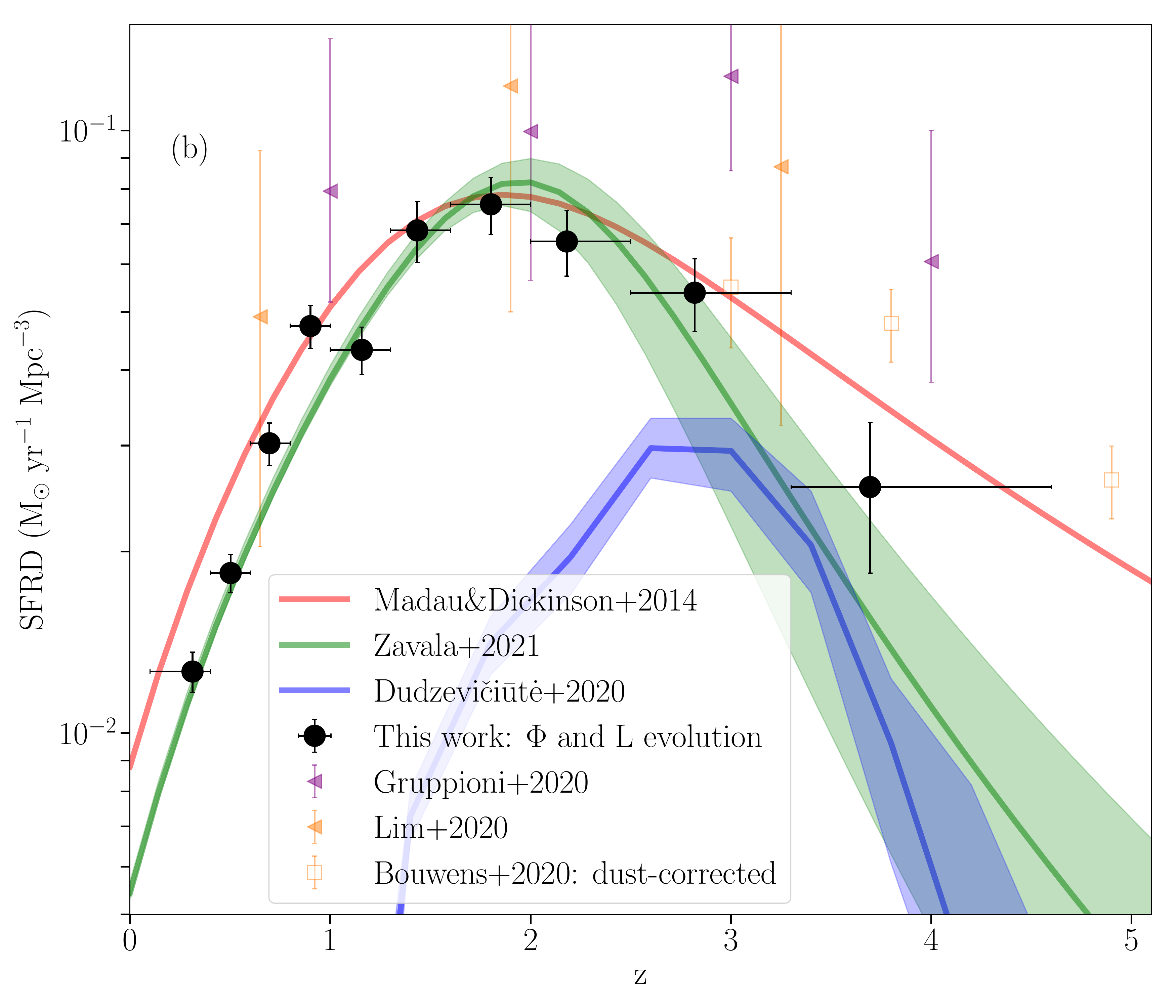}
        \label{fig:SFRD_IR}
    }
    \subfigure
    {
        \includegraphics[width=1.8\columnwidth]{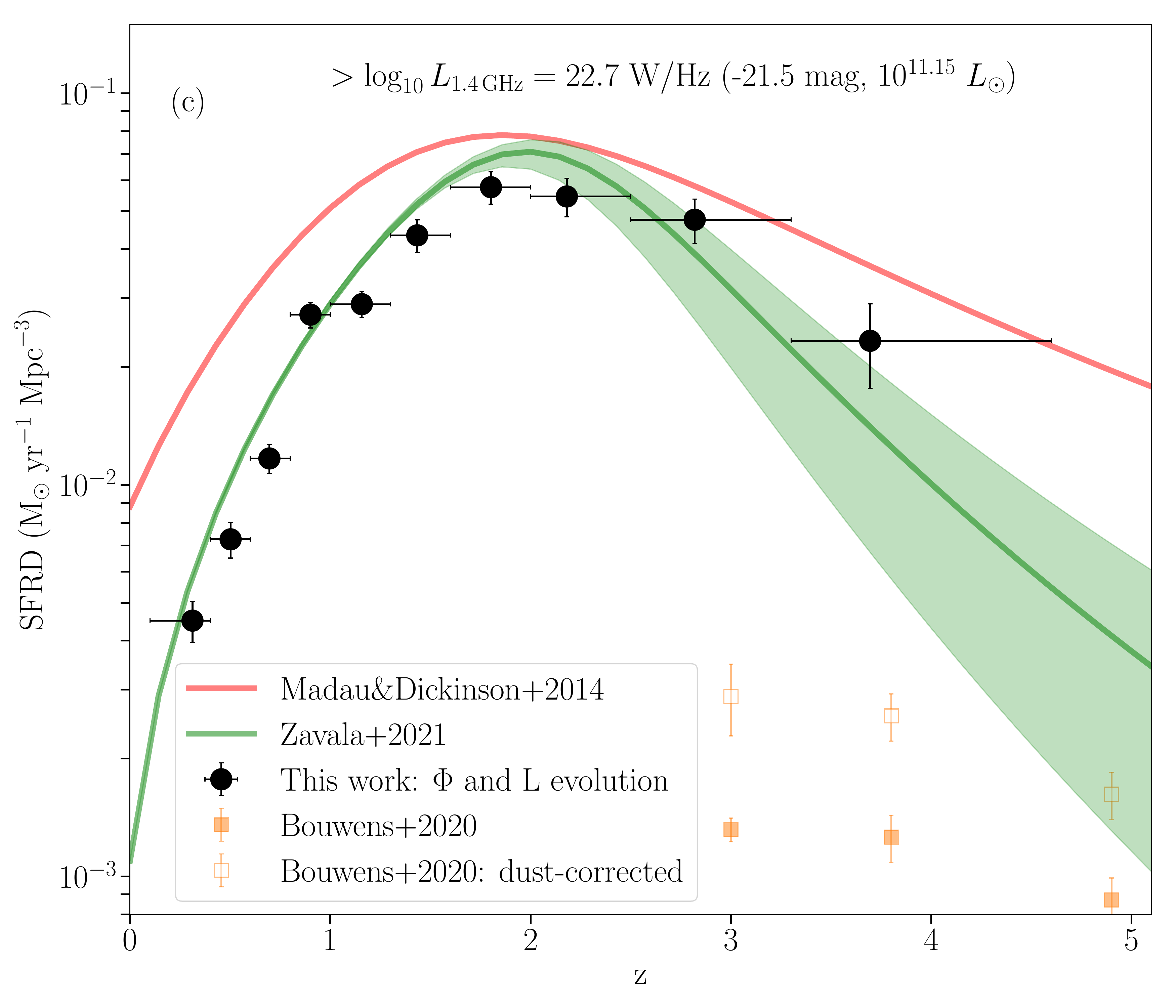}
        \label{fig:SFRD_UV}
    }
    \caption{Cosmic star formation rate density (SFRD) history. Our SFRD history is shown with filled circles in all panels and is obtained from the combined COSMOS-XS + VLA-COSMOS 3 GHZ data-sets (assuming density+luminosity evolution). The study of \citet{Madau_2014} is shown as a red line in all panels. All data shown for comparison are indicated in the legend of each panel; see text for details. The comparison of radio- and UV-based SFRDs, integrated down to the same limit, in panel (c), shows that the UV-based SFRD from \citet{Bouwens_2020} falls $\sim \, 1\, \text{dex}$ below the radio SFRD at $z \, \gtrsim 2.8$. This suggests that the bulk of the star formation contributed by high-luminosity sources at high redshifts is not accounted for by dust corrections.}
    \label{fig:all_SFRD_lit}
\end{figure*}

In Fig.~\ref{fig:all_SFRD_lit}, we show the redshift evolution of the cosmic star formation density derived from this work compared with work in the literature derived at different wavelengths. The study of \citet{Madau_2014} is shown in all panels for ease of comparison. Below $z \, < \, 2$, our data agree well with the compilation from \citet{Madau_2014}, although we observe some scatter in our SFRD estimates around $z \, \sim \, 0.9$ which is likely due to cosmic variance (see Appendix~\ref{sec:Cosmic variance}). Our SFRD turns over at $z \, \sim \, 1.8$ and falls more rapidly than \citet{Madau_2014} out to high-redshift. 

In Fig.~\ref{fig:SFRD_radio} we show our derived SFRD compared to radio observations. \citet{Smolcic_2009_a} derived the SFRD out to $z \, = \, 1.3$ from VLA imaging at $\mathrm{1.4 \, GHz}$. They assumed pure luminosity evolution for the local LF, a non-evolving FIR-radio correlation established by \citet{Bell_2003} and integrated over the full luminosity range. We find a good match with the SFRD derived by \citet{Smolcic_2009_a} despite the different assumptions, though we note that they are only sensitive to lower redshifts ($z\lesssim1$) where the different assumptions have a smaller effect. \citet{Karim_2011} performed stacking on mass selected galaxies and find a rise up to $z \, \sim \, 3$. This rise is mainly due to the fact that they use a non-evolving FIR-radio correlation established by \citet{Bell_2003}. Because this correlation does not evolve towards a lower $q_\text{TIR}$ value at high redshift, the resulting SFRD will be higher at higher $z$ as discussed in Section~\ref{sec:IR-radio conversion}. Finally, as discussed in Section~\ref{sec:VLA-COSMOS 3 GHz Large Project}, \citet{Novak_2017} used the VLA-COSMOS $\mathrm{3 \, GHz}$ Large Project to derive the SFRD up to $z \, = \, 6$. They assumed pure luminosity evolution and an evolving FIR-radio correlation $q_{\text{TIR}}(z)$ derived by \citet{Delhaize_2017}. Below $z \, \sim \, 2$, our data agree well with the SFRD derived by \citet{Novak_2017}. However, our SFRD declines towards a much lower value than \citet{Novak_2017} found for $z \, \gtrsim \, 2.2$. This is due to the fitted density evolution, as discussed in Section~\ref{sec:evolution parameters}, and can also be seen from Fig~\ref{fig:SFRD_q}. The offset would be even larger if we would have used a similar FIR-radio correlation as \citet{Novak_2017} used. This can be seen from Fig.~\ref{fig:SFRD_D_L}, which shows the large offset between the SFRD we calculate assuming the FIR-radio correlation from \citet{Delhaize_2017} compared to that of \citet{Delvecchio_2021}.

In Fig.~\ref{fig:SFRD_IR}, we compare our measurement of the SFRD to results from recent FIR observations from \citet{Gruppioni_2020} and \citet{Lim_2020}. \citet{Gruppioni_2020} derive the dust-obscured SFRD using the serendipitously detected sources in the ALPINE survey. In this case, the SFRD is derived from an extrapolation of the FIR LF, where the LF (shown in Fig.~\ref{fig:radio_LF_IR}) is integrated down to $\log_{10}(L_{\text{IR}}/L_{\odot}) \, = \, 8$. \citet{Lim_2020} derive the SFRD by integrating the FIR LF shown in Fig.~\ref{fig:radio_LF_IR} inferred using SCUBA-2 $450 \, \mathrm{\mu m}$ observations. They used the integration limits of $L_{\text{min}} \, = \, 0.03 L_{\star}$ and $L_{\text{max}} \, = \, 10^{13.5} L_{\odot}$. This integration is necessary in both studies since the data only constrains a small part of the LF, as can be seen in Fig.~\ref{fig:radio_LF_IR}. 

The first thing that stands out from Fig.~\ref{fig:SFRD_IR} are the large error bars found in the studies of \citet{Gruppioni_2020} and \citet{Lim_2020}, which are due to the small sample of sources considered in these studies. The observations by \citet{Lim_2020} are still in agreement with our observations within these error margins. The next thing to note is that both FIR studies find a higher SFRD over the whole redshift range compared to the radio SFRD. This cannot be explained by the different integration limits, which should result in a higher radio SFRD as this is computed over the full luminosity range. However, \citet{Zavala_2021} suggest that the SFRD found by \citet{Gruppioni_2020} may be unusually high due to possible clustering of the serendipitous targets. 

Fig.~\ref{fig:SFRD_IR} also shows results from recent sub-mm observations from \citet{Dudzeviciute_2020} and \citet{Zavala_2021}. \citet{Dudzeviciute_2020} used the AS2UDS sample from the $\sim 1 \text{deg}^2$ SCUBA-2 survey to derive the SFR from {\sc{magphys}} fits. The SFRD is then found from an extrapolation of the ${\rm 870\mu m}$ flux limit of 3.6 mJy to 1 mJy (equivalent to $L_{\text{IR}} \, \approx \, 10^{12} L_{\odot}$) using the slope from the sub-millimeter counts in \citet{Hatsukade_2018}. Because of the area covered by this survey, it is likely to be much more representative than smaller volume studies such as \citet{Gruppioni_2020}. The curve from \citet{Dudzeviciute_2020} does not match our radio SFRD at $z \, \lesssim \, 3$ because this curve does not represent the total SFRD but shows the SMG contribution. The study by \citet{Dudzeviciute_2020} demonstrates that the activity of SMGs peaks at $z \sim 3$, suggesting that more massive and obscured galaxies are more active at earlier times. At $z \, \sim \, 4$ the curve is roughly consistent with our data. \citet{Zavala_2021} used the results from the MORA survey to search for DSFGs at $2 \, \mathrm{mm}$. The number counts from the survey are combined with the number counts at 1.2 and $3 \, \mathrm{mm}$ to place constraints on the evolution of the FIR LF by making use of the evolution model of \citet{Casey_2018}. The SFRD is then found by integrating the best-fit FIR LF with an integration interval of $\log_{10}(L_{\text{IR}}/L_{\odot}) \, = \, [9, 13.8]$. The curve from \citet{Zavala_2021} is consistent with our data despite the different integration limits.

In addition, Fig.~\ref{fig:SFRD_IR} shows the results from the UV observations from \citet{Bouwens_2020}. They make use of ALMA observations for a sample of galaxies in the HUDF at $1.5 \, < \, z \, < \, 10$ to provide improved constraints on the IRX-$\beta$ relation. \citet{Bouwens_2020} integrate their UV LFs from $0.03 \times L_{\star}$ to $\to \infty$ in order to derive the SFRD. The radio SFRD matches the UV SFRD at $z \, \sim \, 3$ and at $z \, > \, 3$ the UV SFRD rises above the radio SFRD. However. it is import to realize that the UV SFRD and radio SFRD compared in Fig.~\ref{fig:SFRD_IR} are derived using different integration limits.

Differing integration limits will have a more substantial effect for the comparison between our radio-based study and UV-based studies, given the different shapes of the derived LFs evident in Fig.~\ref{fig:radio_LF_UV}. To investigate the impact of the integration limits, in Fig.~\ref{fig:SFRD_UV}, we compare our radio-based results with the FIR-based study from \citet{Zavala_2021} and the UV-based study from \citet{Bouwens_2020}, but now using a consistent integration limit across all studies except for the compilation from \citet{Madau_2014} which remains unchanged for ease of comparison. In particular, \citet{Bouwens_2020} originally integrate their UV LFs from $0.03 \times L_{\star}$ to $\to \infty$ in order to derive the SFRD. However, a fairer comparison of the radio- and UV-based SFRDs necessitates that they be integrated down to the same limit. As the radio observations do not reach the faint luminosities the UV observations reach, we have chosen the integration limit as the luminosity limit reached by the radio observations between $z \, = \, 1.15$ and $z \, = \, 1.43$, which is $\log_{10} L_{1.4 \, \text{GHz}} \, = \, 22.7 \, \text{W} \text{Hz}^{-1}$. This corresponds to a luminosity limit of $-21.5 \, \text{mag}$ ($0.038 \, L^{\star}_{z=3}$) for the UV LF. For the \citet{Zavala_2021} FIR-based study, this corresponds to an FIR luminosity limit of of $10^{11.15} \, L_{\odot}$.

Below $z \, < \, 2$, Fig.~\ref{fig:SFRD_UV} shows that the radio data now falls below the SFRD from \citet{Madau_2014}, which can be explained by the limit that has been set for the integration of the radio LF. For $z \, \gtrsim \, 2.2$, the difference between our radio-based SFRD and \citet{Madau_2014} becomes similar to what was found in Fig.~\ref{fig:SFRD_radio} and Fig.~\ref{fig:SFRD_IR}. The \citet{Zavala_2021} curve appears similarly affected by the new integration limits, now falling below the \citet{Madau_2014} compilation, but continuing to follow the radio-based SFRD reasonably well. 

In contrast, the UV-based SFRD from \citet{Bouwens_2020} falls $\sim \, 1\, \text{dex}$ below the radio SFRD at $z \, \gtrsim 2.8$. This result is very different from a naive comparison between the radio and UV-based SFRDs using their respective nominal integration limits, which would result in a reasonable match of the SFRDs even at the high redshift end. However, this can be explained by a ``conspiracy'' between the amount in which different sources contribute to the LFs at the different wavelengths. Observations in the UV find that the faint-end slope of the UV LF at high redshift is very steep, and the bulk of the luminosity at high redshift is thus coming from faint sources, as can be seen in Fig.~\ref{fig:radio_LF_UV}. Our radio observations, on the other hand, suggest a much shallower faint-end slope, but they instead find a significant amount of star formation in high-luminosity sources that is missed by UV observations. When the integration limit is thus fixed to avoid extrapolating the radio LFs significantly below our detection limit, we find a significant discrepancy in the resulting SFRDs. Fig.~\ref{fig:SFRD_UV} shows that this is true even when UV observations are corrected for dust. In particular, \citet{Bouwens_2020} make use of improved constraints on the IRX-$\beta$ relation. This $\sim \, 1\, \text{dex}$ discrepancy in the resulting SFRDs therefore suggests that the bulk of the star formation contributed by high-luminosity sources at high redshifts is not accounted for by dust corrections. As discussed in Section~\ref{sec:UV}, including ``optically dark'' sources would only increase this discrepancy further. 

\section{Summary \& Conclusions}
\label{sec:Conclusions}
We studied a 3\,GHz-selected sample of star-forming galaxies (SFGs) identified in the ultra-deep, multi-band COSMOS-XS survey. Using the deep multi-wavelength data available in the COSMOS field, and selecting SFGs based on the FIR-radio correlation, we identify $\sim$1300 SFGs with redshifts out to $z\sim4.6$. We use this SFG sample to study the evolution of the radio luminosity function (LF) with redshift. 

We fit our radio LFs with a modified-Schechter function evolved in luminosity (pure luminosity evolution). By fixing the faint and bright end shape of the radio LFs to the local values, we find a strong trend in redshift for the luminosity parameter of $\alpha_L \propto (3.40 \pm 0.11) - (0.48 \pm 0.06)z$. This evolution agrees with what has been reported in previous radio-based studies \citep[e.g.,][]{Novak_2017}. 

We then combined the ultra-deep COSMOS-XS data-set with the shallower VLA-COSMOS $\mathrm{3 \, GHz}$ large project data-set over the wider COSMOS field. This combination increases our dynamic range to include both the faintest and brightest sources, allowing us to simultaneously constrain the density and luminosity evolution. Doing so, we find evidence for significant density evolution over the observed redshift range.

In order to compare our radio LFs to FIR LFs, we converted FIR luminosities to radio luminosities using a redshift-dependent FIR-radio correlation. We find that our LFs agree well with the FIR LFs at $z \, < \, 2$. At $z \, > \, 2$ our LFs are systematically lower than \citet{Gruppioni_2013}, which we attribute at least partly to AGN contamination. In addition, we find that the radio data is most consistent with the dust-poor model from \citet{Casey_2018}. 

We also compare the radio LFs to the UV LFs of \citet{Metha_2017}, \citet{Ono_2018} and \citet{Bouwens_2021}, which are based on UV rest-frame observations of Lyman break galaxies. By fitting the local LF to the UV and UV+radio LFs and integrating down to $0.03 \, L^{\star}_{z=3}$, we find evidence for a significant underestimation of the UV LF by $21.6\% \, \pm \, 14.3 \, \%$ at high redshift ($3.3\, < \, z \, < \, 4.6$). We attribute this underestimation to appreciable star formation in highly dust-obscured galaxies. 

We integrate the derived radio LFs with joint density+luminosity evolution to determine the cosmic star formation rate density (SFRD). We find the radio-derived SFRD to be consistent with the established behavior at low redshift, where it increases strongly with redshift out to $z \, \sim \, 1.8$. The radio-based SFRD then declines more rapidly out to high-redshift than previous radio-based estimates, and is more consistent with the recent FIR-based estimated from \citet{Zavala_2021}.

In order to more directly compare the radio-based SFRD derived here with the recent UV-based SFRD from \citet{Bouwens_2020}, and to avoid extrapolating far below the radio detection limit, we integrate both LFs down to a consistent limit ($0.038 \, L^{\star}_{z=3}$). This direct comparison reveals that the discrepancy between the radio and UV LFs discussed above translates to an even more significant ($\sim$1\,dex) discrepancy between the radio- and UV-based SFRDs at high redshifts ($z \, > \, 3$). This discrepancy persists even when the UV observations are corrected for dust obscuration assuming the latest dust corrections. The discrepancy would only increase with the inclusion of ``optically dark'' sources, which will be discussed further in a future paper. 

\section*{Acknowledgements}
The authors wish to thank Mara Salvato for providing us with the COSMOS spectroscopic master catalog. We thank Ivan Delvecchio, Mara Salvato and Vasily Kokorev for helpful comments on the manuscript.
The National Radio Astronomy Observatory is a facility of the National Science Foundation operated under cooperative agreement by Associated Universities, Inc. D.vdV. and J.H. acknowledge support of the VIDI research programme with project number 639.042.611, which is (partly) financed by the Netherlands Organisation for Scientific Research (NWO). H.S.B.A. acknowledges support from NAOJ ALMA Scientific Research Grant Code 2021-19A. I.S acknowledges support from STFC (ST/T000244/1). D.R. acknowledges support from the National Science Foundation under grant number AST-1614213. D.R. also acknowledges support from the Alexander von Humboldt Foundation through a Humboldt Research Fellowship for Experienced Researchers. This research made use of ASTROPY, a community developed core Python package for astronomy (\citealt{astropy:2013, astropy:2018}) hosted at http://www.astropy.org/, matplotlib (\citealt{Hunter_2007}), numpy (\citealt{Walt_2011}), scipy (\citealt{Jones_2001}), and of TOPCAT (\citealt{Taylor_2005}).

\appendix
\section{Cosmic variance}
\label{sec:Cosmic variance}

\begin{figure*}
\centering
\includegraphics[width=\textwidth]{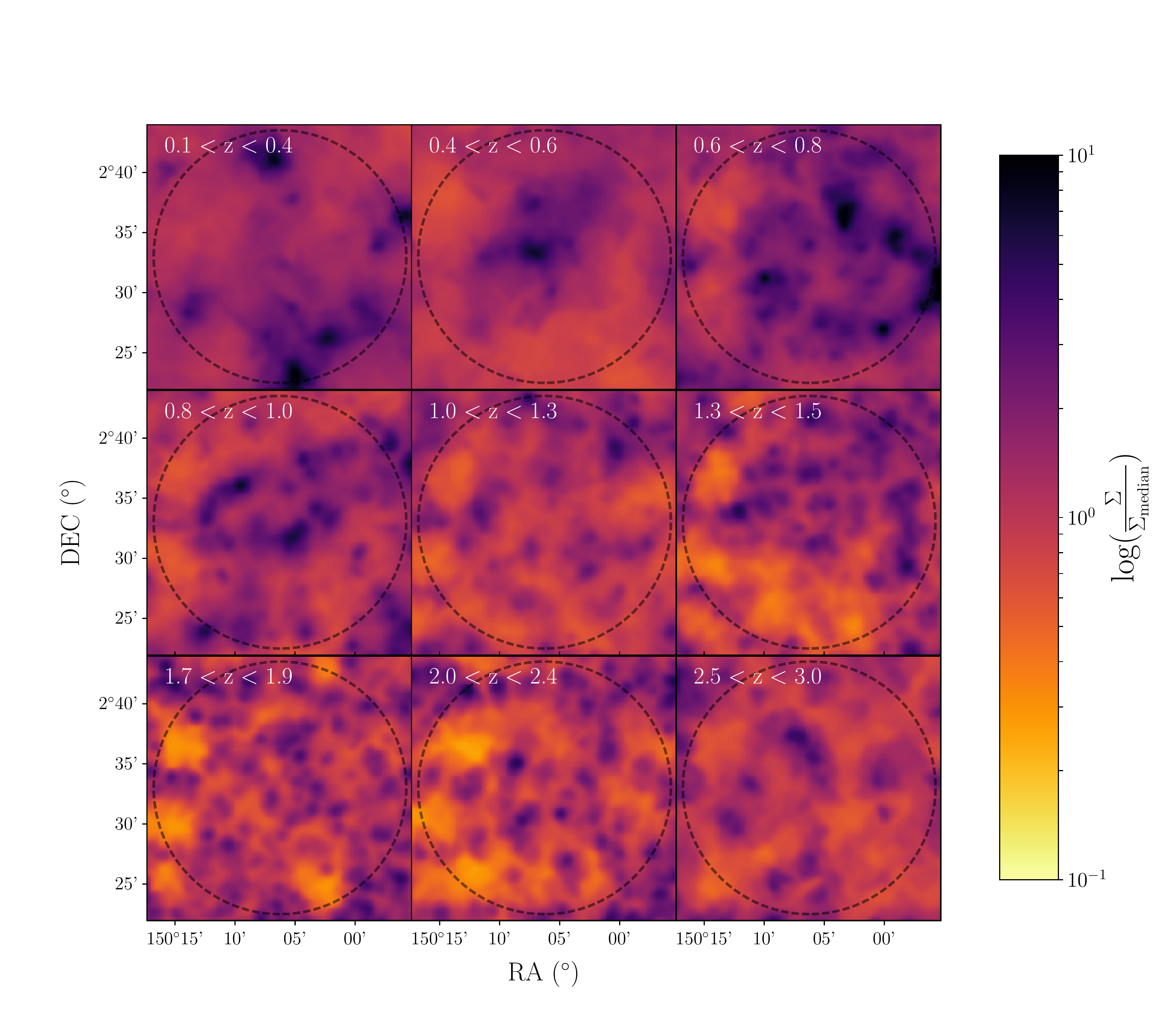}
\caption{Density maps as a function of redshift in the COSMOS-XS field of view constructed by \citet{Scoville_2013} with the Voronoi technique. The images were made by summing the derived overdensities measured from the individual redshift slices. The colorbar corresponds to the density unit per Mpc$^2$ divided by the median density over the whole COSMOS field. The redshift range is given in each panel and the COSMOS-XS area is shown with the dashed circle.}
\label{fig:overdensities}
\end{figure*}

\def\baselinestretch{1.0}
\begin{deluxetable}{cc}
	\tabletypesize{\footnotesize}
	\tablewidth{0pt}
	\tablecaption{Median overdensity over the COSMOS-XS survey area. The overdensity parameter $o$ is defined as the surface density normalized to the median surface density in that redshift range. The error margins are derived via a bootstrap analysis.}
	
	\tablehead{
		\colhead{Redshift range} &
		\colhead{$o$} 
	}
	
	\startdata
	 
	\vspace{-1.0ex}\\
	0.1 $< \, z \, <$ 0.4 & 1.56 $\pm {0.006}$ \\
    0.4 $< \, z \, <$ 0.6  & 1.25 $\pm {0.004}$ \\
    0.6 $< \, z \, <$ 0.8  & 2.04 $\pm {0.007}$ \\
    0.8 $< \, z \, <$ 1.0  & 1.30 $\pm {0.007}$ \\
    1.0 $< \, z \, <$ 1.3  & 1.02 $\pm {0.004}$ \\
    1.3 $< \, z \, <$ 1.6  & 1.22 $\pm {0.007}$ \\
    1.6 $< \, z \, <$ 2.0  & 1.10 $\pm {0.004}$ \\
    2.0 $< \, z \, <$ 2.5  & 0.89 $\pm {0.005}$ \\
    2.5 $< \, z \, <$ 3.0  & 0.99 $\pm {0.003}$
	\enddata
    
	\label{tab:overdensity}
\end{deluxetable}

We need to consider whether our single pointing of $350 \, \mathrm{arcmin^2}$ covers over-densities that will affect our LF measurements. In particular, the COSMOS field contains a very complex structure located in an extremely narrow redshift slice at $z \, \sim \, 0.73$ \citep{Iovino_2016}. This structure includes a rich X-ray cluster \citep{Finoguenov_2007} and a number of groups \citep{Knobel_2012}. Our field of view covers part of this structure and this can also be seen in Fig.~\ref{fig:covarage} from the large number of sources detected in the redshift slice $0.6 \, < \, z \, < \, 0.8$. In other redshift slices we also cover several X-ray clusters and groups. At $z \, \sim \, 0.5$, $z \, \sim \, 0.9$, and $z \, \sim \, 1.25$ our field of view covers X-ray clusters described by \citet{Finoguenov_2007} and at $z \, \sim \,  0.35$, $z \, \sim \, 0.5$, and $z \, \sim \, 0.8$ our field of view covers groups described by \citet{Knobel_2012}. We also cover part of an under-density or void at $2.0 \, < \, z \, < \, 2.5$ as found by \citet{Krolewski_2018}. 

\citet{Scoville_2013} studied the large-scale structures using a $K_s$-band selected sample of galaxies in the COSMOS field. They estimated the environmental densities within 127 redshift slices out to $z \, < \, 3$ using a Voronoi-based algorithm. Using the established density maps, we are able to estimate the median overdensity in our pointing in the redshift ranges considered. Fig.~\ref{fig:overdensities} shows the density maps as a function of redshift in the COSMOS-XS field of view. The images were made by summing the derived over-densities measured from the individual redshift slices. Table~\ref{tab:overdensity} lists the over-density factors defined as the surface density normalized to the median surface density in that redshift range. If we assume radio galaxies follow the distribution of the $K_s$-band selected galaxies, we can use the over-density factors calculated to scale the measured LFs. Cosmic variance affects, to first order, the measured overall number density and will thus move the radio LF up and down relative to the full COSMOS field. The shape of the LF would be left unchanged. We introduce the over-density factor in Equation~\ref{eq:LF_vmax} as:
\begin{align}
	\Phi(L,z) \, = \, \frac{1}{\Delta \log_{10} L} \sum_{i} \frac{1}{V_{\text{max}, i} \times w_{i}(z) \times o_{i}(z)} \, ,
	\label{eq:LF_overdensity}
\end{align}
where $V_{\text{max}}$ is the comoving volume over which the ith galaxy could be observed, $\Delta \log_{10} L$ is the size of the luminosity bin, $w_{i}$ is the completeness correction factor of the $i$th galaxy and $o_{i}$ is the over-density correction factor of the $i$th galaxy as tabulated in Table~\ref{tab:overdensity}. The equation of the error of the LF in each redshift and luminosity bin (Eq.~\ref{eq:LF_error}) then becomes:

\begin{align}
	\sigma_{\Phi}(L,z) \, = &\, \frac{1}{\Delta \log_{10} L} \notag \\
	&\sqrt{\sum_{i} \left(\frac{1}{\frac{\Omega}{4 \pi} \times V_{\text{max}, i} \times w_{i}(z) \times o_{i}(z)}\right)^{2}} \, . 
	\label{eq:LF_error_overdensity}
\end{align}

The derived over-density corrections are subsequently applied to the derived LF in each redshift and luminosity bin. As the environmental densities are only constrained to $z \, < \, 3$, we do not apply any correction factor for the last redshift bin considered ($3.3 \, < \, z \, < \, 4.6$).

\section{Luminosity functions of star-forming galaxies}
\label{sec:LF of SFGs}
Table~\ref{tab:LF_values} gives the luminosity functions of star-forming galaxies in the COSMOS-XS survey obtained with the $V_{\text{max}}$ method.
\pagebreak
\def\baselinestretch{1.1}
\begin{longtable}{ccc}
	\tabletypesize{\footnotesize}
	\tablecolumns{3}
	\tablecaption{Luminosity functions of star-forming galaxies obtained with the $V_{\text{max}}$ method.}
	
	\tablehead{
		\colhead{Redshift} &
		\colhead{$\log_{10}(L_{1.4 \, \text{ GHz}}/\text{W}\text{Hz}^{-1})$} &
		\colhead{$\log_{10}(\Phi/\text{Mpc}^{-3}\text{dex}^{-1})$} 
	\\}
	
    \vspace{-2.0ex} \\
0.1 $ < \, z \, < $ 0.4  &  20.88 $^{+  0.11 }_{-  0.46 }$ &  $-2.4$ $^{+  0.21 }_{-  0.34 }$ \\
  &  21.13 $^{+  0.16 }_{-  0.13 }$ &  $-2.25$ $^{+  0.11 }_{-  0.11 }$ \\
  &  21.5 $^{+  0.07 }_{-  0.22 }$ &  $-2.31$ $^{+  0.08 }_{-  0.08 }$ \\
  &  21.74 $^{+  0.12 }_{-  0.17 }$ &  $-2.5$ $^{+  0.07 }_{-  0.07 }$ \\
  &  21.94 $^{+  0.2 }_{-  0.08 }$ &  $-2.5$ $^{+  0.07 }_{-  0.07 }$ \\
  &  22.23 $^{+  0.2 }_{-  0.08 }$ &  $-2.9$ $^{+  0.1 }_{-  0.1 }$ \\
  &  22.5 $^{+  0.22 }_{-  0.07 }$ &  $-3.31$ $^{+  0.16 }_{-  0.23 }$ \\
  &  22.88 $^{+  0.12 }_{-  0.17 }$ &  $-3.21$ $^{+  0.14 }_{-  0.2 }$ \\
\\
\pagebreak
\hline \\
0.4 $ < \, z \, < $ 0.6  &  21.68 $^{+  0.12 }_{-  0.05 }$ &  $-2.58$ $^{+  0.19 }_{-  0.29 }$ \\
  &  21.9 $^{+  0.07 }_{-  0.1 }$ &  $-2.56$ $^{+  0.13 }_{-  0.13 }$ \\
  &  22.02 $^{+  0.12 }_{-  0.05 }$ &  $-2.54$ $^{+  0.1 }_{-  0.1 }$ \\
  &  22.24 $^{+  0.07 }_{-  0.1 }$ &  $-2.58$ $^{+  0.09 }_{-  0.09 }$ \\
  &  22.42 $^{+  0.06 }_{-  0.11 }$ &  $-2.84$ $^{+  0.11 }_{-  0.11 }$ \\
  &  22.57 $^{+  0.09 }_{-  0.08 }$ &  $-3.04$ $^{+  0.13 }_{-  0.13 }$ \\
  &  22.68 $^{+  0.14 }_{-  0.03 }$ &  $-3.38$ $^{+  0.19 }_{-  0.29 }$ \\
  &  22.93 $^{+  0.07 }_{-  0.1 }$ &  $-3.78$ $^{+  0.28 }_{-  0.57 }$ \\
  &  23.14 $^{+  0.03 }_{-  0.14 }$ &  $-3.78$ $^{+  0.28 }_{-  0.57 }$ \\
\\
\hline \\
0.6 $ < \, z \, < $ 0.8  &  22.05 $^{+  0.09 }_{-  0.15 }$ &  $-2.47$ $^{+  0.1 }_{-  0.1 }$ \\
  &  22.26 $^{+  0.11 }_{-  0.12 }$ &  $-2.46$ $^{+  0.06 }_{-  0.06 }$ \\
  &  22.51 $^{+  0.09 }_{-  0.14 }$ &  $-2.82$ $^{+  0.06 }_{-  0.06 }$ \\
  &  22.72 $^{+  0.11 }_{-  0.12 }$ &  $-2.95$ $^{+  0.06 }_{-  0.06 }$ \\
  &  22.91 $^{+  0.16 }_{-  0.08 }$ &  $-3.2$ $^{+  0.08 }_{-  0.08 }$ \\
  &  23.14 $^{+  0.16 }_{-  0.07 }$ &  $-3.36$ $^{+  0.1 }_{-  0.1 }$ \\
  &  23.41 $^{+  0.12 }_{-  0.12 }$ &  $-3.95$ $^{+  0.19 }_{-  0.29 }$ \\
  &  23.85 $^{+  0.14 }_{-  0.32 }$ &  $-4.49$ $^{+  0.24 }_{-  0.42 }$ \\
\\
\hline \\
0.8 $ < \, z \, < $ 1.0  &  22.31 $^{+  0.06 }_{-  0.18 }$ &  $-2.39$ $^{+  0.1 }_{-  0.1 }$ \\
  &  22.51 $^{+  0.1 }_{-  0.14 }$ &  $-2.4$ $^{+  0.06 }_{-  0.06 }$ \\
  &  22.69 $^{+  0.14 }_{-  0.09 }$ &  $-2.77$ $^{+  0.07 }_{-  0.07 }$ \\
  &  22.9 $^{+  0.18 }_{-  0.06 }$ &  $-2.92$ $^{+  0.07 }_{-  0.07 }$ \\
  &  23.15 $^{+  0.16 }_{-  0.07 }$ &  $-3.3$ $^{+  0.1 }_{-  0.1 }$ \\
  &  23.37 $^{+  0.18 }_{-  0.06 }$ &  $-3.7$ $^{+  0.16 }_{-  0.23 }$ \\
  &  23.76 $^{+  0.25 }_{-  0.22 }$ &  $-4.55$ $^{+  0.28 }_{-  0.57 }$ \\
  &  24.21 $^{+  0.04 }_{-  0.19 }$ &  $-4.27$ $^{+  0.28 }_{-  0.57 }$ \\
\\
\hline \\
1.0 $ < \, z \, < $ 1.3  &  22.58 $^{+  0.07 }_{-  0.13 }$ &  $-2.86$ $^{+  0.14 }_{-  0.14 }$ \\
  &  22.78 $^{+  0.07 }_{-  0.13 }$ &  $-2.64$ $^{+  0.07 }_{-  0.07 }$ \\
  &  22.95 $^{+  0.1 }_{-  0.1 }$ &  $-2.83$ $^{+  0.07 }_{-  0.07 }$ \\
  &  23.15 $^{+  0.09 }_{-  0.11 }$ &  $-3.22$ $^{+  0.09 }_{-  0.09 }$ \\
  &  23.39 $^{+  0.05 }_{-  0.15 }$ &  $-3.46$ $^{+  0.11 }_{-  0.11 }$ \\
  &  23.51 $^{+  0.14 }_{-  0.06 }$ &  $-3.88$ $^{+  0.17 }_{-  0.26 }$ \\
  &  23.72 $^{+  0.12 }_{-  0.08 }$ &  $-4.06$ $^{+  0.21 }_{-  0.34 }$ \\
  &  23.95 $^{+  0.09 }_{-  0.11 }$ &  $-4.36$ $^{+  0.28 }_{-  0.57 }$ \\
  &  24.17 $^{+  0.07 }_{-  0.13 }$ &  $-4.19$ $^{+  0.24 }_{-  0.42 }$ \\
\\
\hline \\
1.3 $ < \, z \, < $ 1.6  &  22.73 $^{+  0.06 }_{-  0.13 }$ &  $-2.63$ $^{+  0.12 }_{-  0.12 }$ \\
  &  22.87 $^{+  0.11 }_{-  0.08 }$ &  $-2.71$ $^{+  0.08 }_{-  0.08 }$ \\
  &  23.06 $^{+  0.11 }_{-  0.08 }$ &  $-2.85$ $^{+  0.07 }_{-  0.07 }$ \\
  &  23.25 $^{+  0.1 }_{-  0.09 }$ &  $-3.12$ $^{+  0.07 }_{-  0.07 }$ \\
  &  23.47 $^{+  0.07 }_{-  0.12 }$ &  $-3.4$ $^{+  0.09 }_{-  0.09 }$ \\
  &  23.62 $^{+  0.1 }_{-  0.08 }$ &  $-3.6$ $^{+  0.11 }_{-  0.11 }$ \\
  &  23.84 $^{+  0.07 }_{-  0.12 }$ &  $-3.99$ $^{+  0.17 }_{-  0.26 }$ \\
  &  23.95 $^{+  0.15 }_{-  0.03 }$ &  $-3.93$ $^{+  0.16 }_{-  0.23 }$ \\
  &  24.18 $^{+  0.11 }_{-  0.08 }$ &  $-4.08$ $^{+  0.19 }_{-  0.29 }$ \\
\\
\hline \\
1.6 $ < \, z \, < $ 2.0  &  22.99 $^{+  0.08 }_{-  0.15 }$ &  $-2.86$ $^{+  0.11 }_{-  0.11 }$ \\
  &  23.21 $^{+  0.09 }_{-  0.14 }$ &  $-3.1$ $^{+  0.09 }_{-  0.09 }$ \\
  &  23.43 $^{+  0.1 }_{-  0.13 }$ &  $-3.31$ $^{+  0.08 }_{-  0.08 }$ \\
  &  23.62 $^{+  0.14 }_{-  0.09 }$ &  $-3.56$ $^{+  0.09 }_{-  0.09 }$ \\
  &  23.83 $^{+  0.16 }_{-  0.07 }$ &  $-3.93$ $^{+  0.13 }_{-  0.13 }$ \\
  &  24.1 $^{+  0.12 }_{-  0.11 }$ &  $-4.26$ $^{+  0.19 }_{-  0.29 }$ \\
  &  24.4 $^{+  0.51 }_{-  0.18 }$ &  $-4.97$ $^{+  0.24 }_{-  0.42 }$ \\
\\
\hline \\
2.0 $ < \, z \, < $ 2.5  &  23.23 $^{+  0.03 }_{-  0.19 }$ &  $-3.2$ $^{+  0.14 }_{-  0.2 }$ \\
  &  23.39 $^{+  0.1 }_{-  0.13 }$ &  $-3.29$ $^{+  0.12 }_{-  0.12 }$ \\
  &  23.65 $^{+  0.06 }_{-  0.16 }$ &  $-3.57$ $^{+  0.11 }_{-  0.11 }$ \\
  &  23.83 $^{+  0.1 }_{-  0.12 }$ &  $-3.8$ $^{+  0.12 }_{-  0.12 }$ \\
  &  24.0 $^{+  0.16 }_{-  0.06 }$ &  $-4.28$ $^{+  0.19 }_{-  0.29 }$ \\
  &  24.21 $^{+  0.17 }_{-  0.05 }$ &  $-4.28$ $^{+  0.19 }_{-  0.29 }$ \\
  &  24.55 $^{+  0.5 }_{-  0.17 }$ &  $-4.85$ $^{+  0.21 }_{-  0.34 }$ \\
\\
\hline \\
2.5 $ < \, z \, < $ 3.3  &  23.53 $^{+  0.07 }_{-  0.13 }$ &  $-3.52$ $^{+  0.14 }_{-  0.19 }$ \\
  &  23.71 $^{+  0.09 }_{-  0.11 }$ &  $-3.73$ $^{+  0.13 }_{-  0.13 }$ \\
  &  23.92 $^{+  0.07 }_{-  0.13 }$ &  $-3.95$ $^{+  0.13 }_{-  0.13 }$ \\
  &  24.11 $^{+  0.08 }_{-  0.12 }$ &  $-4.07$ $^{+  0.13 }_{-  0.13 }$ \\
  &  24.31 $^{+  0.07 }_{-  0.12 }$ &  $-4.39$ $^{+  0.17 }_{-  0.26 }$ \\
  &  24.46 $^{+  0.12 }_{-  0.08 }$ &  $-4.87$ $^{+  0.28 }_{-  0.57 }$ \\
  &  24.7 $^{+  0.46 }_{-  0.13 }$ &  $-5.04$ $^{+  0.21 }_{-  0.34 }$ \\
\\
\hline \\
3.3 $ < \, z \, < $ 4.6  &  23.67 $^{+  0.23 }_{-  0.13 }$ &  $-3.97$ $^{+  0.21 }_{-  0.34 }$ \\
  &  24.1 $^{+  0.17 }_{-  0.19 }$ &  $-4.5$ $^{+  0.16 }_{-  0.23 }$ \\
  &  24.4 $^{+  0.59 }_{-  0.13 }$ &  $-4.99$ $^{+  0.15 }_{-  0.21 }$ \\

	\label{tab:LF_values}
\end{longtable}

\section{Posterior distributions}
\label{sec:cornerplot}
\begin{figure*}
    \centering
    \subfigure
    {
        \includegraphics[width=0.63\columnwidth]{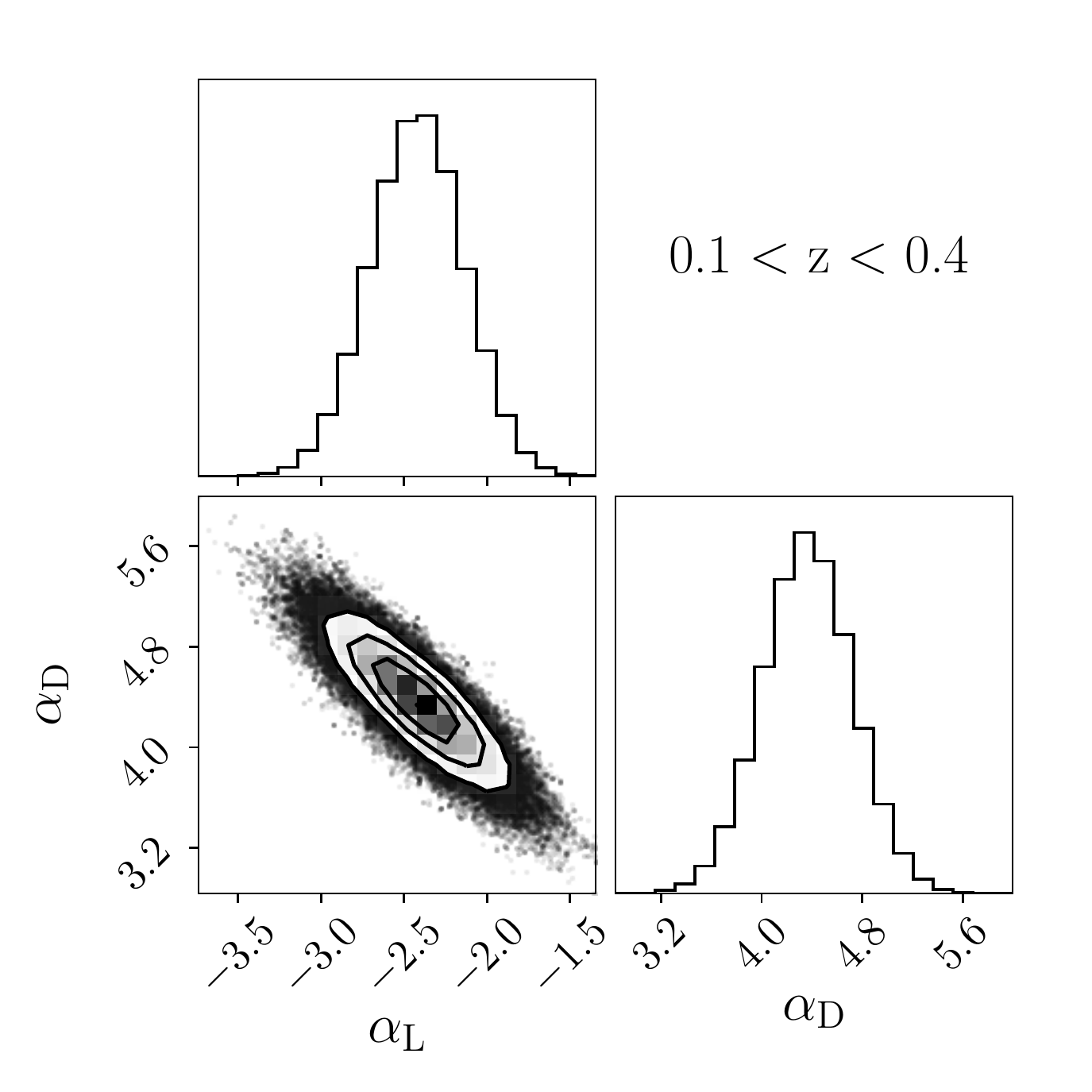}
    }
    \subfigure
    {
        \includegraphics[width=0.63\columnwidth]{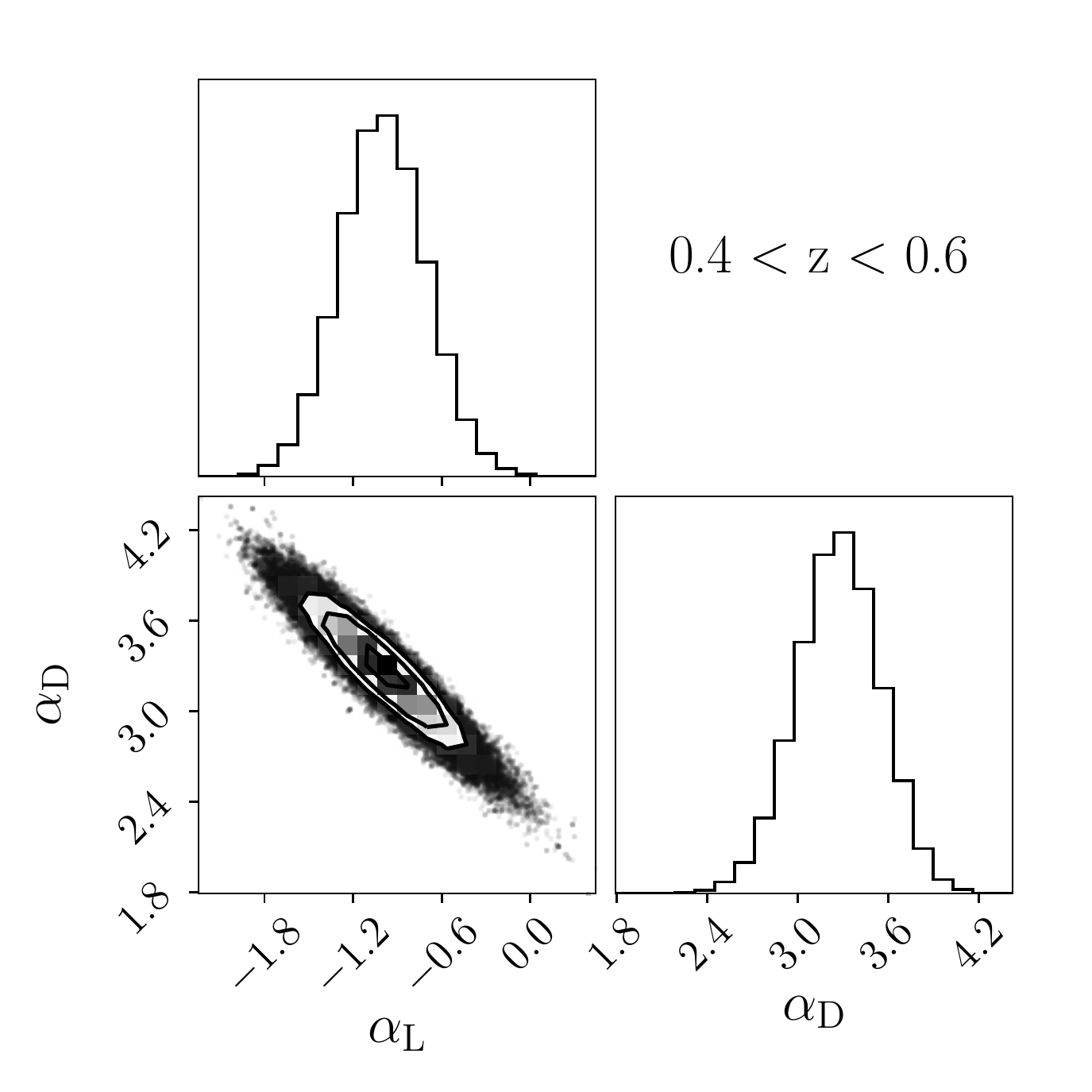}
    }
    \subfigure
    {
        \includegraphics[width=0.63\columnwidth]{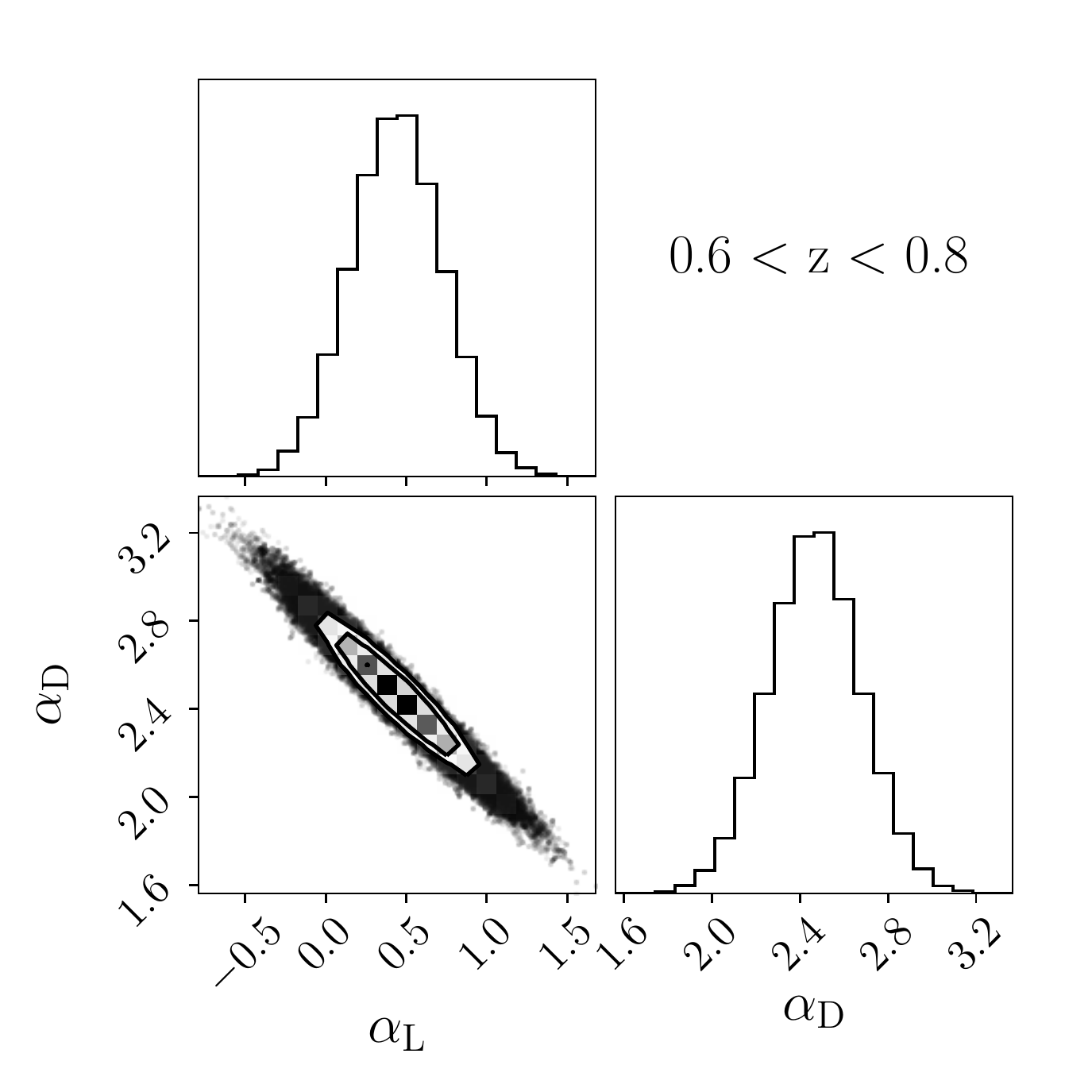}
    }
    \subfigure
    {
        \includegraphics[width=0.63\columnwidth]{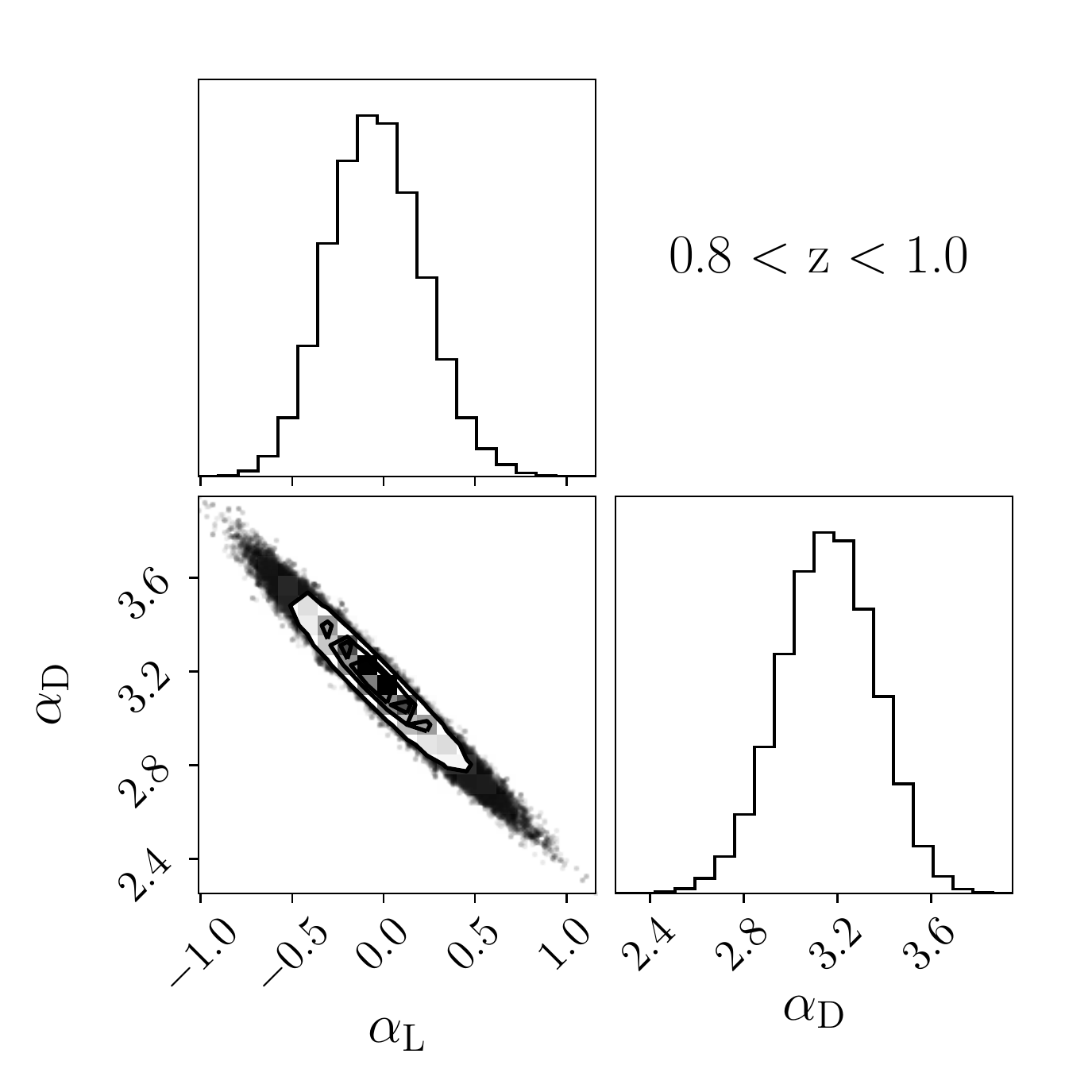}
    }
    \subfigure
    {
        \includegraphics[width=0.63\columnwidth]{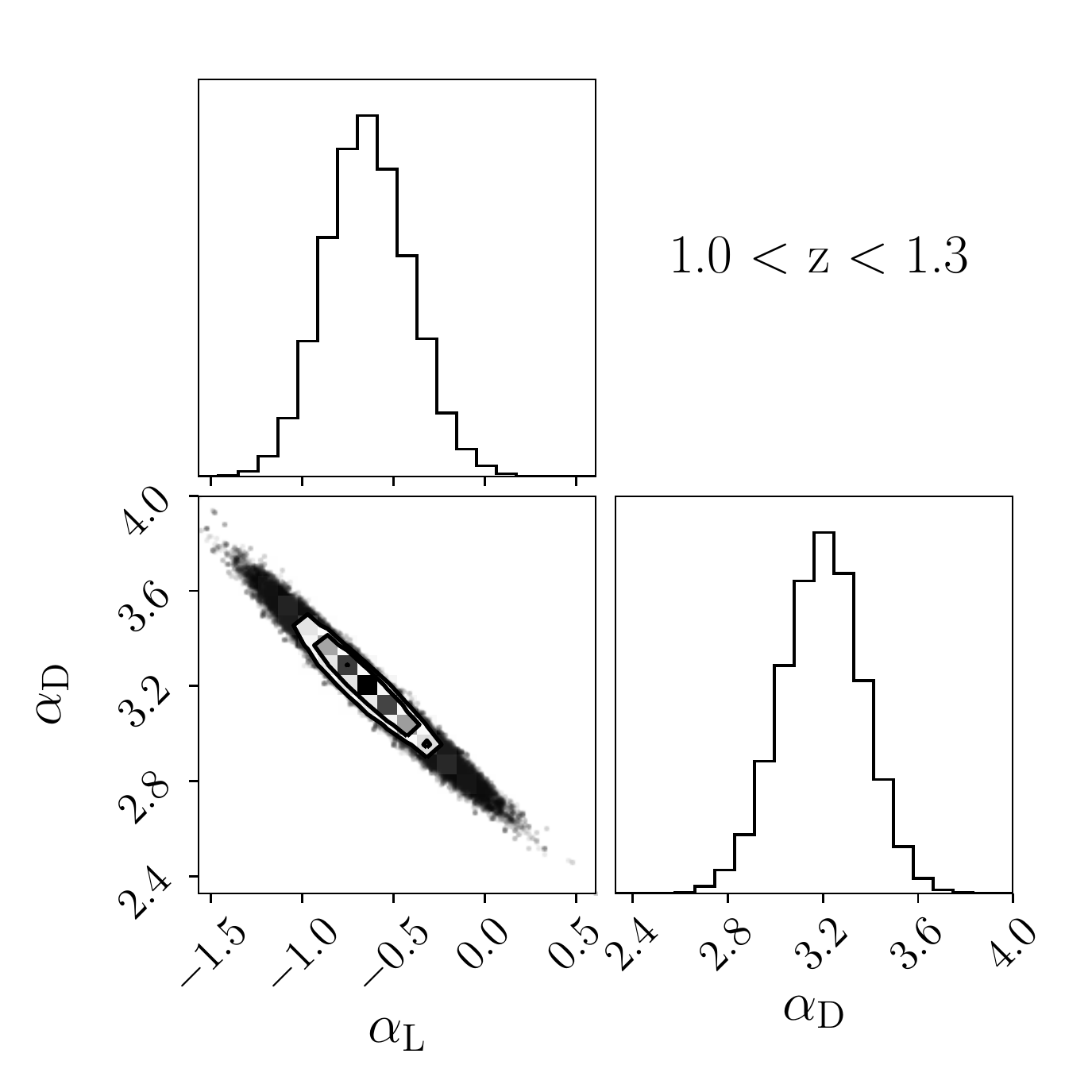}
    }
    \subfigure
    {
        \includegraphics[width=0.63\columnwidth]{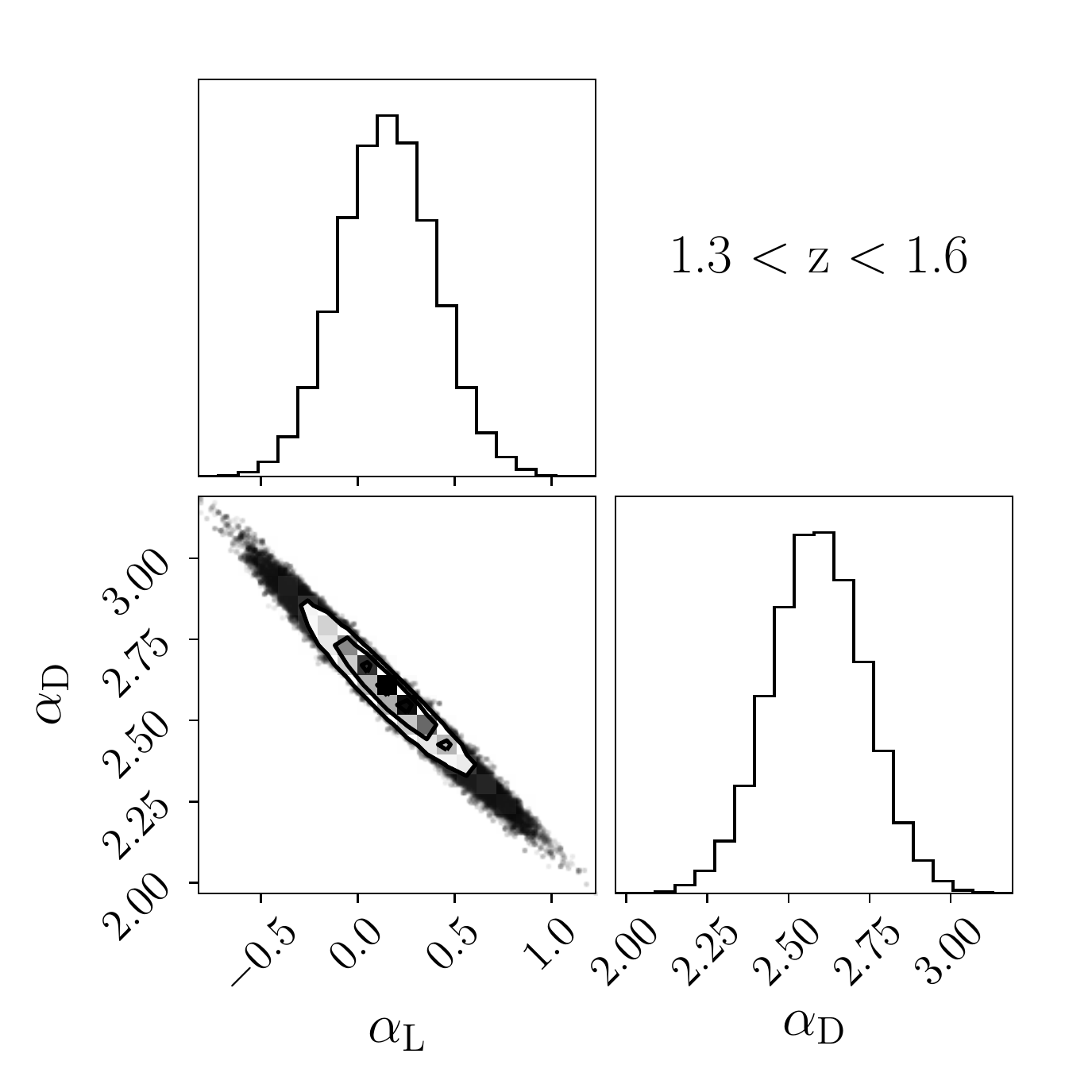}
    }
    \subfigure
    {
        \includegraphics[width=0.63\columnwidth]{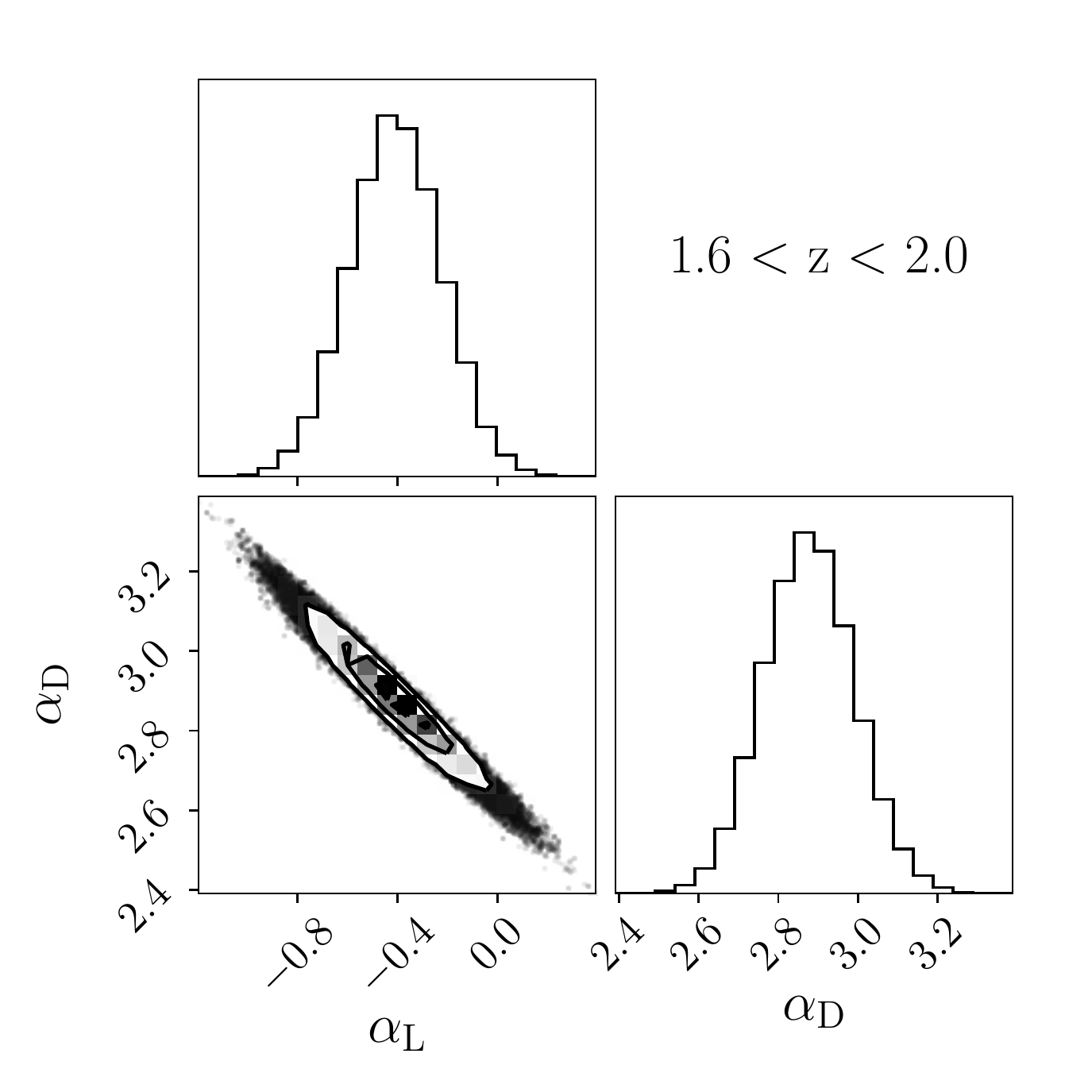}
    }
    \subfigure
    {
        \includegraphics[width=0.63\columnwidth]{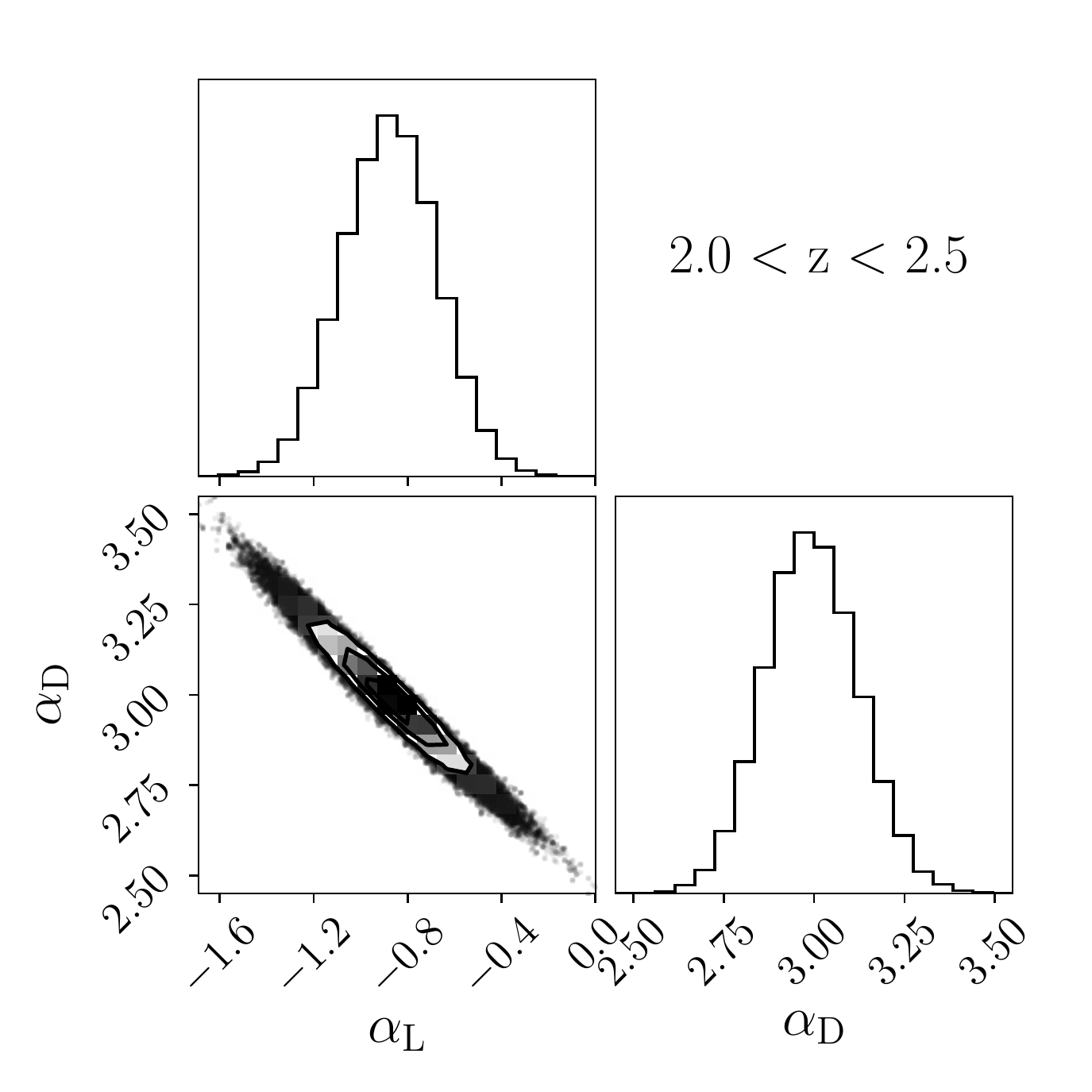}
    }
    \subfigure
    {
        \includegraphics[width=0.63\columnwidth]{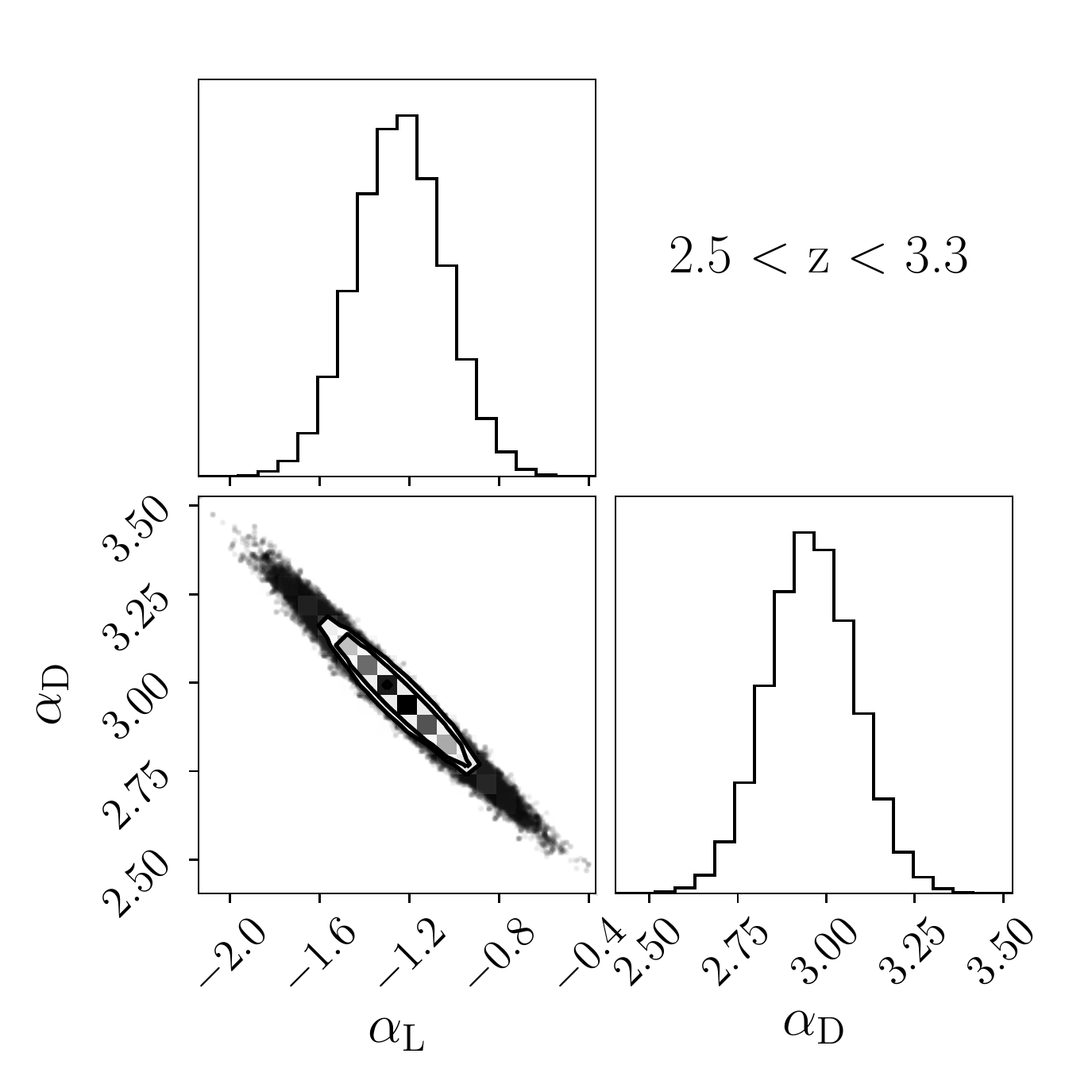}
    }
    \subfigure
    {
        \includegraphics[width=0.63\columnwidth]{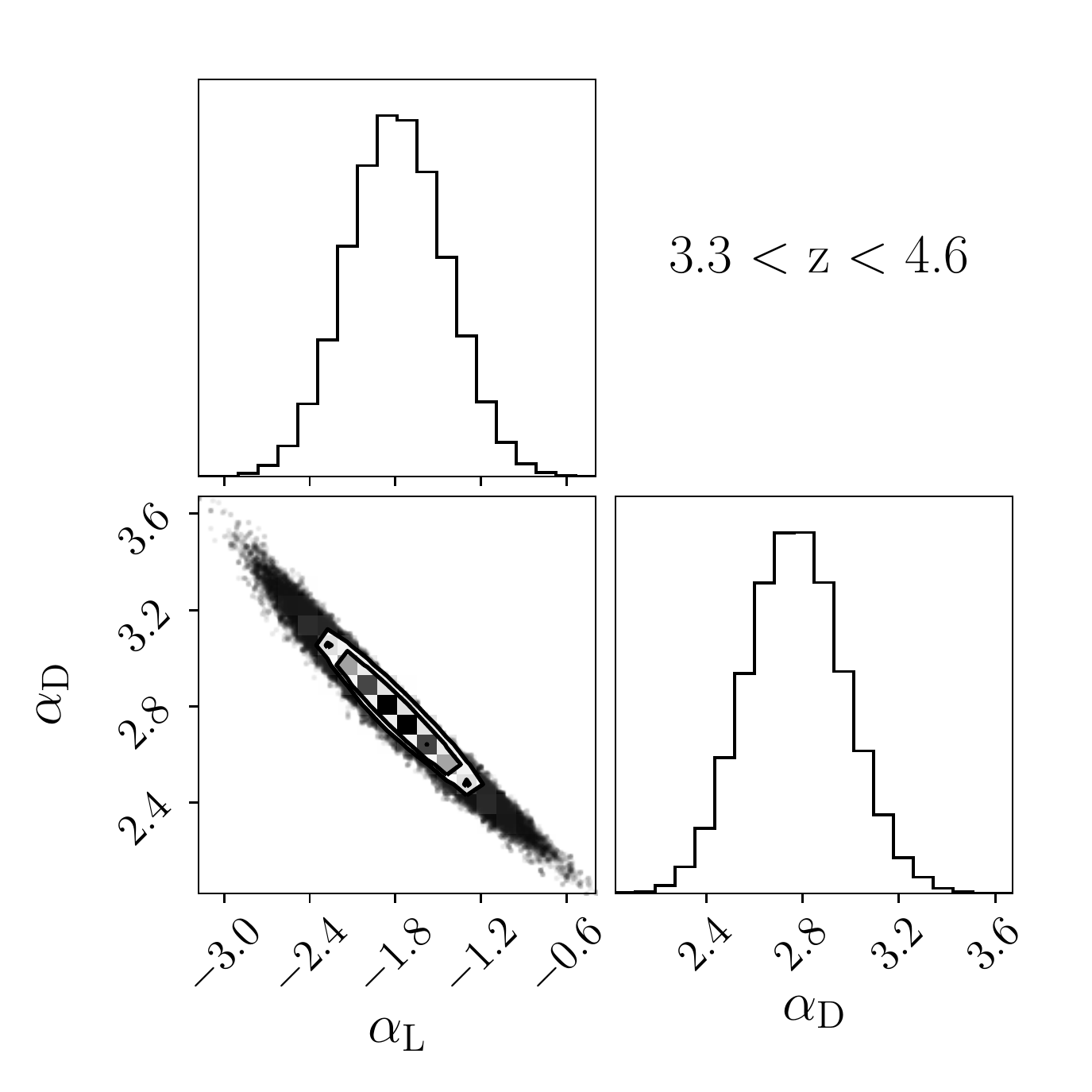}
    }
    \caption{Corner plots showing the two dimensional posterior probability distributions of $\alpha_{\text{L}}$ and $\alpha_{\text{D}}$ for the density+luminosity evolution fitted to the COSMOS-XS + VLA-COSMOS 3 GHz samples. The marginalized distributions for each parameter is shown independently in the histograms. For all redshift bins the parameters have well-defined peaks.}
    \label{fig:all_corner_plots}
\end{figure*}

Fig.~\ref{fig:all_corner_plots} shows the two dimensional posterior probability distributions of $\alpha_{\text{L}}$ and $\alpha_{\text{D}}$ for the density+luminosity evolution fitted to the combination of the COSMOS-XS survey and the VLA-COSMOS $\mathrm{3 \, GHz}$ large project. The marginalized distributions for each parameter is shown independently in the histograms.

\newpage
\bibliographystyle{apj}
\bibliography{ref}

%%%%%%%%%%%%%%%%%%%%%%%%%%%%%%%%%%%%%%%%%%%%%%%%%%
%TC:endignore
% Don't change these lines
%\bsp	% typesetting comment
\label{lastpage}
\end{document}